\documentclass[aps,nofootinbib,preprintnumbers]{revtex4}
\usepackage{eurosym}
\usepackage{CJK}
\usepackage{lipsum}
\usepackage{amsfonts}
\usepackage{amsmath}
\usepackage{amssymb}
\usepackage[english]{babel}
\usepackage{graphicx}
\usepackage{epsfig}
\usepackage{bm}
\usepackage{verbatim}
\usepackage[utf8]{inputenc}
\usepackage{booktabs}
\usepackage{multirow}
\usepackage{subfig}
\usepackage{slashed}
\usepackage[colorlinks=true,urlcolor=red,citecolor=red]{hyperref}
\usepackage[font=small]{caption}
\usepackage{float}
\usepackage{placeins}
\usepackage[utf8]{inputenc}
\usepackage{bbm}

\newcommand{\mathsym}[1]{{}}

\newcommand{\be}{\begin{equation}}
\newcommand{\ee}{\end{equation}}
\newcommand{\bea}{\begin{eqnarray}}
\newcommand{\eea}{\end{eqnarray}}
\newcommand{\ben}{\begin{enumerate}}
\newcommand{\een}{\end{enumerate}}
\newcommand{\bit}{\begin{itemize}}
\newcommand{\eit}{\end{itemize}}
\newcommand{\bde}{\begin{widetext}}
\newcommand{\ede}{\end{widetext}}

\newcommand{\bc}{\begin{center}}
\newcommand{\ec}{\end{center}}

\DeclareUnicodeCharacter{2212}{-}
\input{tcilatex}
\begin{document}

\title{Universal Inverse seesaw mechanism as a source of the SM fermion mass
hierarchy}
\author{A. E. C\'arcamo Hern\'andez}
\email{antonio.carcamo@usm.cl}
\affiliation{{Universidad T\'ecnica Federico Santa Mar\'{\i}a, Casilla 110-V, Valpara%
\'{\i}so, Chile}}
\affiliation{{Centro Cient\'{\i}fico-Tecnol\'ogico de Valpara\'{\i}so, Casilla 110-V,
Valpara\'{\i}so, Chile}}
\affiliation{{Millennium Institute for Subatomic physics at high energy frontier -
SAPHIR, Fernandez Concha 700, Santiago, Chile}}
\author{D. T. Huong}
\email{dthuong@iop.vast.vn}
\affiliation{Institute of Physics, VAST, 10 Dao Tan, Ba Dinh, Hanoi, Vietnam}
\author{Ivan Schmidt}
\email{ivan.schmidt@usm.cl}
\affiliation{{Universidad T\'ecnica Federico Santa Mar\'{\i}a, Casilla 110-V, Valpara%
\'{\i}so, Chile}}
\affiliation{{Centro Cient\'{\i}fico-Tecnol\'ogico de Valpara\'{\i}so, Casilla 110-V,
Valpara\'{\i}so, Chile}}
\affiliation{{Millennium Institute for Subatomic Physics at High-Energy Frontier
(SAPHIR), Fern\'andez Concha 700, Santiago, Chile}}
\date{\today }

\begin{abstract}
We build a renormalizable theory where the inverse seesaw mechanism explains
the pattern of SM fermion masses. To the best of our knowledge, our model
corresponds to the first implementation of the inverse seesaw mechanism for
the charged fermion sector. In our theory, the inverse seesaw mechanism is
implemented at the tree and one-loop levels in order to generate the masses
for the second and first families of the SM charged fermions, respectively.
The third family of SM charged fermions obtain tree-level masses from the
Higgs doublets $\phi _{1}$ (for the top quark) and $\phi _{2}$ (for the
bottom quark and tau lepton). The masses of the active light neutrinos are
generated from a two-loop level inverse seesaw mechanism. Our model
successfully explains the observed SM fermion mass hierarchy, the tiny
masses of the active light neutrinos, contains the necessary means for
efficient leptogenesis and is in accordance with the constraints resulting
from meson oscillations, as well as with the measured values of the observed
dark matter relic density and of the muon and electron anomalous magnetic
moments.
\end{abstract}

\maketitle

\section{\label{intro}Introduction}

Despite the remarkable success of the Standard Model (SM) in describing the
strong and electroweak interactions with a high degree of accuracy, as
confirmed by the experiments at the Large Hadron Collider (LHC), it does not
address several questions. One of them is the unexplained hierarchy of the
SM fermion masses, which extends over a range of 13 orders of magnitude,
from the light active neutrino mass scale up to the top quark mass. Other
unaddressable issues of the SM are the number of SM fermion families, the
current amount of dark matter relic density, lepton asymmetry of the
Universe, and the muon and electron anomalous magnetic moments.

Models with extended symmetries, enlarged particle content, and radiative
seesaw mechanisms, are frequently used to tackle the limitations of the SM. 
\cite{Davidson:1987mh, Davidson:1989bx, Davidson:1987tr, Balakrishna:1988ks, Ma:1988fp, Ma:1989ys, Ma:1990ce, Ma:1998dn,
Kitabayashi:2000nq, Ma:2006km, Dong:2006gx, Chang:2006aa, Gu:2007ug,
Ma:2008cu, Sierra:2008wj, Nardi:2011jp, Huong:2012pg, Restrepo:2013aga,
Ma:2013yga, Ma:2013mga, Hernandez:2013mcf, Hernandez:2013dta, Okada:2013iba,
Sierra:2014rxa, Campos:2014lla, Boucenna:2014ela, Hernandez:2015hrt,
Aranda:2015xoa, Restrepo:2015ura, Longas:2015sxk, Fraser:2015zed,
Fraser:2015mhb, Okada:2015bxa, Wang:2015saa, Sierra:2016qfa,
Arbelaez:2016mhg, Nomura:2016emz, Nomura:2016fzs, Kownacki:2016hpm,
Kownacki:2016pmx, Nomura:2016ezz, Camargo-Molina:2016yqm,
Camargo-Molina:2016bwm, vonderPahlen:2016cbw, Bonilla:2016diq, Gu:2016xno,
Das:2017ski, Nomura:2017emk, Nomura:2017vzp, Nomura:2017tzj, Wang:2017mcy,
Bernal:2017xat, CarcamoHernandez:2017kra, CarcamoHernandez:2017cwi,
Ma:2017kgb, Cepedello:2017eqf, Dev:2018pjn, CarcamoHernandez:2018hst,
Rojas:2018wym, Nomura:2018vfz, Reig:2018mdk, Bernal:2018aon, Calle:2018ovc,
Aranda:2018lif, Cepedello:2018rfh, CarcamoHernandez:2018vdj, Ma:2018zuj,
Ma:2018bow, Li:2018aov, Arnan:2019uhr, Ma:2019byo, Ma:2019iwj,
CarcamoHernandez:2019xkb, Ma:2019yfo, Nomura:2019yft, Nomura:2019vqc,
Nomura:2019jxj, CentellesChulia:2019gic, Bonilla:2018ynb, Pramanick:2019oxb,
Arbelaez:2019wyz, Avila:2019hhv, CarcamoHernandez:2019cbd,
CarcamoHernandez:2019lhv, Arbelaez:2019ofg, CarcamoHernandez:2020pnh,
CarcamoHernandez:2020ney, CarcamoHernandez:2020pxw,
CarcamoHernandez:2020owa, CarcamoHernandez:2020ehn,
Hernandez:2021uxx,CarcamoHernandez:2021iat,CarcamoHernandez:2021qhf,CarcamoHernandez:2021tlv,Abada:2021yot,Bonilla:2021ize,Hernandez:2021zje}%
. Furthermore, several extensions of the SM have been constructed to explain
the experimental value of the muon anomalous magnetic moment \cite%
{Kiritsis:2002aj,Appelquist:2004mn,Giudice:2012ms,Omura:2015nja,Falkowski:2018dsl,Crivellin:2018qmi,Allanach:2015gkd,Padley:2015uma,Chen:2016dip,Raby:2017igl,Chiang:2017tai,Chen:2017hir,Megias:2017dzd,Davoudiasl:2018fbb,Liu:2018xkx,CarcamoHernandez:2019xkb,Nomura:2019btk,Kawamura:2019rth,Bauer:2019gfk,Botella:2018gzy,Han:2018znu,Wang:2018hnw,Dutta:2018fge,Badziak:2019gaf,Endo:2019bcj,Hiller:2019mou,CarcamoHernandez:2019ydc,CarcamoHernandez:2019lhv,Kawamura:2019hxp,Sabatta:2019nfg,Chen:2020tfr,CarcamoHernandez:2020pxw,Iguro:2019sly,Li:2020dbg,Arbelaez:2020rbq,Hiller:2020fbu,Jana:2020pxx,deJesus:2020ngn,deJesus:2020upp,Hati:2020fzp,Botella:2020xzf,Dorsner:2020aaz,Calibbi:2020emz,Dinh:2020pqn,Jana:2020joi,Chun:2020uzw,Chua:2020dya,Daikoku:2020nhr,Banerjee:2020zvi,Chen:2020jvl,Bigaran:2020jil,Kawamura:2020qxo,Endo:2020mev,Iguro:2020rby,Yin:2020afe,Chen:2021rnl,Athron:2021iuf,Arcadi:2021cwg,Das:2021zea,Yin:2021yqy,Yin:2021mls,Chiang:2021pma,Escribano:2021css,Zhang:2021gun,Yang:2021duj,Li:2021lnz,Hernandez:2021uxx,Hernandez:2021tii,Hernandez:2021mxo,CarcamoHernandez:2021iat,CarcamoHernandez:2021tlv,CarcamoHernandez:2021qhf,Bonilla:2021ize,Saez:2021qta,Yu:2021suw,Chowdhury:2021tnm,Zhang:2021dgl,Perez-Martinez:2021zjj,Jueid:2021avn}%
, an anomaly not explained by the SM and recently confirmed by the Muon $g-2$
experiment at FERMILAB \cite{Abi:2021gix}.

Intending to address the drawbacks as mentioned earlier of the SM, we
propose an extension of the Two Higgs Doublet Model (2HDM) with enlarged
particle spectrum and symmetries, which allows for a successful
implementation of the inverse seesaw mechanism to explain the SM fermion
mass hierarchy. Unlike most of the works considered in the literature, in
the proposed theory the inverse seesaw mechanism is implemented not only for
the neutrino sector, but also for the charged fermion sector. Specifically,
the charged fermions of the first and the second families receive masses via
one-loop and tree-levels inverse seesaw mechanisms, respectively. For the
third family, the charged fermions obtain a mass at tree-level, namely the
t-quark mass depends on the VEV of the Higgs doublet, $\phi _{1}$, while the
masses of both tau-lepton and b-quark depend on the VEV of $\phi _{2}$. The
light active neutrinos gain masses at two loop level. 
The content of this paper goes as follows. In Sec. \ref{model} we describe
our proposed model. The implications of the model in the SM fermion mass
hierarchy are discussed in section \ref{fermionmasses}, while we study the
new contributions to the muon and electron anomalous magnetic moments in Sec.%
\ref{gminus2sec}. Meson mixings are analyzed in Sec.\ref{FCNC}, and the
constraints of our model in dark matter and leptogenesis are discussed in
Secs.\ref{DM} and \ref{leptogenesis}. We conclude in section \ref%
{conclusions}.

\section{The model}

\label{model} We start this section by explaining the reasoning behind the
inclusion of extra scalars, fermions, and symmetries required for the
implementation of tree and one-loop level inverse seesaw mechanisms to
generate the masses of the second and first families of SM charged fermions,
respectively, and a two-loop level inverse seesaw mechanism to produce the
tiny active neutrino masses. In our model, which is an extended 2HDM theory,
the top quark mass will arise at tree level from a renormalizable Yukawa
interaction involving a $SU\left( 2\right) $ Higgs doublet, i.e., $\phi _{1}$%
. In contrast, the bottom quark and tau lepton will get tree-level masses
from the second $SU\left( 2\right) $ Higgs doublet $\phi _{2}$. Such
required Yukawa interactions, which generate the tree level masses for the
top and bottom quarks as well as for the tau lepton, are: 
\begin{equation}
\overline{q}_{3L}\widetilde{\phi }_{1}u_{iR},\hspace{1.5cm}\overline{q}%
_{3L}\phi _{2}d_{iR},\hspace{1.5cm}\overline{l}_{iL}\phi _{2}l_{3R},\hspace{%
1.5cm}i=1,2,3
\end{equation}%
The $SU\left( 2\right) $ Higgs doublets $\phi _{1}$ and $\phi _{2}$ need to
be distinguished by a symmetry, which can be a $U\left( 1\right) _{X}$\
local symmetry, assumed to be non universal in the quark
sector, as it will be shown below. Such symmetry will distinguish the left
handed quark doublets $q_{nL}$ ($n=1,2$) from the $q_{3L}$. Furthermore, a
discrete symmetry $Z_{4}$ is required to allow only the above given Yukawa
interactions, thus allowing to forbid the operators: 
\begin{equation}
\overline{q}_{nL}\widetilde{\phi }_{2}u_{iR},\hspace{1.5cm}\overline{q}%
_{nL}\phi _{1}d_{iR},\hspace{1.5cm}\overline{l}_{iL}\phi _{2}l_{nR},\hspace{%
1.5cm}i=1,2,3,\hspace{1.5cm}n=1,2
\end{equation}
which would give rise to Dirac type masses for the first two generations of
SM fermions. Furthermore, the $Z_{4}$ discrete symmetry, together with the $%
U\left( 1\right) _{X}$ gauge symmetry, will also permit the implementation
of tree and one-loop level inverse seesaw mechanisms to generate the masses
for the second and first generation of SM charged fermions, respectively.
The advantage of the local $U\left( 1\right) _{X}$\ gauge symmetry, with
respect to a cyclic symmetry, is that it allows more freedom in the particle
assignments. The presence of this symmetry means that heavy non SM fermions
can be produced via a Drell-Yan portal mediated by a heavy $Z^{\prime }$
gauge boson. The masses of the active light neutrinos, obtained from an
inverse seesaw mechanism when the full neutrino mass matrix is expressed in
the basis $\left( \nu _{L},\nu _{R}^{C},N_{R}^{C}\right) $, has the
following structure: 
\begin{equation}
M_{\nu }=\left( 
\begin{array}{ccc}
0_{3\times 3} & m_{\nu D} & 0_{3\times 3} \\ 
m_{\nu D}^{T} & 0_{3\times 3} & M \\ 
0_{3\times 3} & M^{T} & \mu%
\end{array}%
\right) ,  \label{Mnufull}
\end{equation}

where $\nu _{iL}$ ($i=1,2,3$) correspond to the active neutrinos, whereas $%
\nu _{iR}$ and $N_{iR}$ ($i=1,2,3$) are the sterile neutrinos. Furthermore,
the entries of the full neutrino mass matrix should fulfill the hierarchy $%
\mu _{ij}<<\left( m_{\nu D}\right) _{ij}<<M_{ij}$ ($i,j=1,2,3$). It is worth
mentioning that in the case $\mu =0$, the light active neutrinos remain
massless and the sterile neutrinos $\nu _{iR}$ and $N_{iR}$ become
degenerate, forming Dirac neutrinos whose corresponding mass matrix is $M$,
and their contribution to the light active neutrino masses vanishes. Thus,
by analogy with the neutrino sector, the implementation of the tree-level
inverse seesaw mechanism for the SM charged fermions requires that its
corresponding mass matrix in the basis $(\overline{f}_{1L},\overline{f}_{2L},%
\overline{f}_{3L},\overline{F}_{L},\overline{\widetilde{F}}_{L})$-$%
(f_{1R},f_{2R},f_{3R},F_{R},\widetilde{F}_{R})$ should have the form:

\begin{equation}
M_{F}=\left( 
\begin{array}{ccc}
0_{3\times 3} & F_{F} & 0_{3\times 1} \\ 
G_{F}^{T} & 0 & X_{F} \\ 
0_{1\times 3} & Y_{F} & m_{F}%
\end{array}%
\right) ,  \label{MF}
\end{equation}

where $f_{i}$ ($i=1,2,3$) correspond to the SM charged fermions, whereas $F$
and $\widetilde{F}$ are the exotic charged fermions. Let us
note that in the limit $m_{F}\rightarrow 0$ ($F=T,D,E$), the model under
consideration has an accidental $U_{\mathbf{\tciFourier }}\left( 1\right) $
symmetry under which the charges of the $f_{iL}$, $f_{iR}$, $%
F_{L}$, $F_{R}$, $\widetilde{F}_{L}$, $\widetilde{F}_{R}$ are given by: 
\begin{equation}
Q_{U_{\mathbf{\tciFourier }}\left( 1\right) }\left( f_{iL}\right) =Q_{U_{%
\mathbf{\tciFourier }}\left( 1\right) }\left( \widetilde{F}_{L}\right)
=Q_{U_{\mathbf{\tciFourier }}\left( 1\right) }\left( F_{R}\right) =a,\hspace{%
0.5cm}\hspace{0.5cm}Q_{U_{\mathbf{\tciFourier }}\left( 1\right) }\left(
f_{iR}\right) =Q_{U_{\mathbf{\tciFourier }}\left( 1\right) }\left( 
\widetilde{F}_{R}\right) =Q_{U_{\mathbf{\tciFourier }}\left( 1\right)
}\left( F_{L}\right) =b,\hspace{0.5cm}\hspace{0.5cm}a\neq b.  \label{U1F}
\end{equation}%

It is worth mentioning that unlike the neutrino sector, in the charged
fermion sector the tree level inverse seesaw mechanism is only implemented
to generate the masses of the second family of SM charged fermions. This is
due to the fact that the third family of SM charged fermions obtain their
masses from renormalizable interactions involving the $SU\left( 2\right) $
Higgs doublets $\phi _{1}$ and $\phi _{2}$, whereas the first family of SM
charged fermions get their masses from radiative corrections, as it will be
shown below. Because of this reason there is only one family of exotic
charged fermions $F$ and $\widetilde{F}$ involved in the tree level inverse
seesaw mechanism, whereas the sterile neutrino spectrum resulting from Eq. (%
\ref{Mnufull}) is composed of three copies of the $\nu _{R}^{C}$ and $%
N_{R}^{C}$ fields. It is worth mentioning that the Dirac type masses of the
first two generations of SM charged fermions that would result from the $%
\overline{q}_{nL}\widetilde{\phi }_{2}u_{iR}$ and $\overline{q}_{nL}\phi
_{1}d_{iR}$ operators will be forbidden by the $Z_{4}$ discrete symmetry.

To generate the charged fermion mass matrix of Eq. (\ref{MF}), one has to
include the following operators: 
\begin{eqnarray}
&&\overline{q}_{nL}\widetilde{\phi }_{2}U_{R},\hspace{1.5cm}\overline{T}%
_{L}\eta U_{R},\hspace{1.5cm}m_{T}\overline{T}_{L}T_{R},\hspace{1.5cm}%
\overline{U}_{L}\chi T_{R},\hspace{1.5cm}\overline{U}_{L}\sigma ^{\ast
}u_{iR},\hspace{1.5cm}i=1,2,3,  \notag \\
&&\overline{q}_{nL}\phi _{1}D_{1R},\hspace{1.5cm}\overline{B}_{L}\eta ^{\ast
}D_{1R},\hspace{1.5cm}m_{B}\overline{B}_{L}B_{R},\hspace{1.5cm}\overline{D}%
_{1L}\chi ^{\ast }B_{R},\hspace{1.5cm}\overline{D}_{1L}\sigma d_{iR},
\label{opcf} \\
&&\overline{l}_{iL}\phi _{2}E_{1R},\hspace{1.5cm}\overline{E}_{2L}\rho
^{\ast }E_{1R},\hspace{1.5cm}m_{E}\overline{E}_{2L}E_{2R},\hspace{1.5cm}%
\overline{E}_{1L}\rho E_{2R},\hspace{1.5cm}\overline{E}_{1L}Sl_{nR},\hspace{%
1.5cm}n=1,2,  \notag
\end{eqnarray}%
Here $q_{iL}$, $l_{iL}$ ($i=1,2,3$) are the $SU\left( 2\right) $ left handed
SM quark and lepton doublets, respectively, whereas $u_{iR}$, $d_{iR}$ and $%
l_{iR}$ stand for the right SM up-type quarks, down type quarks, and
right-handed leptons. Furthermore, the SM fermion sector has to be extended
to include the exotic fermions: up type quarks $U$, $T$, down type quarks $%
D_{1}$, $D_{2}$, $B$, charged exotic leptons $E_{1}$, $E_{2}$, as well as
the right-handed Majorana neutrinos $\nu _{iR}$ ($i=1,2,3$), $N_{iR}$, $\Psi
_{mR}$ and $\Omega _{nR}^{C}$, in singlet representations under $SU\left(
2\right) _{L}$, with appropriate $U\left( 1\right) _{X}$ charges (to be
specified below), which will allow the implementation of inverse seesaw
mechanisms to explain the SM charged fermion mass hierarchy and the tiny
values of the light active neutrino masses, respectively, in a way
consistent with the cancellation of chiral anomalies. Up to this point, the
first generation of SM charged fermions remain massless up to loop
corrections. In order to generate the masses for the first generation of SM
charged fermions, the one-loop level corrections to the SM charged fermion
mass matrices should have different vertices used for generating the
resulting tree-level mass matrices arising from the inverse seesaw. The most
economical solution to this problem requires considering one-loop level
corrections involving electrically charged scalars running in the internal
lines of the loop. So we have to introduce extra electrically charged scalar
singlets in the scalar spectrum, generating the following interactions. 
\begin{equation}
\overline{U}_{L}\zeta _{1}^{+}d_{iR},\hspace{1.5cm}\overline{D}_{1L}\zeta
_{1}^{-}u_{iR},\hspace{1.5cm}\overline{N_{iR}^{C}}\zeta _{2}^{+}l_{nR},%
\hspace{1.5cm}i=1,2,3,\hspace{1.5cm}n=1,2.
\end{equation}

Then, the up quark mass can be generated at one-loop level from the first
four operators of the second line of Eq. (\ref{opcf}), as well as from the $%
\overline{D}_{1L}\zeta _{1}^{-}u_{iR}$ interaction. Similarly, one-loop
diagrams which generate a mass for down quark, arise from the first four
operators of the first line of Eq. (\ref{opcf}), as well as from the $%
\overline{U}_{L}\zeta _{1}^{+}d_{iR}$ interaction. In order to close the
one-loop Feynman diagrams giving rise to the first generation SM charged
fermions, the following scalar interactions are required: 
\begin{equation}
\varepsilon _{ab}\phi _{1}^{a}\phi _{2}^{b}\zeta _{3}^{-}\sigma ^{\ast },%
\hspace{1.5cm}\zeta _{1}^{+}\zeta _{3}^{-}S^{2},\hspace{1.5cm}\zeta
_{2}^{+}\zeta _{3}^{-}\sigma ^{\ast },\hspace{1.5cm}a,b=1,2.
\label{scalarsop}
\end{equation}

Moreover, the operators required for the implementation of the two loop
level inverse seesaw mechanism that produces the tiny active neutrino masses
are: 
\begin{equation}
\overline{l}_{iL}\widetilde{\phi }_{2}\nu _{jR},\hspace{0.6cm}\nu
_{iR}\sigma ^{\ast }\overline{N_{kR}^{C}},\hspace{0.6cm}\overline{N}%
_{rR}\Psi _{mR}^{C}\varphi _{1},\hspace{0.6cm}\overline{\Psi }_{nR}\varphi
_{2}\Omega _{mR}^{C},\hspace{0.6cm}\overline{\Omega }_{nR}\Omega
_{mR}^{C}\eta ,\hspace{0.6cm}i,j,k,r=1,2,3,\hspace{0.6cm}n,m=1,2,
\label{opnl}
\end{equation}%
It is worth mentioning that the first two operators of Eq. (\ref{opnl}),
together with the $\overline{N_{iR}^{C}}\zeta _{2}^{+}l_{nR}$, $\varepsilon
_{mn}\phi _{1}^{m}\phi _{2}^{n}\zeta _{3}^{-}\sigma ^{\ast }$ and $\zeta
_{2}^{+}\zeta _{3}^{-}\sigma ^{\ast }$ interactions, are crucial for
generating a nonvanishing one-loop level electron mass.

Implementing the above-described inverse seesaw mechanism requires to add a
preserved $Z_{2}$ symmetry, under which the right-handed Majorana neutrinos $%
\Psi _{nR}$, the exotic down type quark fields $D_{2L}$, $D_{2R}$\ and the
gauge scalar singlets $\varphi _{1}$ and $\varphi _{2}$ are charged, whereas
the remaining fields are neutral. Additionally, we have extended the scalar
sector of our 2HDM theory by including the electrically neutral gauge
singlet scalars $\sigma $, $\chi $, $\eta $, $\rho $, $S$, $\varphi _{1}$, $%
\varphi _{2}$, as well as the electrically charged gauge singlet scalars $%
\zeta _{1}^{+}$, $\zeta _{2}^{+}$ and $\zeta _{3}^{+}$. Notice that the
electrically neutral gauge singlet scalars $\sigma $ and $S$ are needed to
generate mixings between the right-handed SM charged fermions and
left-handed heavy fermionic fields. On the other hand, the singlet scalars $%
\chi $, $\eta $, and $\rho $ generate mixings between the heavy charged
exotic fermionic fields. Those mixings are crucial for implementing a
tree-level universal seesaw mechanism that produces the masses for the
second generation of SM charged fermions. As mentioned earlier, we will be
able to implement seesaw mechanisms useful for explaining the SM fermion
mass hierarchy by suitable charge assignments, specified below. Due to those
mentioned above exotic charged fermion spectrum, the SM charged fermion mass
matrices would feature a proportionality between rows and columns, thus
implying that the first generation SM charged fermions would be massless at
tree level. The one-loop level corrections to these matrices involving
different vertices that the ones used to implement the tree level universal
seesaw will break such proportionality, thus giving rise to one-loop level
masses for the up and down quarks, as well as for the electron. In order for
this to happen, the electrically charged scalar fields $\zeta _{1}^{+}$, $%
\zeta _{2}^{+}$ and $\zeta _{3}^{+}$ have to be introduced in the scalar
spectrum. \ Furthermore, the inert electrically neutral scalar singlets $%
\varphi _{1}$ and $\varphi _{2}$ are needed to generate the Majorana $\mu $
term at the two-loop level. Thus, such scalar and exotic charged fermion
spectrum is the minimal required so that no massless charged SM-fermions
would appear in the model, provided that one loop level corrections are
taken into account. Furthermore, as mentioned earlier, exotic neutral lepton
content is the minimal one required to generate the masses for two active
light neutrinos, as required from the neutrino oscillation experimental
data. 
\begin{table}[tbp]
\begin{tabular}{|c|c|c|c|c|c|c|}
\hline
& $SU\left( 3\right) _{C}$ & $SU\left( 2\right) _{L}$ & $U\left( 1\right)
_{Y}$ & $U\left( 1\right) _{X}$ & $Z_{2}$ & $Z_{4}$ \\ \hline
$\phi_{1}$ & $1$ & $2$ & $\frac{1}{2}$ & $\frac{1}{3}$ & $0$ & 
$-1$ \\ \hline
$\phi_{2}$ & $1$ & $2$ & $\frac{1}{2}$ & $\frac{2}{3}$ & $0$ & 
$1$ \\ \hline
$\sigma$ & $1$ & $1$ & $0$ & $\frac{1}{3}$ & $0$ & $-1$ \\ \hline
$\chi$ & $1$ & $1$ & $0$ & $\frac{2}{3}$ & $0$ & $-2$ \\ \hline
$\eta$ & $1$ & $1$ & $0$ & $-1$ & $0$ & $2$ \\ \hline
$\rho$ & $1$ & $1$ & $0$ & $2$ & $0$ & $-1$ \\ \hline
$S$ & $1$ & $1$ & $0$ & $0$ & $0$ & $1$ \\ \hline
$\zeta^+_1$ & $1$ & $1$ & $1$ & $\frac{2}{3}$ & $0$ & $-1$ \\ 
\hline
$\zeta^+_2$ & $1$ & $1$ & $1$ & $1$ & $0$ & $0$ \\ \hline
$\zeta^+_3$ & $1$ & $1$ & $1$ & $\frac{2}{3}$ & $0$ & $1$ \\ 
\hline
$\varphi _{1}$ & $1$ & $1$ & $0$ & $1$ & $1$ & $0$ \\ \hline
$\varphi _{2}$ & $1$ & $1$ & $0$ & $0$ & $1$ & $1$ \\ \hline
\end{tabular}
\caption{Scalar assignments under $SU\left( 3\right) _{C}\times SU\left(
2\right) _{L}\times U\left( 1\right) _{Y}\times U\left( 1\right) _{X}\times
Z_{2}\times Z_{4}$.}
\label{scalars}
\end{table}
\begin{table}[tbp]
\begin{tabular}{|c|c|c|c|c|c|c|}
\hline
& $SU\left( 3\right) _{C}$ & $SU\left( 2\right) _{L}$ & $U\left( 1\right)
_{Y}$ & $U\left( 1\right) _{X}$ & $Z_{2}$ & $Z_{4}$ \\ \hline
$q_{nL}$ & $3$ & $2$ & $\frac{1}{6}$ & $0$ & $0$ & $1$ \\ \hline
$q_{3L}$ & $3$ & $2$ & $\frac{1}{6}$ & $\frac{1}{3}$ & $0$ & $0$ \\ \hline
$u_{iR}$ & $3$ & $1$ & $\frac{2}{3}$ & $\frac{2}{3}$ & $0$ & $1$ \\ \hline
$d_{iR}$ & $3$ & $1$ & $-\frac{1}{3}$ & $-\frac{1}{3}$ & $0$ & $-1$ \\ \hline
$U_{L}$ & $3$ & $1$ & $\frac{2}{3}$ & $\frac{1}{3}$ & $0$ & $2$ \\ \hline
$U_{R}$ & $3$ & $1$ & $\frac{2}{3}$ & $\frac{2}{3}$ & $0$ & $2$ \\ \hline
$T_{L}$ & $3$ & $1$ & $\frac{2}{3}$ & $-\frac{1}{3}$ & $0$ & $0$ \\ \hline
$T_{R}$ & $3$ & $1$ & $\frac{2}{3}$ & $-\frac{1}{3}$ & $0$ & $0$ \\ \hline
$D_{1L}$ & $3$ & $1$ & $-\frac{1}{3}$ & $0$ & $0$ & $2$ \\ \hline
$D_{1R}$ & $3$ & $1$ & $-\frac{1}{3}$ & $-\frac{1}{3}$ & $0$ & $2$ \\ \hline
$D_{2L}$ & $3$ & $1$ & $-\frac{1}{3}$ & $0$ & $1$ & $-1$ \\ \hline
$D_{2R}$ & $3$ & $1$ & $-\frac{1}{3}$ & $-\frac{1}{3}$ & $1$ & $0$ \\ \hline
$B_{L}$ & $3$ & $1$ & $-\frac{1}{3}$ & $\frac{2}{3}$ & $0$ & $0$ \\ \hline
$B_{R}$ & $3$ & $1$ & $-\frac{1}{3}$ & $\frac{2}{3}$ & $0$ & $0$ \\ \hline
\end{tabular}%
\caption{Quark assignments under $SU\left( 3\right) _{C}\times SU\left(
2\right) _{L}\times U\left( 1\right) _{Y}\times U\left( 1\right) _{X}\times
Z_{2}\times Z_{4}$. Here $i=1,2,3$.}
\label{quarks}
\end{table}
\begin{table}[tbp]
\begin{tabular}{|c|c|c|c|c|c|c|}
\hline
& $SU\left( 3\right) _{C}$ & $SU\left( 2\right) _{L}$ & $U\left( 1\right)
_{Y}$ & $U\left( 1\right) _{X}$ & $Z_{2}$ & $Z_{4}$ \\ \hline
$l_{iL}$ & $1$ & $2$ & $-\frac{1}{2}$ & $-\frac{1}{3}$ & $0$ & $1$ \\ \hline
$l_{nR}$ & $1$ & $1$ & $-1$ & $-1$ & $0$ & $1$ \\ \hline
$l_{3R}$ & $1$ & $1$ & $-1$ & $-1$ & $0$ & $0$ \\ \hline
$E_{1L}$ & $1$ & $1$ & $-1$ & $-1$ & $0$ & $-2$ \\ \hline
$E_{1R}$ & $1$ & $1$ & $-1$ & $-1$ & $0$ & $0$ \\ \hline
$E_{2L}$ & $1$ & $1$ & $-1$ & $1$ & $0$ & $1$ \\ \hline
$E_{2R}$ & $1$ & $1$ & $-1$ & $1$ & $0$ & $1$ \\ \hline
$\nu _{iR}^{C}$ & $1$ & $1$ & $0$ & $-\frac{1}{3}$ & $0$ & $0$ \\ \hline
$N_{iR}$ & $1$ & $1$ & $0$ & $0$ & $0$ & $-1$ \\ \hline
$\Psi _{nR}$ & $1$ & $1$ & $0$ & $1$ & $1$ & $1$ \\ \hline
$\Omega _{nR}$ & $1$ & $1$ & $0$ & $-1$ & $0$ & $0$ \\ \hline
\end{tabular}%
\caption{Lepton assigments under $SU\left( 3\right) _{C}\times SU\left(
2\right) _{L}\times U\left( 1\right) _{Y}\times U\left( 1\right) _{X}\times
Z_{2}\times Z_{4}$. Here $i=1,2,3$ and $n=1,2$.}
\label{leptons}
\end{table}
\begin{figure}[th]
\includegraphics[width=0.45\textwidth]{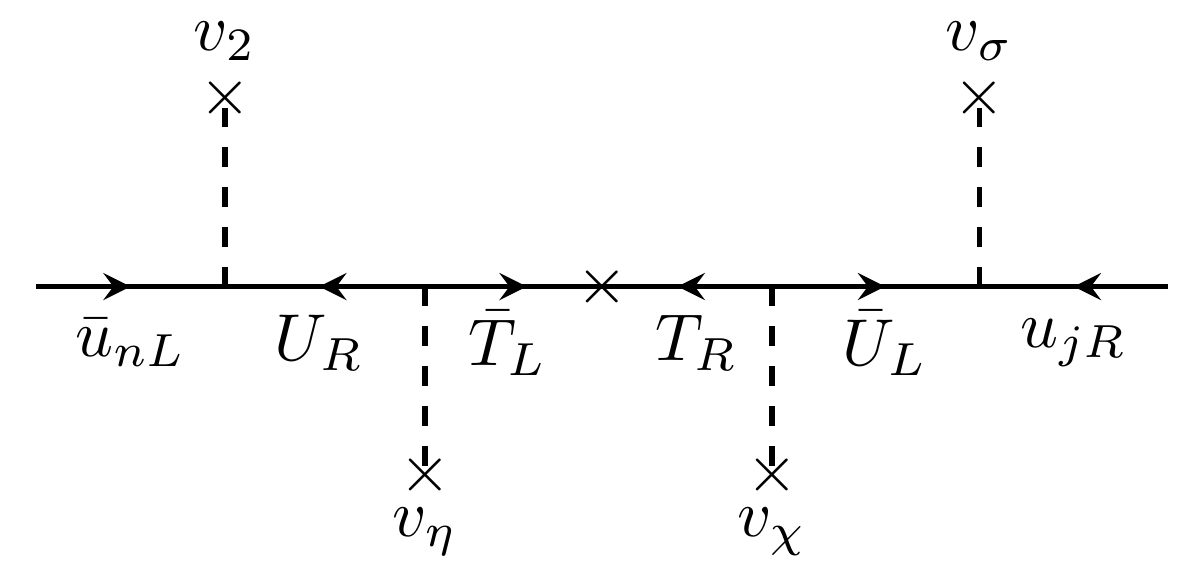}%
\includegraphics[width=0.45\textwidth]{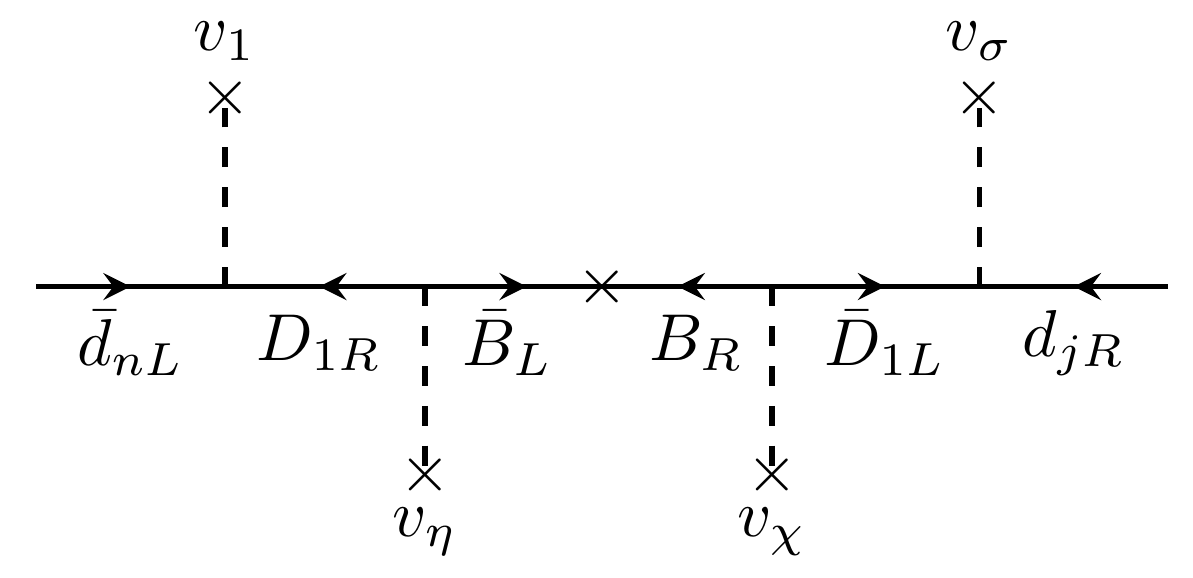}\newline
\includegraphics[width=0.45\textwidth]{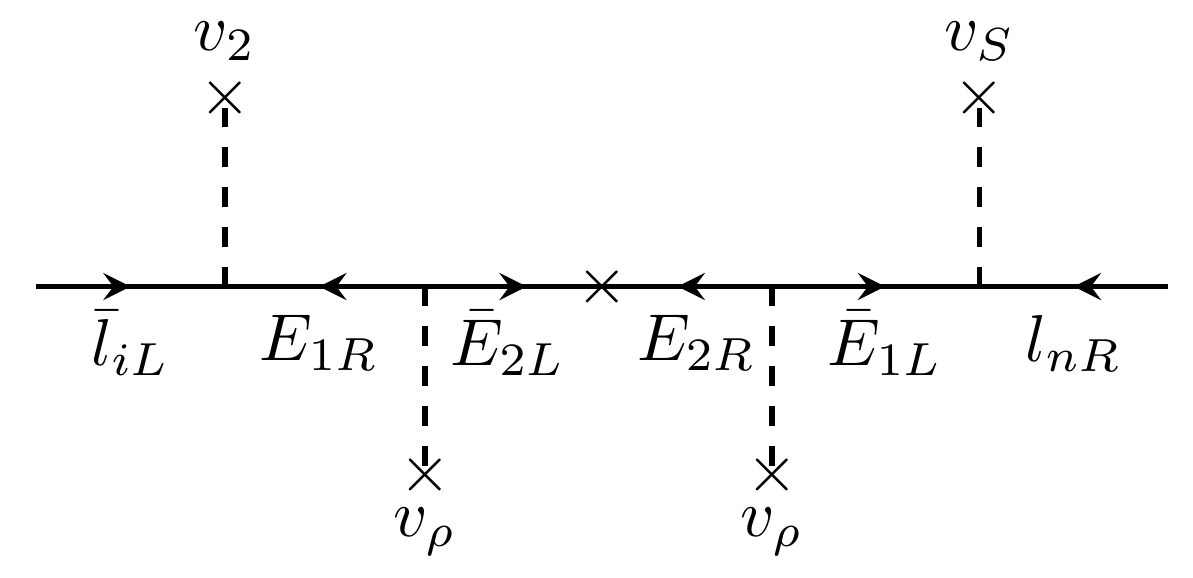}%
\includegraphics[width=0.45\textwidth]{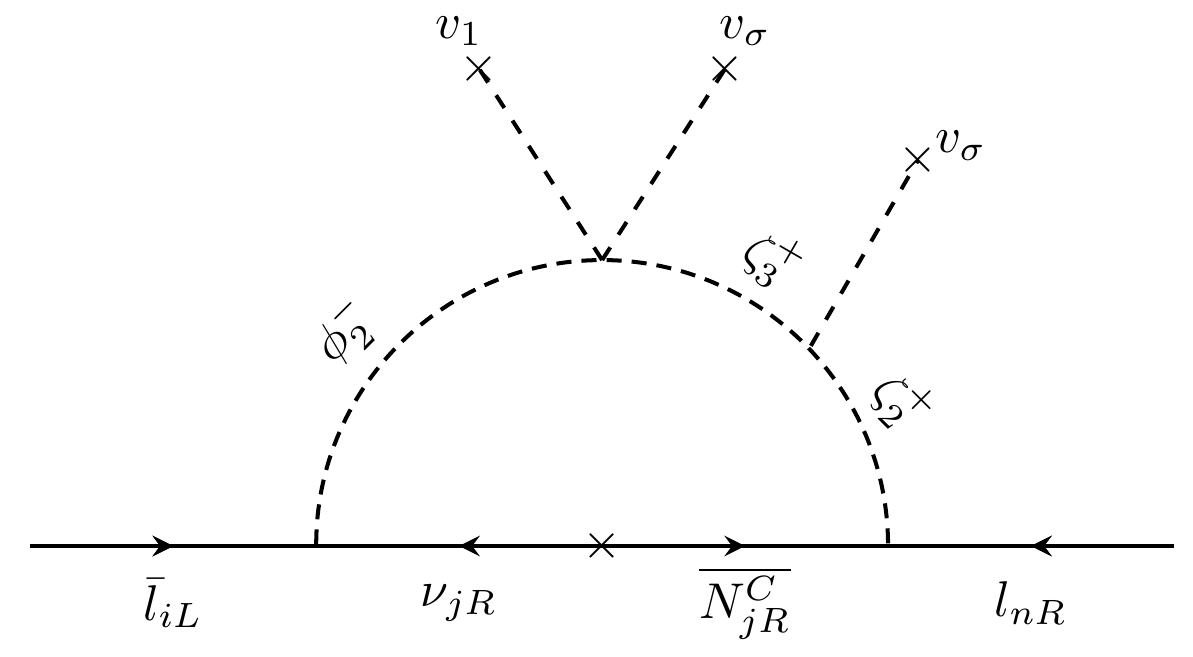}\newline
\includegraphics[width=0.45\textwidth]{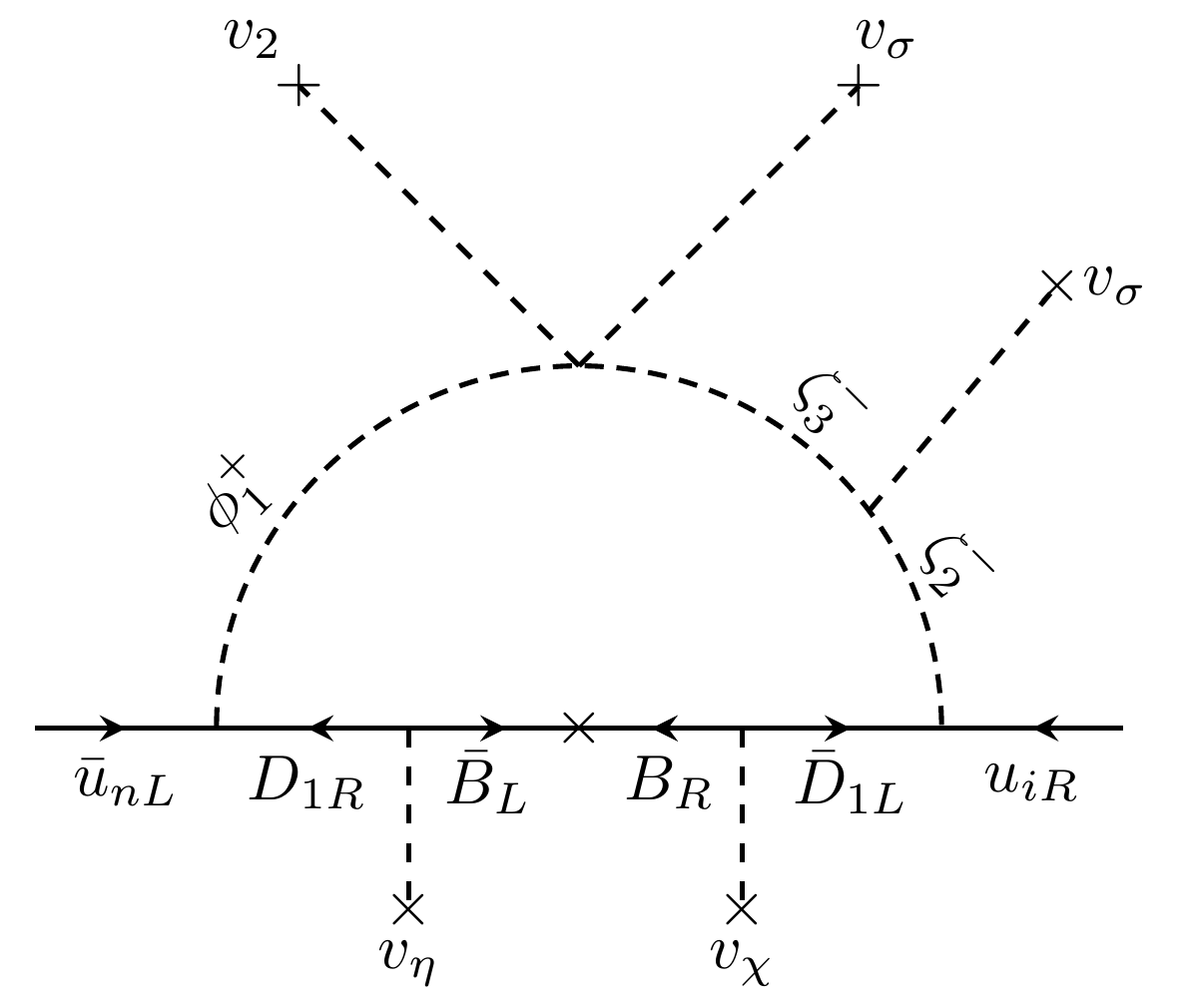}%
\includegraphics[width=0.45\textwidth]{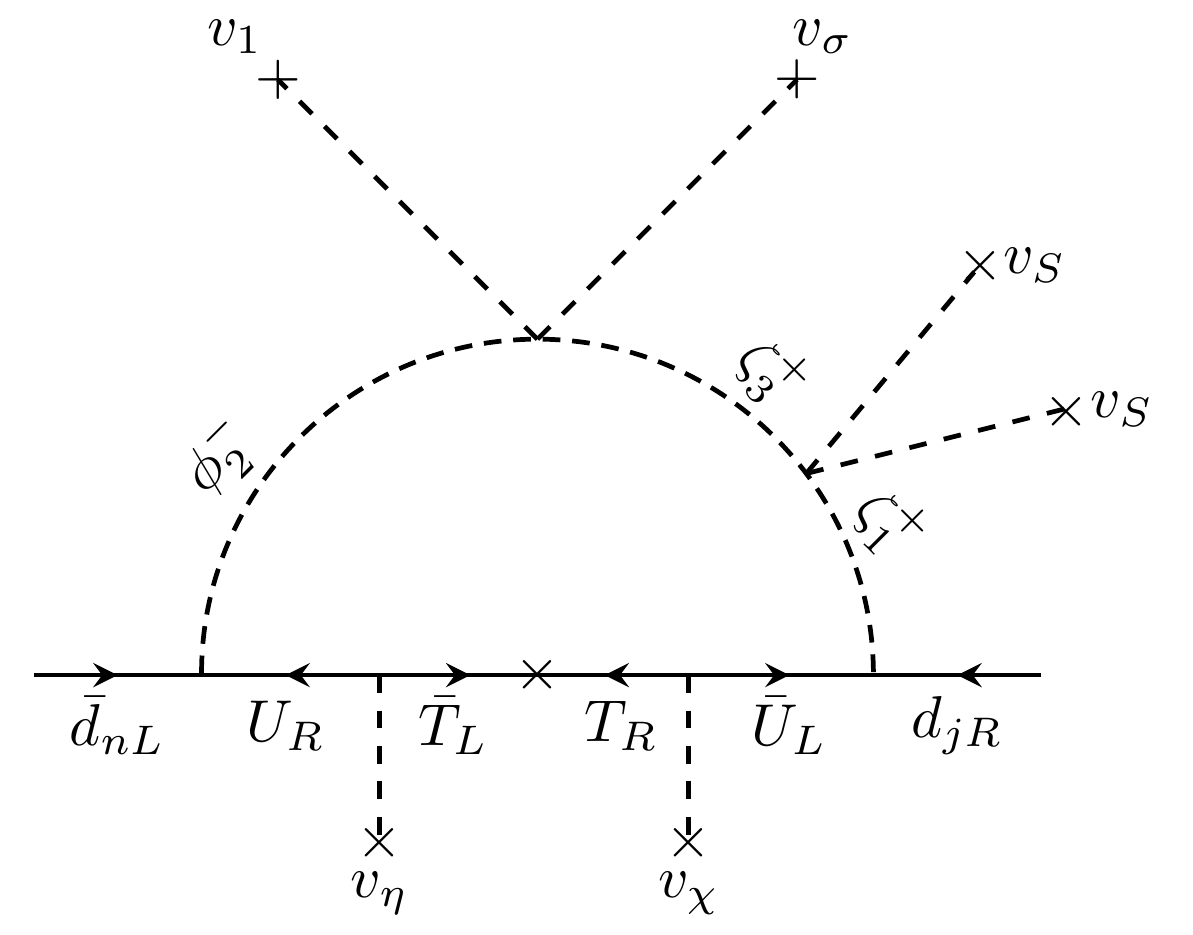}
\caption{Feynman diagrams contributing to the entries of the SM charged
fermion mass matrices. Here, $n=1,2$ and $i,j=1,2,3$.}
\label{Diagramschargedfermions}
\end{figure}

Our proposed model corresponds to an extension of a 2HDM based on the $%
SU\left( 3\right) _{C}\times SU\left( 2\right) _{L}\times U\left( 1\right)
_{Y}\times U\left( 1\right) _{X}$ gauge symmetry, supplemented by the $%
Z_{2}\times Z_{4}$ discrete group. The $SU\left( 2\right) _{L}\times U\left(
1\right) _{Y}\times U\left( 1\right) _{X}$ gauge symmetry, as well as the $%
Z_{4}$ discrete symmetry, are spontaneously broken, whereas the $Z_{2}$
symmetry is preserved, which allows the implementation of a two loop level
inverse seesaw mechanism to produce the tiny active neutrino masses. The
scalar sector of our extended 2HDM model is composed of two Higgs doublets
(having different $U\left( 1\right) _{X}$ and $Z_{4}$ charges) plus several
electrically neutral gauge singlet scalars. In our model one Higgs doublet,
i.e., $\phi _{1}$, provides a tree level mass to the top quark, whereas the
other one, i.e., $\phi _{2}$, generates tree level masses for the bottom
quark and tau lepton. The second and first families of SM charged fermions
get their masses from tree and one loop level inverse seesaw mechanisms,
respectively. The tiny active neutrino masses arise from a two loop level
inverse seesaw mechanism. The scalar, quark and lepton content with their
assignments under the $SU(3)_{C}\times SU(2)_{L}\times U(1)_{Y}\times
U\left( 1\right) $ group are shown in Tables \ref{scalars}, \ref{quarks} and %
\ref{leptons}, respectively.

With the particle content previously specified, we have the following
relevant Yukawa terms invariant under the symmetries of the model: 
\begin{eqnarray}
-\mathcal{L}_{Y}^{\left( q\right) } &=&\sum_{i=1}^{3}y_{i}^{\left( u\right) }%
\overline{q}_{3L}\widetilde{\phi }_{1}u_{iR}+\sum_{i=1}^{3}y_{i}^{\left(
d\right) }\overline{q}_{3L}\phi _{2}d_{iR}+\sum_{n=1}^{2}x_{n}^{\left(
U\right) }\overline{q}_{nL}\widetilde{\phi }_{2}U_{R}+\sum_{n=1}^{2}x_{n}^{%
\left( D\right) }\overline{q}_{nL}\phi _{1}D_{1R}  \notag \\
&&+z_{D}\overline{D}_{2L}\sigma D_{2R}+\sum_{i=1}^{3}x_{i}^{\left( u\right) }%
\overline{U}_{L}\sigma ^{\ast }u_{iR}+z_{T}\overline{U}_{L}\chi T_{R}+z_{U}%
\overline{T}_{L}\eta U_{R}+m_{T}\overline{T}_{L}T_{R}  \notag \\
&&+\sum_{i=1}^{3}x_{i}^{\left( d\right) }\overline{D}_{1L}\sigma d_{iR}+z_{B}%
\overline{D}_{1L}\chi ^{\ast }B_{R}+z_{D}\overline{B}_{L}\eta ^{\ast
}D_{1R}+m_{B}\overline{B}_{L}B_{R}  \notag \\
&&+\sum_{i=1}^{3}w_{i}^{\left( u\right) }\overline{D}_{1L}\zeta
_{1}^{-}u_{iR}+\sum_{i=1}^{3}w_{i}^{\left( d\right) }\overline{U}_{L}\zeta
_{1}^{+}d_{iR}+H.c,  \label{Lyq}
\end{eqnarray}%
\begin{eqnarray}
-\mathcal{L}_{Y}^{\left( l\right) } &=&\sum_{i=1}^{3}y_{i}^{\left( l\right) }%
\overline{l}_{iL}\phi _{2}l_{3R}+\sum_{i=1}^{3}y_{i}^{\left( E\right) }%
\overline{l}_{iL}\phi _{2}E_{1R}+\sum_{n=1}^{2}x_{n}^{\left( l\right) }%
\overline{E}_{1L}Sl_{nR}+y_{E}\overline{E}_{1L}\rho E_{2R}+x_{E}\overline{E}%
_{2L}\rho ^{\ast }E_{1R}+m_{E}\overline{E}_{2L}E_{2R}  \notag \\
&&+\sum_{i=1}^{3}\sum_{j=1}^{3}y_{ij}^{\left( \nu \right) }\overline{l}_{iL}%
\widetilde{\phi }_{2}\nu _{jR}+\sum_{i=1}^{3}\sum_{n=1}^{2}z_{in}^{\left(
l\right) }\overline{N_{iR}^{C}}\zeta
_{2}^{+}l_{nR}+\sum_{i=1}^{3}\sum_{j=1}^{3}y_{ij}^{\left( N\right) }\nu
_{iR}\sigma ^{\ast }\overline{N_{jR}^{C}}+\sum_{i=1}^{3}\sum_{n=1}^{2}\left(
x_{N}\right) _{in}\overline{N}_{iR}\Psi _{nR}^{C}\varphi _{1}  \notag \\
&&+\sum_{n=1}^{2}\sum_{m=1}^{2}\left( x_{\Psi }\right) _{nm}\overline{\Psi }%
_{nR}\varphi _{2}\Omega _{mR}^{C}+\sum_{n=1}^{2}\sum_{m=1}^{2}\left(
y_{\Omega }\right) _{nm}\overline{\Omega }_{nR}\Omega _{mR}^{C}\eta +h.c
\label{Lyl}
\end{eqnarray}%
\begin{figure}[th]
\includegraphics[width=0.5\textwidth]{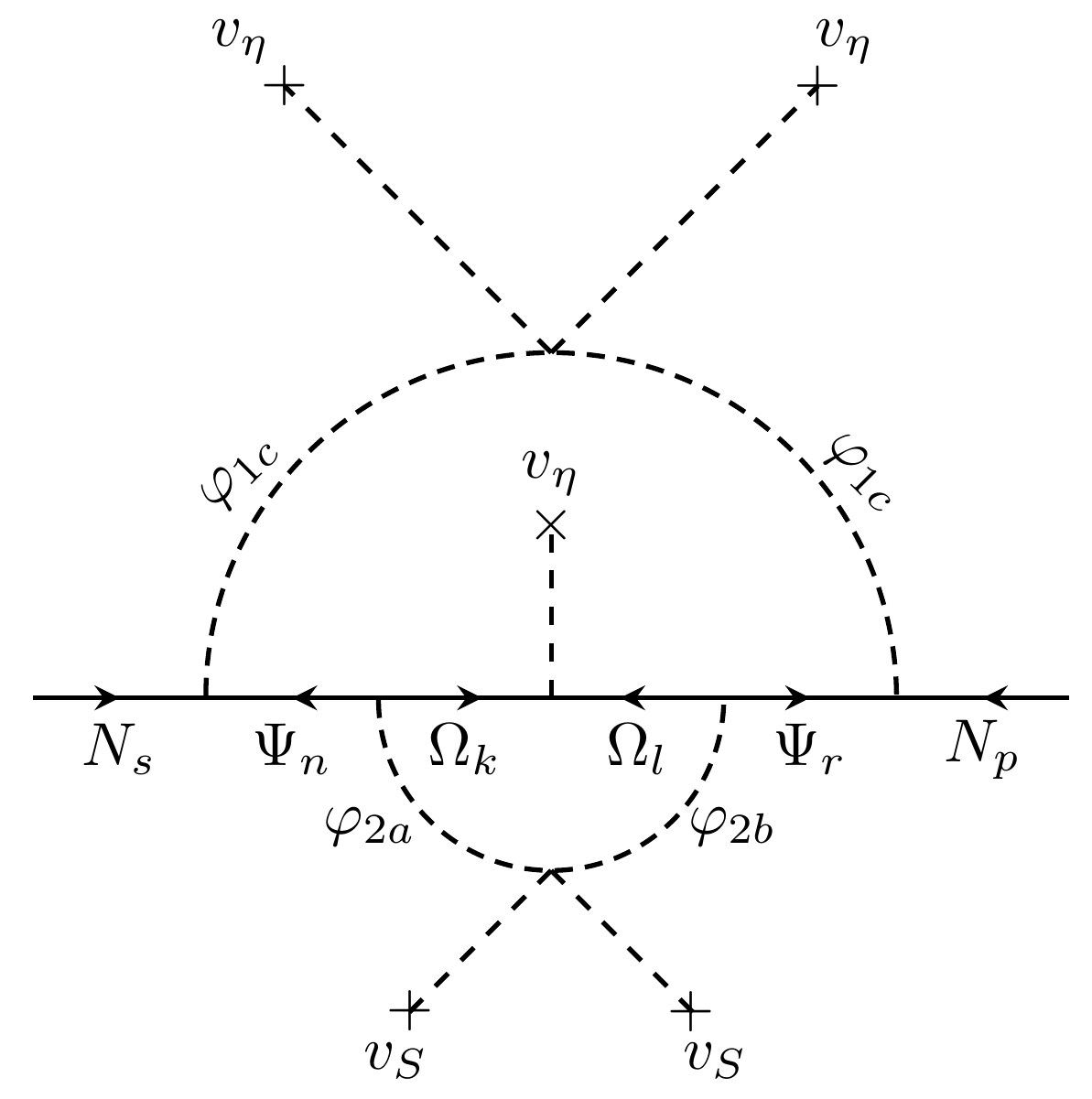}
\caption{Two-loop Feynman diagram contributing to the Majorana neutrino mass
submatrix $\protect\mu $. Here $n,k,l,r=1,2$, $s,p=1,2,3$, $a,b,c=R,I$, with 
$\protect\varphi _{nR}$ and $\protect\varphi _{nI}$ corresponding to the CP
even and CP odd parts of the scalar field $\protect\varphi _{n}$,
respectively.}
\label{Neutrinoloopdiagram}
\end{figure}

\newpage

\section{Fermion mass matrices}

\label{fermionmasses} From the charged fermion Yukawa terms we find that the
up and down type quark mass matrices in the basis $(\overline{u}_{1L},%
\overline{u}_{2L},\overline{u}_{3L},\overline{U}_{L},\overline{T}_{L})$-$%
(u_{1R},u_{2R},u_{3R},U_{R},T_{R})$ and $(\overline{d}_{1L},\overline{d}%
_{2L},\overline{d}_{3L},\overline{D}_{1L},\overline{B}_{L})$-$%
(d_{1R},d_{2R},d_{3R},D_{1R},B_{R})$ are respectively given by: 
\begin{eqnarray}
M_{U} &=&\left( 
\begin{array}{ccc}
C_{U}+\Delta _{U} & F_{U} & 0_{3\times 1} \\ 
G_{U}^{T} & 0 & X_{U} \\ 
0_{1\times 3} & Y_{U} & m_{T}%
\end{array}%
\right) ,\hspace{0.1cm}\hspace{0.1cm}\left( F_{U}\right) _{3}=0,\hspace{0.1cm%
}\hspace{0.1cm}\left( F_{U}\right) _{n}=x_{n}^{\left( U\right) }\frac{v_{2}}{%
\sqrt{2}},\hspace{0.5cm}\left( G_{U}^{T}\right) _{i}=x_{i}^{\left( u\right) }%
\frac{v_{\sigma }}{\sqrt{2}},  \notag \\
C_{U} &=&\left( 
\begin{array}{ccc}
0 & 0 & 0 \\ 
0 & 0 & 0 \\ 
y_{1}^{\left( u\right) } & y_{2}^{\left( u\right) } & y_{3}^{\left( u\right)
}%
\end{array}%
\right) \frac{v_{1}}{\sqrt{2}},\hspace{0.5cm}\hspace{0.5cm}X_{U}=z_{T}\frac{%
v_{\chi }}{\sqrt{2}},\hspace{0.5cm}\hspace{0.5cm}Y_{U}=z_{U}\frac{v_{\eta }}{%
\sqrt{2}},\hspace{0.5cm}\hspace{0.5cm}i=1,2,3,\hspace{0.5cm}\hspace{0.5cm}%
n=1,2,  \label{MU}
\end{eqnarray}%
\begin{eqnarray}
M_{D} &=&\left( 
\begin{array}{ccc}
C_{D}+\Delta _{D} & F_{D} & 0_{3\times 1} \\ 
G_{D}^{T} & 0 & X_{D} \\ 
0_{1\times 3} & Y_{D} & m_{D}%
\end{array}%
\right) ,\hspace{0.1cm}\hspace{0.1cm}\left( F_{D}\right) _{3}=0,\hspace{0.1cm%
}\hspace{0.1cm}\left( F_{D}\right) _{n}=x_{n}^{\left( D\right) }\frac{v_{1}}{%
\sqrt{2}},\hspace{0.5cm}\left( G_{D}^{T}\right) _{i}=x_{i}^{\left( d\right) }%
\frac{v_{\sigma }}{\sqrt{2}},  \notag \\
C_{D} &=&\left( 
\begin{array}{ccc}
0 & 0 & 0 \\ 
0 & 0 & 0 \\ 
y_{1}^{\left( d\right) } & y_{2}^{\left( d\right) } & y_{3}^{\left( d\right)
}%
\end{array}%
\right) \frac{v_{2}}{\sqrt{2}},\hspace{0.5cm}\hspace{0.5cm}X_{D}=z_{B}\frac{%
v_{\chi }}{\sqrt{2}},\hspace{0.5cm}\hspace{0.5cm}Y_{D}=z_{D}\frac{v_{\eta }}{%
\sqrt{2}},\hspace{0.5cm}\hspace{0.5cm}i=1,2,3,\hspace{0.5cm}\hspace{0.5cm}%
n=1,2,  \label{MD}
\end{eqnarray}%
Furthermore, notice that due to the $Z_{2}$ symmetry, the exotic $D_{2}$
down type quark does not mix with the remaining down type quarks. This
exotic $D_{2}$ quark gets a tree level mass equal to $z_{D}\frac{v_{\sigma }%
}{\sqrt{2}}$ , which is at the $v_{\sigma }$ scale. Furthermore, the charged
lepton mass matrix, written in the basis $(\overline{l}_{1L},\overline{l}%
_{2L},\overline{l}_{3L},\overline{E}_{1L},\overline{E}_{2L})$ versus $%
(l_{1R},l_{2R},l_{3R},E_{1R},E_{2R})$, takes the form: 
\begin{eqnarray}
M_{E} &=&\left( 
\begin{array}{ccc}
C_{E}+\Delta _{E} & F_{E} & 0_{3\times 1} \\ 
G_{E}^{T} & 0 & X_{E} \\ 
0_{1\times 3} & Y_{E} & m_{E}%
\end{array}%
\right) ,\hspace{0.1cm}\hspace{0.1cm}\left( F_{E}\right) _{i}=y_{i}^{\left(
E\right) }\frac{v_{2}}{\sqrt{2}},\hspace{0.5cm}\left( G_{E}^{T}\right)
_{n}=x_{n}^{\left( l\right) }\frac{v_{S}}{\sqrt{2}},\hspace{0.5cm}\left(
G_{E}^{T}\right) _{3}=0,  \notag \\
C_{E} &=&\left( 
\begin{array}{ccc}
0 & 0 & y_{1}^{\left( l\right) } \\ 
0 & 0 & y_{2}^{\left( l\right) } \\ 
0 & 0 & y_{3}^{\left( l\right) }%
\end{array}%
\right) \frac{v_{2}}{\sqrt{2}},\hspace{0.5cm}\hspace{0.5cm}X_{E}=y_{E}\frac{%
v_{\rho }}{\sqrt{2}},\hspace{0.5cm}\hspace{0.5cm}Y_{E}=x_{E}\frac{v_{\rho }}{%
\sqrt{2}},\hspace{0.5cm}\hspace{0.5cm}i=1,2,3,\hspace{0.5cm}\hspace{0.5cm}%
n=1,2,  \label{ME}
\end{eqnarray}%
Here we assume that the entries of the charged fermion mass matrices fulfill
the hierarchy: 
\begin{eqnarray}
m_{T},m_{D} &<&<<\left( F_{Q}\right) _{n},\left( G_{Q}^{T}\right)
_{i}<X_{Q}\sim Y_{Q}\sim \mathcal{O}(\text{TeV}),\hspace{0.5cm}\hspace{0.5cm}%
Q=U,D,  \notag \\
m_{E} &<&<<\left( F_{E}\right) _{i},\left( G_{E}^{T}\right) _{n}<X_{E}\sim
Y_{E}\sim \mathcal{O}(\text{TeV}),\hspace{0.5cm}\hspace{0.5cm}i=1,2,3,%
\hspace{0.5cm}\hspace{0.5cm}n=1,2.
\end{eqnarray}%
where $\Delta _{U}$, $\Delta _{D}$ and $\Delta _{E}$ correspond to the
one-loop level corrections to the SM charged fermion mass matrices. The
one-loop level Feynman diagrams, generating the $\Delta _{U}$, $\Delta _{D}$
and $\Delta _{E}$ submatrices, are shown in Figure \ref%
{Diagramschargedfermions}.

The hierarchy, as mentioned above, allows for the implementation of the
inverse seesaw mechanisms at the tree and one-loop level to generate the
masses of the second and first families of SM charged fermions,
respectively. Thus, the resulting SM charged fermion mass matrices are given
by:%
\begin{eqnarray}
\widetilde{M}_{U} &=&C_{U}+\frac{m_{T}}{X_{U}Y_{U}}F_{U}G_{U}^{T}+\Delta
_{U},  \label{Mu} \\
\widetilde{M}_{D} &=&C_{D}+\frac{m_{D}}{X_{D}Y_{D}}F_{D}G_{D}^{T}+\Delta
_{D},  \label{Md} \\
\widetilde{M}_{E} &=&C_{E}+\frac{m_{E}}{X_{E}Y_{E}}F_{E}G_{E}^{T}+\Delta
_{E}.  \label{Ml}
\end{eqnarray}%
where the second and third terms of Eqs. (\ref{Mu}), (\ref{Md}) and (\ref{Ml}%
) correspond to the tree and one-loop levels, which contribute to the SM
charged fermion mass matrices arising from the inverse seesaw mechanism. The
first term in Eqs. (\ref{Mu}), (\ref{Md}) and (\ref{Ml}) corresponds to the
dominant contribution to these matrices, arising from the renormalizable
Yukawa interactions involving the $SU\left( 2\right) $ scalar doublets $%
H_{1} $ (for the up type quark sector) and $H_{2}$ (for the down type quark
and charged lepton sector), which generate the masses for the third family
of SM charged fermions. Furthermore, the resulting physical charged exotic
fermion mass spectrum is composed of two nearly degenerate heavy charged
fermions with masses at the $\mathcal{O}($TeV$)$ scale and a small mass
splitting of the order of $m_{F}$ ($F=U,D,E$). The subGeV mass scale of the
second family of SM charged fermions, arising from a tree-level inverse
seesaw mechanism, can naturally be explained by considering $\left(
F_{Q}\right) _{n}\left( G_{Q}^{T}\right) _{i}\sim \left( F_{E}\right)
_{i}\left( G_{E}^{T}\right) _{n}\sim \mathcal{O}(10^{-2}$TeV$^{2})$, $%
m_{F}\sim \mathcal{O}(10^{-1}$GeV$) $. Therefore, the inverse seesaw
mechanism presented here can naturally explain the SM fermion mass hierarchy.

The one loop level contributions to the SM charged fermion mass
matrices are given by: 
\begin{eqnarray}
\Delta _{U} &=&\frac{m_{\widetilde{T}}}{16\pi ^{2}}\sum_{i=1}^{3}\left( 
\begin{array}{ccc}
r_{1i}^{\left( T\right) }w_{1i}^{\left( T\right) } & r_{1i}^{\left( T\right)
}w_{2i}^{\left( T\right) } & r_{1i}^{\left( T\right) }w_{3i}^{\left(
T\right) } \\ 
r_{2i}^{\left( T\right) }w_{1i}^{\left( T\right) } & r_{2i}^{\left( T\right)
}w_{2i}^{\left( T\right) } & r_{2i}^{\left( T\right) }w_{3i}^{\left(
T\right) } \\ 
0 & 0 & 0%
\end{array}%
\right) \frac{m_{H_{i}^{\pm }}^{2}}{m_{H_{i}^{\pm }}^{2}-m_{\widetilde{T}%
}^{2}}\ln \left( \frac{m_{H_{i}^{\pm }}^{2}}{m_{\widetilde{T}}^{2}}\right) ,
\label{DeltaU} \\
\Delta _{D} &=&\frac{m_{\widetilde{B}}}{16\pi ^{2}}\sum_{i=1}^{3}\left( 
\begin{array}{ccc}
r_{1i}^{\left( B\right) }w_{1i}^{\left( B\right) } & r_{1i}^{\left( B\right)
}w_{2i}^{\left( B\right) } & r_{1i}^{\left( B\right) }w_{3i}^{\left(
B\right) } \\ 
r_{2i}^{\left( B\right) }w_{1i}^{\left( B\right) } & r_{2i}^{\left( B\right)
}w_{2i}^{\left( B\right) } & r_{2i}^{\left( B\right) }w_{3i}^{\left(
B\right) } \\ 
0 & 0 & 0%
\end{array}%
\right) \frac{m_{H_{i}^{\pm }}^{2}}{m_{H_{i}^{\pm }}^{2}-m_{\widetilde{B}%
}^{2}}\ln \left( \frac{m_{H_{i}^{\pm }}^{2}}{m_{\widetilde{B}}^{2}}\right) ,
\label{DeltaD} \\
\Delta _{E} &=&\frac{m_{\widetilde{E}}}{16\pi ^{2}}\sum_{i=1}^{3}\left( 
\begin{array}{ccc}
r_{1i}^{\left( E\right) }w_{1i}^{\left( E\right) } & r_{1i}^{\left( E\right)
}w_{2i}^{\left( E\right) } & r_{1i}^{\left( E\right) }w_{3i}^{\left(
E\right) } \\ 
r_{2i}^{\left( E\right) }w_{1i}^{\left( E\right) } & r_{2i}^{\left( E\right)
}w_{2i}^{\left( E\right) } & r_{2i}^{\left( E\right) }w_{3i}^{\left(
E\right) } \\ 
0 & 0 & 0%
\end{array}%
\right) \frac{m_{H_{i}^{\pm }}^{2}}{m_{H_{i}^{\pm }}^{2}-m_{\widetilde{E}%
}^{2}}\ln \left( \frac{m_{H_{i}^{\pm }}^{2}}{m_{\widetilde{E}}^{2}}\right) ,
\label{DeltaE}
\end{eqnarray}
The experimental values of the SM quark masses \cite%
{Xing:2020ijf,Zyla:2020zbs} and Cabbibo-Kobayashi-Maskawa (CKM) parameters
are: 
\begin{eqnarray}
&&m_{u}(MeV)=1.24\pm 0.22,\hspace{3mm}m_{d}(MeV)=2.69\pm 0.19,\hspace{3mm}%
m_{s}(MeV)=53.5\pm 4.6,  \notag  \label{eq:Qsector-observables} \\
&&m_{c}(GeV)=0.63\pm 0.02,\hspace{3mm}m_{t}(GeV)=172.9\pm 0.4,\hspace{3mm}%
m_{b}(GeV)=2.86\pm 0.03,\hspace{3mm}  \notag \\
&&\sin \theta _{12}=0.2245\pm 0.00044,\hspace{3mm}\sin \theta
_{23}=0.0421\pm 0.00076,\hspace{3mm}\sin \theta _{13}=0.00365\pm 0.00012, 
\notag \\
&&J=\left( 3.18\pm 0.15\right) \times 10^{-5}\,,
\end{eqnarray}%
which can be well reproduced for the following benchmark point:%
\begin{eqnarray}
m_{H_{j}^{\pm }} &=&m_{H^{\pm }}=1.5\mbox{TeV},\hspace{1cm}v_{1}\simeq 244.2%
\mbox{GeV},\hspace{1cm}v_{2}\simeq 30\mbox{GeV},\hspace{1cm}v_{\sigma
}\simeq 6\mbox{TeV},\hspace{1cm}v_{\eta }=v_{\chi }\simeq 5\mbox{TeV}, 
\notag \\
m_{\widetilde{T}} &\simeq &1.7\mbox{TeV},\hspace{1cm}m_{\widetilde{B}}\simeq
1.3\mbox{TeV},\hspace{1cm}m_{T}\simeq 9.7\mbox{GeV},\hspace{1cm}m_{D}=1.6%
\mbox{GeV},\hspace{1cm}j=1,2,3,  \notag \\
y_{1}^{\left( u\right) } &\simeq &0.585,\hspace{0.8cm}y_{2}^{\left( u\right)
}\simeq 0.717,\hspace{0.8cm}y_{3}^{\left( u\right) }\simeq -0.368,\hspace{%
0.8cm}y_{1}^{\left( d\right) }\simeq 0.0713,\hspace{0.8cm}y_{2}^{\left(
d\right) }\simeq 0.060,\hspace{0.8cm}y_{3}^{\left( d\right) }\simeq -0.067 
\notag \\
x_{1}^{\left( U\right) } &\simeq &-4.566-1.209i,\hspace{0.8cm}x_{2}^{\left(
U\right) }\simeq -1.177+4.256i,\hspace{0.8cm}x_{1}^{\left( D\right) }\simeq
-0.732-0.005i,\hspace{0.8cm}x_{2}^{\left( D\right) }\simeq 0.292+0.883i, 
\notag \\
x_{1}^{\left( u\right) } &\simeq &-0.238,\hspace{1cm}x_{2}^{\left( u\right)
}\simeq 0.141,\hspace{1cm}x_{3}^{\left( u\right) }\simeq -0.114,\hspace{1cm}%
z_{T}=z_{U}\simeq -0.481,\hspace{1cm}z_{D}=z_{B}\simeq -0.441,\hspace{1cm} 
\notag \\
x_{1}^{\left( d\right) } &\simeq &-0.038-0.158i,\hspace{1cm}x_{2}^{\left(
d\right) }\simeq -0.037-0.035i,\hspace{1cm}x_{3}^{\left( d\right) }\simeq
0.040+0.145i,\hspace{1cm}  \notag \\
r_{1j}^{\left( T\right) } &\simeq &0.139,\hspace{1cm}r_{2j}^{\left( T\right)
}\simeq 0.087,\hspace{1cm}r_{1j}^{\left( B\right) }\simeq 0.009-0.038i,%
\hspace{1cm}r_{2i}^{\left( B\right) }\simeq 0.083+0.081i,\hspace{1cm}%
w_{3i}^{\left( B\right) }\simeq 0.012  \notag \\
w_{1j}^{\left( T\right) } &\simeq &0.011,\hspace{1cm}w_{2j}^{\left( T\right)
}\simeq -0.014,\hspace{1cm}w_{3j}^{\left( T\right) }\simeq 0.009,\hspace{1cm}%
w_{1i}^{\left( B\right) }\simeq -0.009,\hspace{1cm}w_{2i}^{\left( B\right)
}\simeq -0.013,  \label{benchmarkquarks}
\end{eqnarray}
Thus, the proposed model can numerically reproduce the existing pattern of
the observed quark spectrum. Furthermore, in the benchmark point shown in
Eq. (\ref{benchmarkquarks}), the exotic up and down type quark masses are
close to about $1.7$ \mbox{TeV} and $1.6$ \mbox{TeV}, respectively, which
are values larger than the ATLAS exclusion limits of $1.6$ \mbox{TeV} and $%
1.42$ \mbox{TeV} \cite{Sinervo:2779463}. Besides that, in the simplified
benchmark scenario considered above, the electrically charged scalars are
assumed to be degenerate, and their masses are set to be equal to $1.5$ %
\mbox{TeV}, a value that exceeds the upper limit of $[80,160]$ \mbox{GeV}
arising from collider searches \cite{Sanyal:2019xcp,CMS:2020osd}.

On the other hand, from the neutrino Yukawa interactions and considering $%
\Omega _{nR}$ ($n=1,2$) as physical neutral leptonic fields, we find the
following neutrino mass terms: 
\begin{equation}
-\mathcal{L}_{mass}^{\left( \nu \right) }=\frac{1}{2}\left( 
\begin{array}{ccc}
\overline{\nu _{L}^{C}} & \overline{\nu _{R}} & \overline{N_{R}}%
\end{array}%
\right) M_{\nu }\left( \nu _{L} \hspace*{0.3cm} \nu _{R}^{C} \hspace*{0.3cm}
N_{R}^{C} \right)^T +\sum_{n=1}^{2}\left( m_{\Omega }\right) _{n}\overline{%
\Omega }_{nR}\Omega _{nR}^{C}+H.c,  \label{Lnu}
\end{equation}%
where $\left( m_{\Omega }\right) _{n}=\left( y_{\Omega }\right) _{n}\frac{%
v_{\eta }}{\sqrt{2}}$ ($n=1,2$)\ and the neutrino mass matrix reads: 
\begin{equation}
M_{\nu }=\left( 
\begin{array}{ccc}
0_{3\times 3} & m_{\nu D} & 0_{3\times 3} \\ 
m_{\nu D}^{T} & 0_{3\times 3} & M \\ 
0_{3\times 3} & M^{T} & \mu%
\end{array}%
\right) ,  \label{Mnu}
\end{equation}%
and the submatrices are: 
\begin{eqnarray}
\left( m_{\nu D}\right) _{ij} &=&y_{ij}^{\left( \nu \right) }\frac{v_{2}}{%
\sqrt{2}},\hspace{0.7cm}\hspace{0.7cm}M_{ij}=y_{ij}^{\left( N\right) }\frac{%
v_{\sigma }}{\sqrt{2}},\hspace{0.7cm}\hspace{0.7cm}i,j,s,p=1,2,3,\hspace{%
0.7cm}\hspace{0.7cm}n,k,r=1,2,  \label{mu1a} \\
\mu_{sp}&=&\sum_{k=1}^{2}\frac{\left( x_{N}\right) _{sn}\left( x_{\Psi
}^{\ast }\right) _{nk}\left( x_{\Psi }^{\dag }\right) _{kr}\left(
x_{N}^{T}\right) _{rp}m_{\Omega _{k}}}{4(4\pi )^{4}}\int_{0}^{1}d\alpha
\int_{0}^{1-\alpha }d\beta \frac{1}{\alpha (1-\alpha )}\Biggl[I\left(
m_{\Omega _{k}}^{2},m_{RR}^{2},m_{RI}^{2}\right) -I\left( m_{\Omega
_{k}}^{2},m_{IR}^{2},m_{II}^{2}\right) \Biggr],  \notag
\end{eqnarray}%
with the loop integral given by \cite{Kajiyama:2013rla}: 
\begin{eqnarray}
I(m_{1}^{2},m_{2}^{2},m_{3}^{2})\!\!\! &=&\!\!\!\frac{m_{1}^{2}m_{2}^{2}\log
\left( \displaystyle\frac{m_{2}^{2}}{m_{1}^{2}}\right)
+m_{2}^{2}m_{3}^{2}\log \left( \displaystyle\frac{m_{3}^{2}}{m_{2}^{2}}%
\right) +m_{3}^{2}m_{1}^{2}\log \left( \displaystyle\frac{m_{1}^{2}}{%
m_{3}^{2}}\right) }{(m_{1}^{2}-m_{2}^{2})(m_{1}^{2}-m_{3}^{2})},  \notag \\
m_{ab}^{2}\!\!\! &=&\!\!\!\frac{\beta m_{\left( \varphi _{1}\right)
_{a}}^{2}+\alpha m_{\left( \varphi _{2}\right) _{b}}^{2}}{\alpha (1-\alpha )}%
\quad (a,b=R:\mathrm{or}:I),
\end{eqnarray}%
The active light neutrino masses are generated from an inverse seesaw
mechanism at the two-loop level, and the physical neutrino mass matrices are
given by: 
\begin{eqnarray}
\widetilde{M}_{\nu } &=&m_{\nu D}\left( M^{T}\right) ^{-1}\mu M^{-1}m_{\nu
D}^{T},\hspace{0.7cm}  \label{M1nu} \\
M_{\nu }^{\left( -\right) } &=&-\frac{1}{2}\left( M+M^{T}\right) +\frac{1}{2}%
\mu ,\hspace{0.7cm} \\
M_{\nu }^{\left( +\right) } &=&\frac{1}{2}\left( M+M^{T}\right) +\frac{1}{2}%
\mu .  \label{neutrino-mass}
\end{eqnarray}%
where $\widetilde{M}_{\nu }$ is the mass matrix for the active light
neutrinos ($\nu _{a}$), whereas $M_{\nu }^{(-)}$ and $M_{\nu }^{(+)}$ are
the mass matrices for sterile neutrinos. In the limit $\mu \rightarrow 0$,
which corresponds to unbroken lepton number, the active light neutrinos
become massless. The smallness of the $\mu $ parameter yields a small mass
splitting for the two pairs of sterile neutrinos, thus implying that the
sterile neutrinos form pseudo-Dirac pairs.


\section{Muon and electron anomalous magnetic moments}

\label{gminus2sec} The experimental data shows that the muon and electron
anomalous magnetic moments deviate significantly from their SM values 
\begin{eqnarray}
\Delta a_{\mu } &=&a_{\mu }^{\mathrm{exp}}-a_{\mu }^{\mathrm{SM}}=\left(
2.51\pm 0.59\right) \times 10^{-9}\hspace{17mm}\mbox{%
\cite{Hagiwara:2011af,Davier:2017zfy,Nomura:2018lsx,Nomura:2018vfz,Blum:2018mom,Keshavarzi:2018mgv,Aoyama:2020ynm,Abi:2021gix}}
\label{eq:a-mu} \\
\Delta a_{e} &=&a_{e}^{\mathrm{exp}}-a_{e}^{\mathrm{SM}}=(-0.88\pm
0.36)\times 10^{-12}\hspace{3mm}\mbox{\cite{Parker:2018vye}},\hspace{3mm}%
(4.8\pm 3.0)\times 10^{-13}\hspace{3mm}\mbox{\cite{Morel:2020dww}}
\label{eq:a-e}
\end{eqnarray}%
where the above given value of $a_{\mu }^{\mathrm{exp}}$ is a combined
result of the BNL E821 experiment \cite{Bennett:2006fi} and the recently
announced FNAL Muon g-2 measurement \cite{Abi:2021gix}, showing the 4.2$%
\sigma $ tension between the SM and experiment. The last positive value for $%
\Delta a_{e}$ corresponds to the recently published new measurement of the
fine-structure constant, with an accuracy of 81 parts per trillion~\cite%
{Morel:2020dww}. In this section, we will analyze the implications of our
model in the muon and electron anomalous magnetic moments.

\begin{figure}[h]
\includegraphics[width=9.1cm, height=8cm]{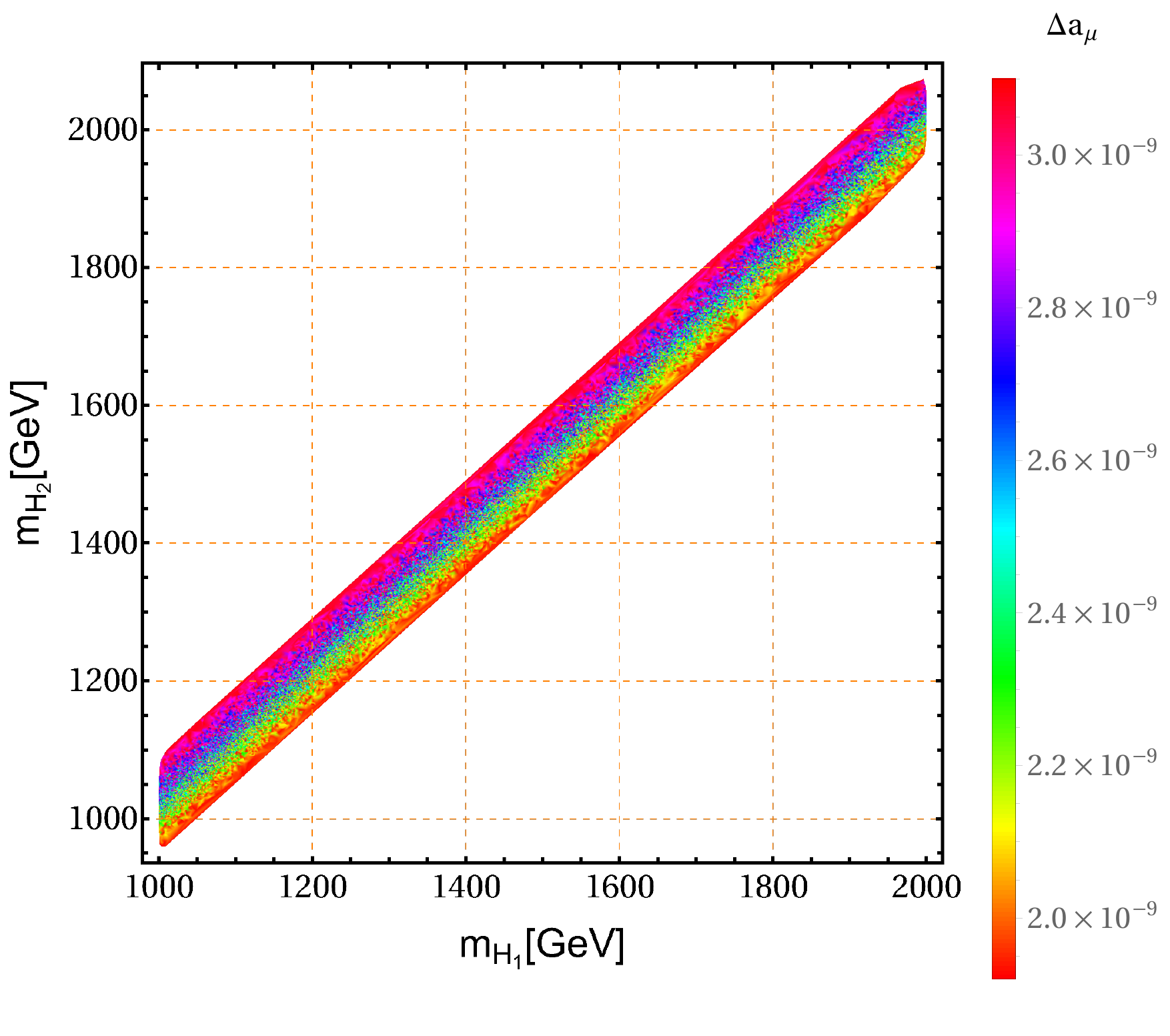}%
\includegraphics[width=9.1cm, height=8cm]{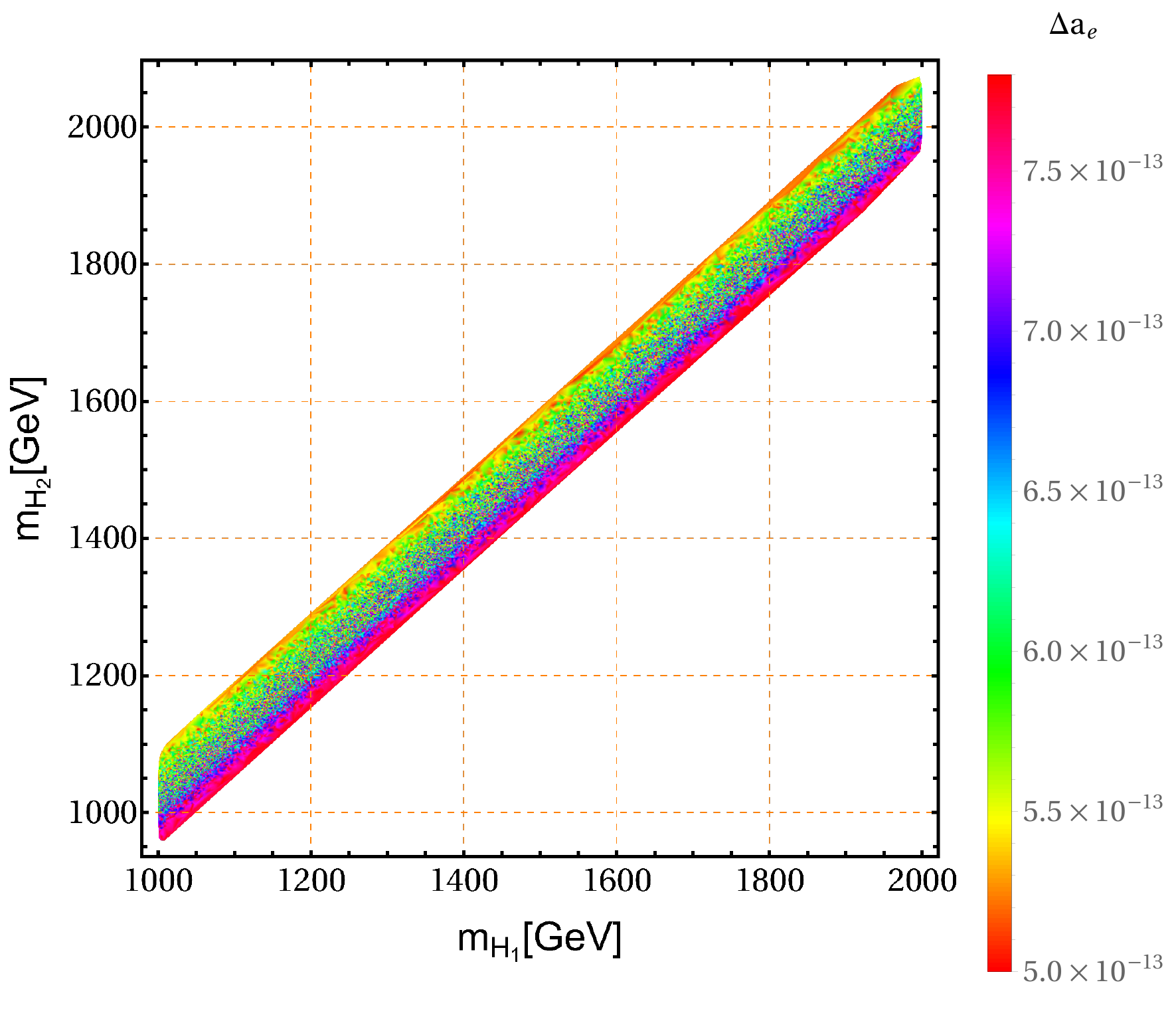}
\caption{Correlation between the masses of the scalars $H_{1}$ and $H_{2}$
consistent with the muon and electron anomalous magnetic moments.}
\label{gminus2}
\end{figure}

Muon and electron anomalous magnetic moments receive contributions from
one-loop diagrams involving the exchange of electrically neutral CP even and
CP odd scalars and charged exotic leptons as well as electrically charged
scalars and sterile neutrinos running in the internal lines of the loop. To
simplify our analysis, we will consider a simplified benchmark scenario
close to the alignment limit, where $\phi _{2R}^{0}$ ($\phi _{2I}^{0}$) and $%
S_{R}$ ($S_{I}$) are mainly composed of two orthogonal combinations
involving two heavy CP even (odd) $H_{1}$ ($A_{1}$), $H_{2}$ ($A_{2}$)
physical scalar fields. We consider the alignment limit in
which the $125$ GeV SM like Higgs boson is mainly composed of the CP even
neutral part of the $SU(2)$ scalar doublet $\phi_1$. This component does not
appear in the neutral scalar contribution to the muon and electron anomalous
magnetic moment shown in Eq. (33). Thus, in the chosen benchmark scenario,
the couplings of the $125$ GeV SM like Higgs boson are very close to the SM
expectation, which is consistent with the experimental data \cite%
{Zyla:2020zbs}. Then, the leading contributions to the muon and electron
anomalous magnetic moments take the form: 
\begin{eqnarray}
\Delta a_{e,\mu } &\simeq &\frac{\func{Re}\left( \alpha _{e,\mu }\beta
_{e,\mu }^{\ast }\right) m_{e,\mu }^{2}}{8\pi ^{2}}\left[ I_{S}^{(e,\mu
)}\left( m_{\widetilde{E}},m_{H_{1}}\right) -I_{S}^{(e,\mu )}\left( m_{%
\widetilde{E}},m_{H_{2}}\right) +I_{P}^{(e,\mu )}\left( m_{\widetilde{E}%
},m_{A_{1}}\right) -I_{P}^{(e,\mu )}\left( m_{\widetilde{E}%
},m_{A_{2}}\right) \right] \sin \theta \cos \theta  \notag \\
&&+\frac{1}{8\pi ^{2}}\sum_{i=1}^{3}\sum_{j=1}^{3}\frac{\func{Re}\left[
\kappa _{e,\mu }^{\left( ij\right) }\left( \gamma _{e,\mu }^{\left(
ij\right) }\right) ^{\ast }\right] m_{e,\mu }m_{\widetilde{N}_{j}}}{%
m_{H_{i}^{\pm }}^{2}}G_{2}\left( \frac{m_{\widetilde{N}_{j}}^{2}}{%
m_{H_{i}^{\pm }}^{2}}\right) ,
\end{eqnarray}%
where $H_{1}\simeq \cos \theta _{S}S_{R}+\sin \theta _{S}\phi _{2R}^{0}$, $%
H_{2}\simeq -\sin \theta _{S}S_{R}+\cos \theta _{S}\phi _{2R}^{0}$, $%
A_{1}\simeq \cos \theta _{P}S_{I}+\sin \theta _{P}\phi _{2I}^{0}$, $%
A_{2}\simeq -\sin \theta _{P}S_{I}+\cos \theta _{P}\phi _{2I}^{0}$, and for
the sake of simplicity we have set $\theta _{S}=\theta _{P}$ and $m_{%
\widetilde{E}}$ is the mass of the nearly degenerate physical charged exotic
leptons. Besides that, $m_{\phi _{i}^{\pm }}$ and $m_{\widetilde{N}_{j}}$ ($%
i,j=1,2,3$)\ are the masses of the physical electrically charged scalars and 
$Z_{2}$ even sterile neutrinos, respectively. Furthermore, the $I_{S\left(
P\right) }\left( m_{\widetilde{E}},m\right) $ and $G_{2}\left( r\right) $
loop functions have the form: \cite%
{Diaz:2002uk,Jegerlehner:2009ry,Kelso:2014qka,Lindner:2016bgg,Kowalska:2017iqv}%
: 
\begin{equation}
I_{S\left( P\right) }^{\left( e,\mu \right) }\left( m_{\widetilde{E}%
},m_{S}\right) =\int_{0}^{1}\frac{x^{2}\left( 1-x\pm \frac{m_{\widetilde{E}}%
}{m_{e,\mu }}\right) }{m_{\mu }^{2}x^{2}+\left( m_{\widetilde{E}%
}^{2}-m_{e,\mu }^{2}\right) x+m_{S,P}^{2}\left( 1-x\right) }dx,\hspace{0.7cm}%
\hspace{0.7cm}G_{2}\left( r\right) =\frac{-1+r^{2}-2r\ln r}{\left(
r-1\right) ^{3}}
\end{equation}%
Requiring that the muon and electron anomalous magnetic moments acquire
values in the ranges shown in Eqs. (\ref{eq:a-mu}) and (\ref{eq:a-e}),
respectively, we display in Figure \ref{gminus2} the correlation between the
masses of the scalars $H_{1}$ and $H_{2}$ consistent with the experimental
data on $\left( g-2\right) _{e,\mu }$. These masses have been
taken to range from $1$ TeV up to $2$ TeV. In contrast, the electrically
charged scalar masses are varied from $0.5$ TeV up to $1.5$ TeV, and the CP
odd scalar masses have been set to be equal to $1$ TeV. Furthermore, the
masses for the heavy vector-like leptons have been varied from $2$ TeV up to 
$3$ TeV, a range of values more extensive than the expected reach of $600$
GeV for the High Luminosity Large Hadron Collider (HL-LHC). The above values
for the scalar masses are consistent with constraints arising from collider
searches \cite{Zyla:2020zbs}, and accommodate a nearly degenerate spectrum
of heavy scalar masses which is favored by electroweak precision tests \cite%
{Hernandez:2015rfa}. Figure \ref{gminus2} shows that our model is consistent with the
experimental values of the muon and electron anomalous magnetic moments. 

\section{Meson oscillations}

\label{FCNC} In this section, we analyze the consequences of our proposed
theory in the $K^{0}-\bar{K}^{0}$, $B_{d}^{0}-\bar{B}_{d}^{0}$ and $%
B_{s}^{0}-\bar{B}_{s}^{0}$ meson oscillations. These meson oscillations are
caused by flavour violating down type quark interactions mediated by the
tree level exchange of electrically neutral CP even and CP odd scalars as
well as by the tree level $Z^{\prime }$ exchange. The $K^{0}-\bar{K}^{0}$, $%
B_{d}^{0}-\bar{B}_{d}^{0}$ and $B_{s}^{0}-\bar{B}_{s}^{0}$ meson
oscillations are described by the following effective Hamiltonians: 
\begin{equation}
\mathcal{H}_{eff}^{\left( K\right) }\mathcal{=}\sum_{j=1}^{3}c_{j}^{\left(
K\right) }\left( \mu \right) \mathcal{O}_{j}^{\left( K\right) }\left( \mu
\right) +\frac{g_{X}^{2}}{9m_{Z^{\prime }}^{2}}\left\vert \left(
V_{DL}^{\ast }\right) _{32}\left( V_{DL}\right) _{31}\right\vert ^{2}%
\mathcal{O}_{4}^{\left( K\right) },
\end{equation}%
\begin{equation}
\mathcal{H}_{eff}^{\left( B_{d}\right) }\mathcal{=}\sum_{j=1}^{3}c_{j}^{%
\left( B_{d}\right) }\left( \mu \right) \mathcal{O}_{j}^{\left( B_{d}\right)
}\left( \mu \right) +\frac{g_{X}^{2}}{9m_{Z^{\prime }}^{2}}\left\vert \left(
V_{DL}^{\ast }\right) _{31}\left( V_{DL}\right) _{33}\right\vert ^{2}%
\mathcal{O}_{4}^{\left( B_{d}\right) },
\end{equation}%
\begin{equation}
\mathcal{H}_{eff}^{\left( B_{s}\right) }\mathcal{=}\sum_{j=1}^{3}c_{j}^{%
\left( B_{s}\right) }\left( \mu \right) \mathcal{O}_{j}^{\left( B_{s}\right)
}\left( \mu \right) +\frac{g_{X}^{2}}{9m_{Z^{\prime }}^{2}}\left\vert \left(
V_{DL}^{\ast }\right) _{32}\left( V_{DL}\right) _{33}\right\vert ^{2}%
\mathcal{O}_{4}^{\left( B_{s}\right) },
\end{equation}

where the operators relevant for the $K^{0}-\bar{K}^{0}$, $B_{d}^{0}-\bar{B}%
_{d}^{0}$ and $B_{s}^{0}-\bar{B}_{s}^{0}$ meson mixings have the form: 
\begin{eqnarray}
\mathcal{O}_{1}^{\left( K\right) } &=&\left( \overline{s}_{R}d_{L}\right)
\left( \overline{s}_{R}d_{L}\right) ,\hspace{0.7cm}\hspace{0.7cm}\mathcal{O}%
_{2}^{\left( K\right) }=\left( \overline{s}_{L}d_{R}\right) \left( \overline{%
s}_{L}d_{R}\right) ,\hspace{0.7cm}  \label{op3f} \\
\mathcal{O}_{3}^{\left( K\right) } &=&\left( \overline{s}_{R}d_{L}\right)
\left( \overline{s}_{L}d_{R}\right) ,\hspace{0.7cm}\hspace{0.7cm}\mathcal{O}%
_{4}^{\left( K\right) }=\left( \overline{s}_{L}\gamma _{\mu }d_{L}\right)
\left( \overline{s}_{L}\gamma ^{\mu }d_{L}\right) , \\
\mathcal{O}_{1}^{\left( B_{d}\right) } &=&\left( \overline{d}%
_{R}b_{L}\right) \left( \overline{d}_{R}b_{L}\right) ,\hspace{0.7cm}\hspace{%
0.7cm}\mathcal{O}_{2}^{\left( B_{d}\right) }=\left( \overline{d}%
_{L}b_{R}\right) \left( \overline{d}_{L}b_{R}\right) ,\hspace{0.7cm} \\
\mathcal{O}_{3}^{\left( B_{d}\right) } &=&\left( \overline{d}%
_{R}b_{L}\right) \left( \overline{d}_{L}b_{R}\right) ,\hspace{0.7cm}\hspace{%
0.7cm}\mathcal{O}_{4}^{\left( B_{d}\right) }=\left( \overline{d}_{L}\gamma
_{\mu }b_{L}\right) \left( \overline{d}_{L}\gamma ^{\mu }b_{L}\right) , \\
\mathcal{O}_{1}^{\left( B_{s}\right) } &=&\left( \overline{s}%
_{R}b_{L}\right) \left( \overline{s}_{R}b_{L}\right) ,\hspace{0.7cm}\hspace{%
0.7cm}\mathcal{O}_{2}^{\left( B_{s}\right) }=\left( \overline{s}%
_{L}b_{R}\right) \left( \overline{s}_{L}b_{R}\right) , \\
\mathcal{O}_{3}^{\left( B_{s}\right) } &=&\left( \overline{s}%
_{R}b_{L}\right) \left( \overline{s}_{L}b_{R}\right) ,\hspace{0.7cm}\hspace{%
0.7cm}\mathcal{O}_{4}^{\left( B_{s}\right) }=\left( \overline{s}_{L}\gamma
_{\mu }b_{L}\right) \left( \overline{s}_{L}\gamma ^{\mu }b_{L}\right) ,
\end{eqnarray}

and the Wilson coefficients are: 
\begin{eqnarray}
c_{1}^{\left( K\right) } &=&\frac{z_{h\overline{s}_{R}d_{L}}^{2}}{m_{h}^{2}}%
+\sum_{j=1}^{2}\frac{z_{H_{j}\overline{s}_{R}d_{L}}^{2}}{M_{H_{j}}^{2}}%
-\sum_{j=1}^{2}\frac{z_{A_{j}\overline{s}_{R}d_{L}}^{2}}{M_{A_{j}}^{2}}, \\
c_{2}^{\left( K\right) } &=&\frac{z_{h\overline{s}_{L}d_{R}}^{2}}{m_{h}^{2}}%
+\sum_{j=1}^{2}\frac{z_{H_{j}\overline{s}_{L}d_{R}}^{2}}{M_{H_{j}}^{2}}%
-\sum_{j=1}^{2}\frac{z_{A_{j}\overline{s}_{L}d_{R}}^{2}}{M_{A_{j}}^{2}},%
\hspace{0.7cm}\hspace{0.7cm} \\
c_{3}^{\left( K\right) } &=&\frac{z_{h\overline{s}_{R}d_{L}}z_{h\overline{s}%
_{L}d_{R}}}{m_{h}^{2}}+\sum_{j=1}^{2}\frac{z_{H_{j}\overline{s}%
_{R}d_{L}}z_{H_{j}\overline{s}_{L}d_{R}}}{M_{H_{j}}^{2}}-\sum_{j=1}^{2}\frac{%
z_{A_{j}\overline{s}_{R}d_{L}}z_{A_{j}\overline{s}_{L}d_{R}}}{M_{A_{j}}^{2}},
\end{eqnarray}%
\begin{eqnarray}
c_{1}^{\left( B_{d}\right) } &=&\frac{z_{h\overline{d}_{R}b_{L}}^{2}}{%
m_{h}^{2}}+\sum_{j=1}^{2}\frac{z_{H_{j}\overline{d}_{R}b_{L}}^{2}}{%
M_{H_{j}}^{2}}-\sum_{j=1}^{2}\frac{z_{A_{j}\overline{d}_{R}b_{L}}^{2}}{%
M_{A_{j}}^{2}}, \\
c_{2}^{\left( B_{d}\right) } &=&\frac{z_{h\overline{d}_{L}b_{R}}^{2}}{%
m_{h}^{2}}+\sum_{j=1}^{2}\frac{z_{H_{j}\overline{d}_{L}b_{R}}^{2}}{%
M_{H_{j}}^{2}}-\sum_{j=1}^{2}\frac{z_{A_{j}\overline{d}_{L}b_{R}}^{2}}{%
M_{A_{j}}^{2}}, \\
c_{3}^{\left( B_{d}\right) } &=&\frac{z_{h\overline{d}_{R}b_{L}}z_{h%
\overline{d}_{L}b_{R}}}{m_{h}^{2}}+\sum_{j=1}^{2}\frac{z_{H_{j}\overline{d}%
_{R}b_{L}}z_{H_{j}\overline{d}_{L}b_{R}}}{M_{H_{j}}^{2}}-\sum_{j=1}^{2}\frac{%
z_{A_{j}\overline{d}_{R}b_{L}}z_{A_{j}\overline{d}_{L}b_{R}}}{M_{A_{j}}^{2}},
\end{eqnarray}%
\begin{eqnarray}
c_{1}^{\left( B_{s}\right) } &=&\frac{z_{h\overline{s}_{R}b_{L}}^{2}}{%
m_{h}^{2}}+\sum_{j=1}^{2}\frac{z_{H_{j}\overline{s}_{R}b_{L}}^{2}}{%
M_{H_{i}}^{2}}-\sum_{j=1}^{2}\frac{z_{A_{j}\overline{s}_{R}b_{L}}^{2}}{%
M_{A_{j}}^{2}}, \\
c_{2}^{\left( B_{s}\right) } &=&\frac{z_{h\overline{s}_{L}b_{R}}^{2}}{%
m_{h}^{2}}+\sum_{j=1}^{2}\frac{z_{H_{j}\overline{s}_{L}b_{R}}^{2}}{%
M_{H_{j}}^{2}}-\sum_{j=1}^{2}\frac{z_{A_{j}\overline{s}_{L}b_{R}}^{2}}{%
M_{A_{j}}^{2}}, \\
c_{3}^{\left( B_{s}\right) } &=&\frac{z_{h\overline{s}_{R}b_{L}}z_{h%
\overline{s}_{L}b_{R}}}{m_{h}^{2}}+\sum_{j=1}^{2}\frac{z_{H_{j}\overline{s}%
_{R}b_{L}}z_{H_{j}\overline{s}_{L}b_{R}}}{M_{H_{j}}^{2}}-\sum_{j=1}^{2}\frac{%
z_{A_{j}\overline{s}_{R}b_{L}}z_{A_{j}\overline{s}_{L}b_{R}}}{M_{A_{j}}^{2}},
\end{eqnarray}%
Furthermore, the following relations have been taken into account: 
\begin{eqnarray}
\widetilde{M}_{f} &=&\left( M_{f}\right) _{diag}=V_{fL}^{\dagger
}M_{f}V_{fR},\hspace{1cm}\hspace{1cm}f_{\left( L,R\right) }=V_{f\left(
L,R\right) }\widetilde{f}_{\left( L,R\right) },  \notag \\
\overline{f}_{iL}\left( M_{f}\right) _{ij}f_{jR} &=&\overline{\widetilde{f}}%
_{kL}\left( V_{fL}^{\dagger }\right) _{ki}\left( M_{f}\right) _{ij}\left(
V_{fR}\right) _{jl}\widetilde{f}_{lR}=\overline{\widetilde{f}}_{kL}\left(
V_{fL}^{\dagger }M_{f}V_{fR}\right) _{kl}\widetilde{f}_{lR}=\overline{%
\widetilde{f}}_{kL}\left( \widetilde{M}_{f}\right) _{kl}\widetilde{f}%
_{lR}=m_{fk}\overline{\widetilde{f}}_{kL}\widetilde{f}_{kR},  \notag \\
k &=&1,2,3\,.
\end{eqnarray}%
Here, $\widetilde{f}_{k\left( L,R\right) }$ and $f_{k\left( L,R\right) }$ ($%
k=1,2,3$) are the SM fermionic fields in the mass and interaction bases,
respectively.

On the other hand, the $K-\bar{K}$, $B_{d}^{0}-\bar{B}_{d}^{0}$ and $%
B_{s}^{0}-\bar{B}_{s}^{0}$\ meson mass differences are given by: 
\begin{equation}
\Delta m_{K}=\Delta m_{K}^{\left( SM\right) }+\Delta m_{K}^{\left( NP\right)
},\hspace{1cm}\Delta m_{B_{d}}=\Delta m_{B_{d}}^{\left( SM\right) }+\Delta
m_{B_{d}}^{\left( NP\right) },\hspace{1cm}\Delta m_{B_{s}}=\Delta
m_{B_{s}}^{\left( SM\right) }+\Delta m_{B_{s}}^{\left( NP\right) },
\label{Deltam}
\end{equation}%
where $\Delta m_{K}^{\left( SM\right) }$, $\Delta m_{B_{d}}^{\left(
SM\right) }$ and $\Delta m_{B_{s}}^{\left( SM\right) }$ stand for the SM
contributions, while $\Delta m_{K}^{\left( NP\right) }$, $\Delta
m_{B_{d}}^{\left( NP\right) }$ and $\Delta m_{B_{s}}^{\left( NP\right) }$
arise from new physics effects.

In the model under consideration, we find the following new physics
contributions to the $K-\bar{K}$, $B_{d}^{0}-\bar{B}_{d}^{0}$ and $B_{s}^{0}-%
\bar{B}_{s}^{0}$ meson mass splittings: 
\begin{equation}
\Delta m_{K}^{\left( NP\right) }=\frac{g_{X}^{2}}{9m_{Z^{\prime }}^{2}}%
\left\vert \left( V_{DL}^{\ast }\right) _{32}\left( V_{DL}\right)
_{31}\right\vert ^{2}f_{K}^{2}B_{K}\eta _{K}m_{K}+\frac{8}{3}f_{K}^{2}\eta
_{K}B_{K}m_{K}\left[ k_{2}^{\left( K\right) }c_{3}^{\left( K\right)
}+k_{1}^{\left( K\right) }\left( c_{1}^{\left( K\right) }+c_{2}^{\left(
K\right) }\right) \right]
\end{equation}%
\begin{equation}
\Delta m_{B_{d}}^{\left( NP\right) }=\frac{g_{X}^{2}}{9m_{Z^{\prime }}^{2}}%
\left\vert \left( V_{DL}^{\ast }\right) _{31}\left( V_{DL}\right)
_{33}\right\vert ^{2}f_{B_{d}}^{2}B_{B_{d}}\eta _{B_{d}}m_{B_{d}}+\frac{8}{3}%
f_{B_{d}}^{2}\eta _{B_{d}}B_{B_{d}}m_{B_{d}}\left[ k_{2}^{\left(
B_{d}\right) }c_{3}^{\left( B_{d}\right) }+k_{1}^{\left( B_{d}\right)
}\left( c_{1}^{\left( B_{d}\right) }+c_{2}^{\left( B_{d}\right) }\right) %
\right]
\end{equation}%
\begin{equation}
\Delta m_{B_{s}}^{\left( NP\right) }=\frac{g_{X}^{2}}{9m_{Z^{\prime }}^{2}}%
\left\vert \left( V_{DL}^{\ast }\right) _{32}\left( V_{DL}\right)
_{33}\right\vert ^{2}f_{B_{s}}^{2}B_{B_{s}}\eta _{B_{s}}m_{B_{s}}+\frac{8}{3}%
f_{B_{s}}^{2}\eta _{B_{s}}B_{B_{s}}m_{B_{s}}\left[ k_{2}^{\left(
B_{s}\right) }c_{3}^{\left( B_{s}\right) }+k_{1}^{\left( B_{s}\right)
}\left( c_{1}^{\left( B_{s}\right) }+c_{2}^{\left( B_{s}\right) }\right) %
\right]
\end{equation}%
In our numerical analysis, we use the following numerical values of the
meson parameters \cite%
{Dedes:2002er,Aranda:2012bv,Khalil:2013ixa,Queiroz:2016gif,Buras:2016dxz,Ferreira:2017tvy,Duy:2020hhk}%
: 
\begin{eqnarray}
\Delta m_{K} &=&\left( 3.484\pm 0.006\right) \times 10^{-12}MeV,\hspace{1.5cm%
}\Delta m_{K}^{\left( SM\right) }=3.483\times 10^{-12}MeV  \notag \\
f_{K} &=&160MeV,\hspace{1.5cm}B_{K}=0.85,\hspace{1.5cm}\eta _{K}=0.57, 
\notag \\
k_{1}^{\left( K\right) } &=&-9.3,\hspace{1.5cm}k_{2}^{\left( K\right) }=30.6,%
\hspace{1.5cm}m_{K}=497.614MeV,\hspace{1.5cm}
\end{eqnarray}%
\begin{eqnarray}
\left( \Delta m_{B_{d}}\right) _{\exp } &=&\left( 3.337\pm 0.033\right)
\times 10^{-10}MeV,\hspace{1.5cm}\Delta m_{B_{d}}^{\left( SM\right)
}=3.582\times 10^{-10}MeV,  \notag \\
f_{B_{d}} &=&188MeV,\hspace{1.5cm}B_{B_{d}}=1.26,\hspace{1.5cm}\eta
_{B_{d}}=0.55,  \notag \\
k_{1}^{\left( B_{d}\right) } &=&-0.52,\hspace{1.5cm}k_{2}^{\left(
B_{d}\right) }=0.88,\hspace{1.5cm}m_{B_{d}}=5279.5MeV,\hspace{1.5cm}
\end{eqnarray}%
\begin{eqnarray}
\left( \Delta m_{B_{s}}\right) _{\exp } &=&\left( 104.19\pm 0.8\right)
\times 10^{-10}MeV,\hspace{1.5cm}\Delta m_{B_{s}}^{\left( SM\right)
}=121.103\times 10^{-10}MeV,  \notag \\
f_{B_{s}} &=&225MeV,\hspace{1.5cm}B_{B_{s}}=1.26,\hspace{1.5cm}\eta
_{B_{s}}=0.55,  \notag \\
k_{1}^{\left( B_{s}\right) } &=&-0.52,\hspace{1.5cm}k_{2}^{\left(
B_{s}\right) }=0.88,\hspace{1.5cm}m_{B_{s}}=5366.3MeV,\hspace{1.5cm}
\end{eqnarray}%
\begin{figure}[h]
\includegraphics[width=0.51\textwidth]{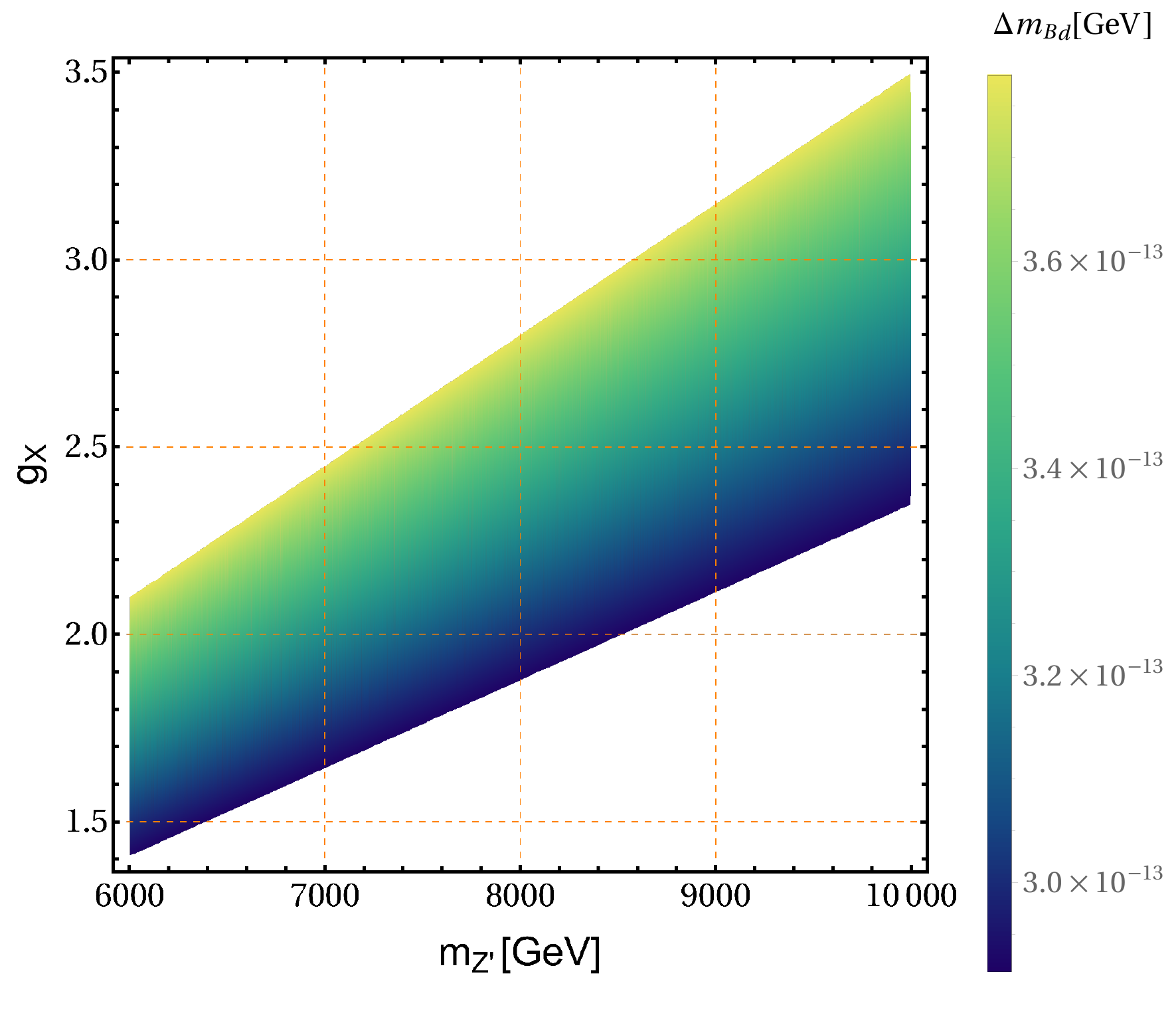}%
\includegraphics[width=0.51\textwidth]{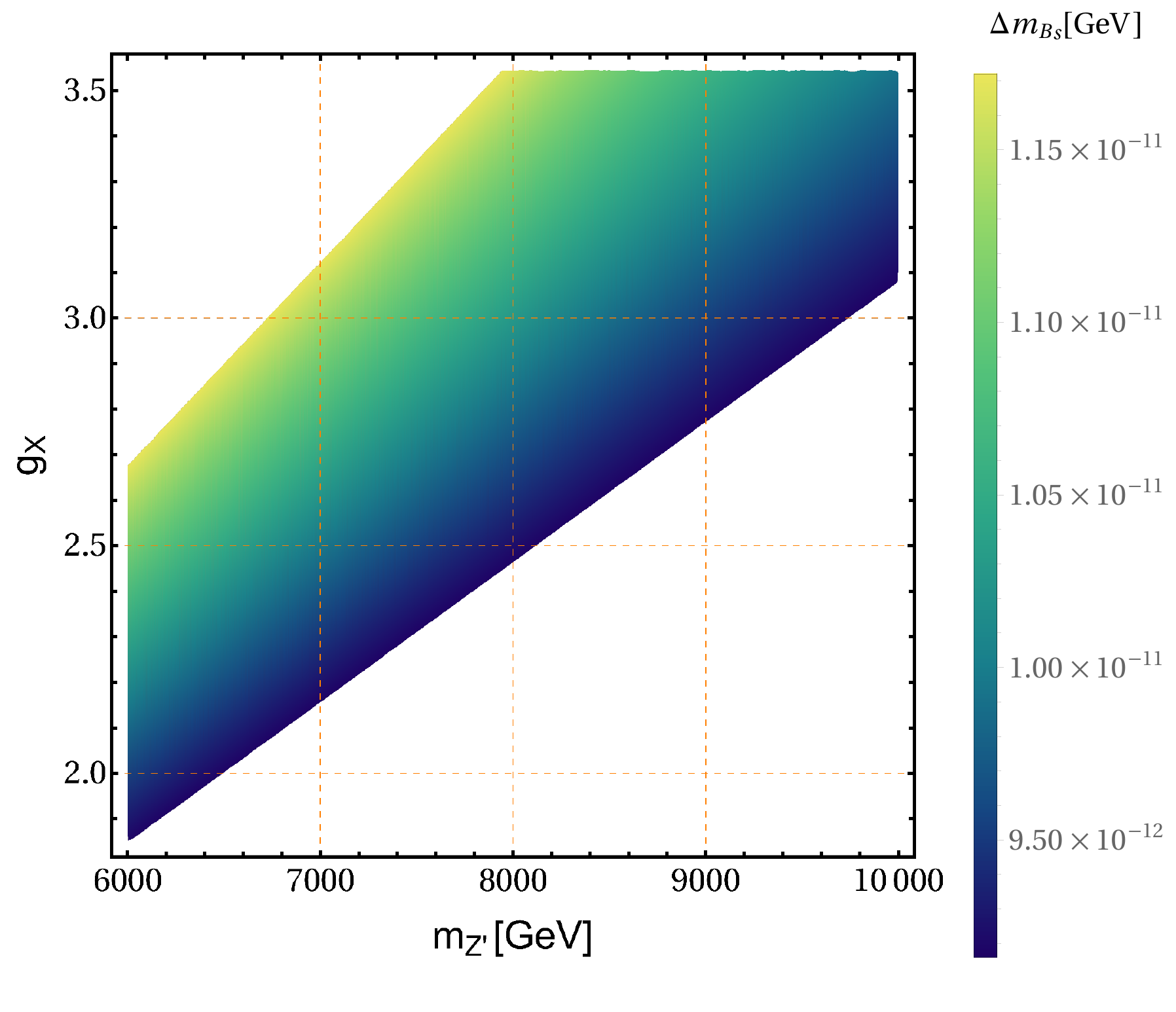} 
\caption{Allowed region in the $m_{Z^{\prime }}-g_{X}$ plane consistent with
the constraint arising from $B_{d}^{0}-\bar{B}_{d}^{0}$ (left-plot) and $%
B_{s}^{0}-\bar{B}_{s}^{0}$ (right-plot) mixings. Here we fix the couplings
of the flavor violating neutral Yukawa interactions as $2\times 10^{-4}$ and 
$10^{-3}$ for the left and right plots, respectively.}
\label{BBbar}
\end{figure}
We plot in Figure \ref{BBbar} the allowed region in the $m_{Z^{\prime
}}-g_{X}$ plane, consistent with the constraint arising from $B_{d}^{0}-\bar{%
B}_{d}^{0}$ (left-plot) and $B_{s}^{0}-\bar{B}_{s}^{0}$ (right-plot)
mixings. In our numerical analysis we have considered a simplified benchmark
scenario where the couplings of the flavor violating neutral Yukawa
interactions responsible for the $B_{d}^{0}-\bar{B}_{d}^{0}$ and $B_{s}^{0}-%
\bar{B}_{s}^{0}$ mixings take values of about $2\times 10^{-4}$ and $10^{-3}$%
, respectively. Furthermore, we have set $M_{H_{1}}=1.2$~TeV, $M_{H_{2}}=1.3$%
~TeV, $M_{A_{1}}=M_{A_{2}}=1$~TeV and the $Z^{\prime }$ mass has been taken
to be in the range $6$ TeV$\leqslant m_{Z^{\prime }}\leqslant $ $10$ TeV. As
seen from Figure \ref{BBbar}, the $B_{d}^{0}-\bar{B}_{d}^{0}$ and $B_{s}^{0}-%
\bar{B}_{s}^{0}$ meson\ oscillations caused by the flavor changing neutral
interactions reach values close to their experimental upper limits, thus
giving rise to the allowed regions in the $g_{X}-m_{Z^{\prime }}$ plane
consistent with these constraints. On the other hand, concerning the $K^{0}-%
\bar{K}^{0}$ mixing, we have numerically checked that in the aforementioned
simplified benchmark scenario and above-described region of parameter space
with a corresponding flavor violating Yukawa coupling of the order of $%
0.5\times 10^{-5}$, the obtained values for the $\Delta m_{K}$ are
consistent with the meson oscillation experimental data. It is worth
mentioning that in our numerical analysis we have considered the case of
real down quark Yukawa couplings, which implies that CP violation in the
quark sector only arises from the up-type quark sector. Therefore, the
constraints that are usually imposed on any possible new contributions to
the $K^{0}-\bar{K}^{0}$, $B_{d}^{0}-\bar{B}_{d}^{0}$ and $B_{s}^{0}-\bar{B}%
_{s}^{0}$ meson oscillations, arising from CP-violating processes, are not
relevant for our case.

\section{Dark matter}

\label{DM} Both the $SU(2)_L \times U(1)_Y \times U(1)_X$ gauge symmetry and
the $Z_4$ discrete symmetry are spontaneously broken, whereas the $Z_2$
symmetry is preserved. This $Z_2$ conservation implies that the particles
carrying a non trivial $Z_2$ charge always couple in pairs, and therefore
the lightest of the electrically neutral $Z_2$ odd particles is a dark
matter candidate. The considered model contains two kinds of candidates: the
fermion singlet $\Psi_{nR}$ and the scalar singlets that are either $%
\varphi_1$ or $\varphi_2$. The fields, $\varphi_1, \Psi_{nR}$, carry the $%
\text{X}$ charge but the $\varphi_2$ does not. Except for the Yukawa
interaction of $\varphi_2$ with new fermions $\Omega_{mR},\Psi_{nR}$, the $%
\varphi_2$ has only quartic scalar interactions. Thus, the $\varphi_2$
mainly annihilates into $W^+W^-, ZZ, t \bar{t}, b \bar{b}, HH $ via a scalar
portal interaction, 
and the relic density is governed by Higgs portal interactions and takes the
form \cite{HuongRelic1} 
\begin{eqnarray}
\Omega h^2 \simeq 0.1 \left(\frac{m_{\varphi_2}}{\lambda_{eff} \times 1.354 
\text{TeV}} \right)^2,
\end{eqnarray}
where $\lambda_{eff}$ is an effective coupling, which depends on all
trilinear Higgs couplings of $\varphi_2$ with the remaining Higgs. In order
to consistently reproduce the experimental value of the dark matter relic
density \cite{HuongRelic2}, $\Omega h^2 =0.1198\pm 0.0026$, the mass $%
m_{\varphi_2}$ has to fulfill the constraint, $m_{\varphi_2} < \lambda_{eff}
\times 1.5 \text{TeV}$. If the effective coupling is in the range $0.5
<\lambda_{eff} < 1.5$, the dark matter mass satisfies $0.75 \text{TeV}%
<m_{\varphi_2}< 2.25 \text{TeV}$. Since $\varphi_2$ is a gauge singlet
scalar, it is electrically neutral, and then it only scatters off in a quark
antiquark pair via SM Higgs portal interaction, which has a rate
proportional to the quartic coupling of $\varphi_2^2 H^2$, denoted as $%
\lambda_D$. The tree-level SM Higgs exchange produces a spin-independent
cross section given by \cite{Directsearch}: 
\begin{eqnarray}
\sigma_{\varphi_2-p,n}\simeq 3.88 \times 10^{-45} \left( \frac{\lambda_{D}}{%
0.5}\right)^2 \left(\frac{2 \text{TeV}}{m_{\varphi_2} } \right)^2 \text{cm}%
^2.
\end{eqnarray}
This scattering cross-section reaches the direct detection limit from the
XENON1T experiment \cite{XENON1T} for dark matter mass around $2 \text{TeV}$
and effective coupling $\lambda_{D}\simeq 0.5.$ Unlike $\varphi_2$, $%
\varphi_1$ carries a $\text{X}-$charge. Thus, if $\varphi_1$ is a dark
matter candidate, it will scatter off a nucleon not only through the
exchange of the Higgs, but also via the exchange of a new neutral gauge
boson. This obeys direct detection limits from the XENON1T experiment \cite%
{XENON1T}, and yields the correct relic density if the dark matter mass is
heavier than $3 \text{TeV}$, (see in \cite{Queiroz}). \newline
\hspace*{0.3cm} Let us now assume that the dark matter is the neutral
fermion, denoted $\Psi$. This is a $SU(2)_L$ singlet 
which does not carry hypercharge, but carries a $U(1)_X$ charge. 
Thus, the interaction of two dark matter candidates $\Psi$ with a new
neutral gauge boson, called $Z_X$, determines the dark matter phenomenology. 
$\Psi_n$ annihilates into SM particles through the exchange of a new gauge
boson. The relic density is given as \cite{Huong2019vej} 
\begin{eqnarray}
\Omega_\Phi h^2 \simeq 0.1 \text{pb}\left( \frac{\alpha}{150 \text{GeV}}%
\right)^{-2} \left(\frac{m_{\Psi}}{2.86 \text{TeV}} \right)^2,  \label{Rel}
\end{eqnarray}
where $\left(\frac{\alpha}{150 \text{GeV}} \right)^2 \simeq 1 \text{pb}$.
This relic density satisfies the experimental value \cite{HuongRelic2} if
and only if $m_\Psi > 3.13 \text{TeV}$. At the tree-level, the dark matter $%
\Psi_N$ scatters off nuclei via the exchange of the gauge boson $\text{Z}_X$%
. In the limit, $m_\Psi > 3.13 \text{TeV}$, the cross-section of this
scattering is predicted to be consistent with the XENON1T experiment (see 
\cite{Abada:2021yot, HuongRelic1, Queiroz}).

\section{Leptogenesis}

\label{leptogenesis} 
\begin{figure}[tbp]
	\centering
	\begin{minipage}{0.5\textwidth}
		{\label{a}\includegraphics[width=\textwidth]{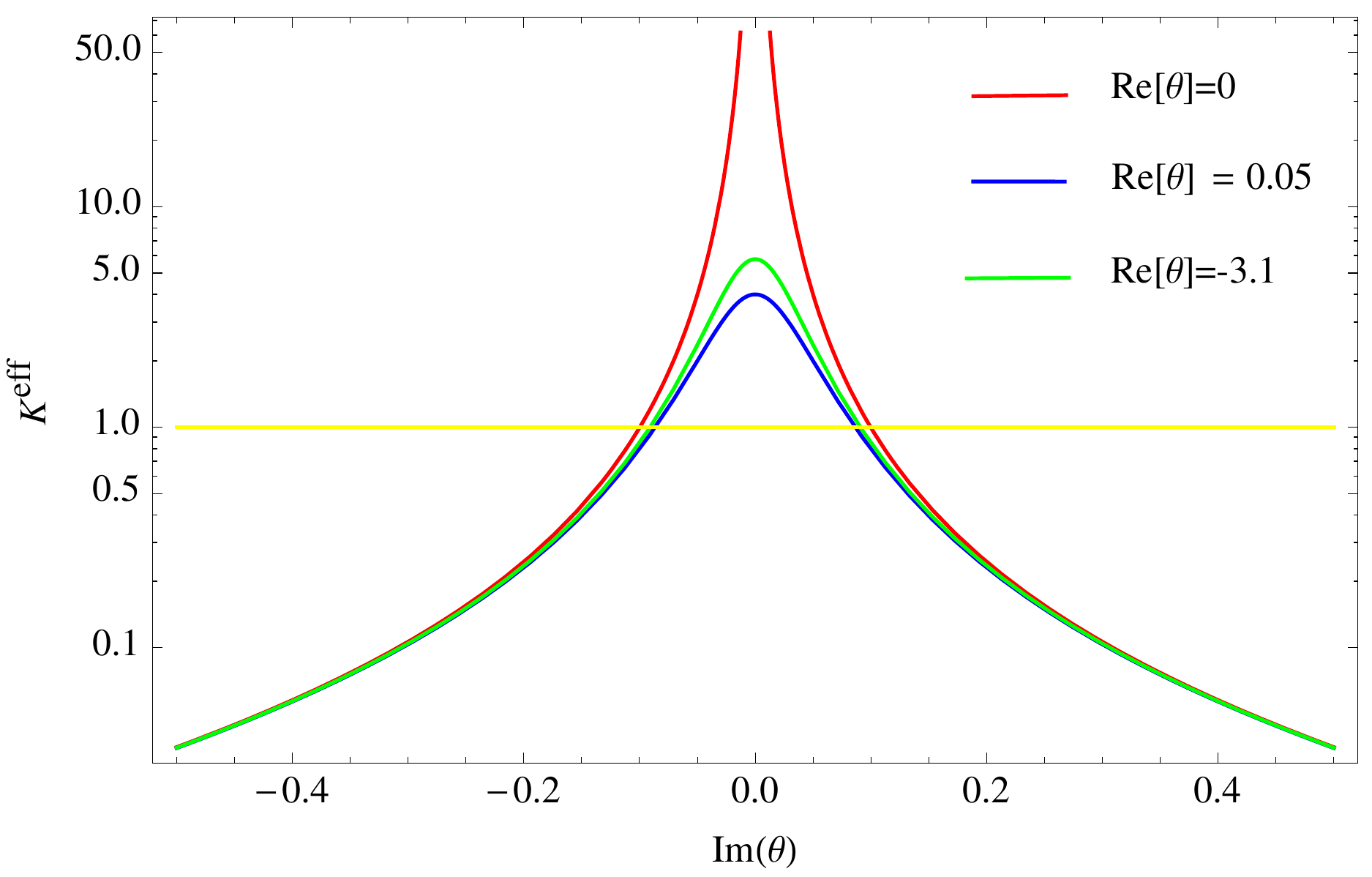}}
	\end{minipage}\hfill 
	\begin{minipage}{0.5\textwidth}
		\includegraphics[width=\textwidth]{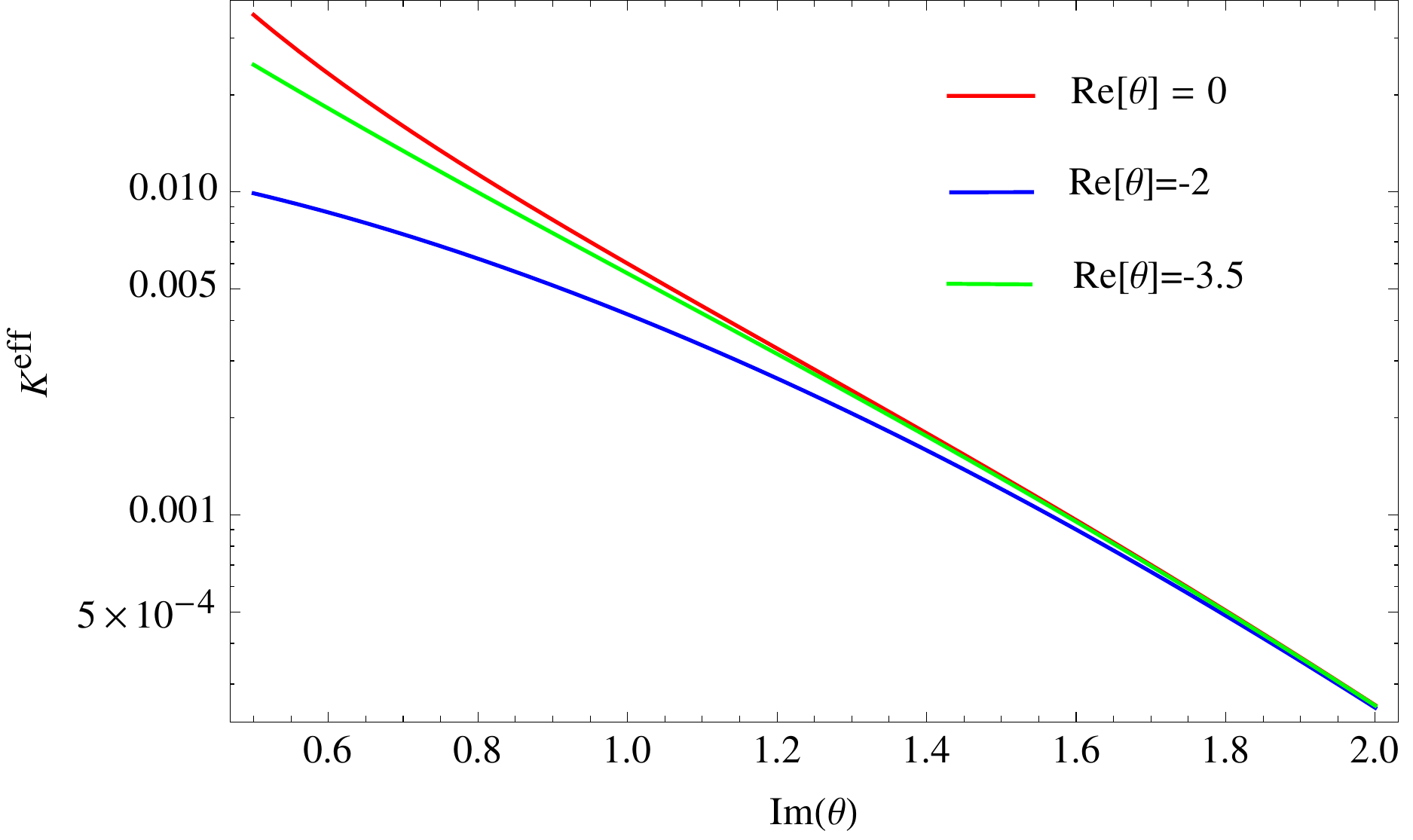}
	\end{minipage}
	\caption{The washout parameter as a function of the $\Im \lbrack \protect%
		\theta ]$. }
	\label{washout}
\end{figure}

\begin{figure}[tbp]
	\centering
	\begin{minipage}{0.5\textwidth}
		{\label{a}\includegraphics[width=\textwidth]{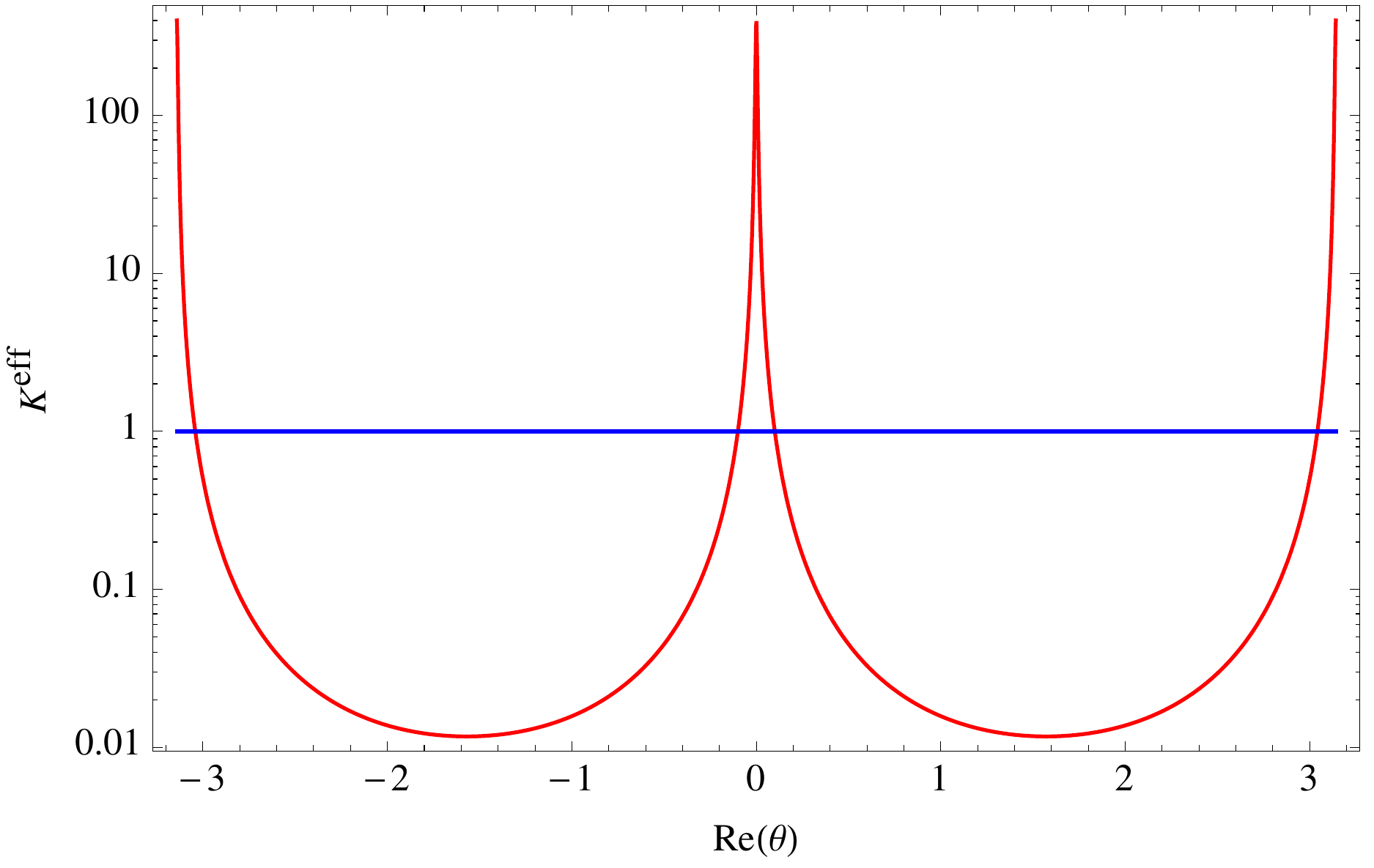}}
	\end{minipage}\hfill 
	\begin{minipage}{0.5\textwidth}
		\includegraphics[width=\textwidth]{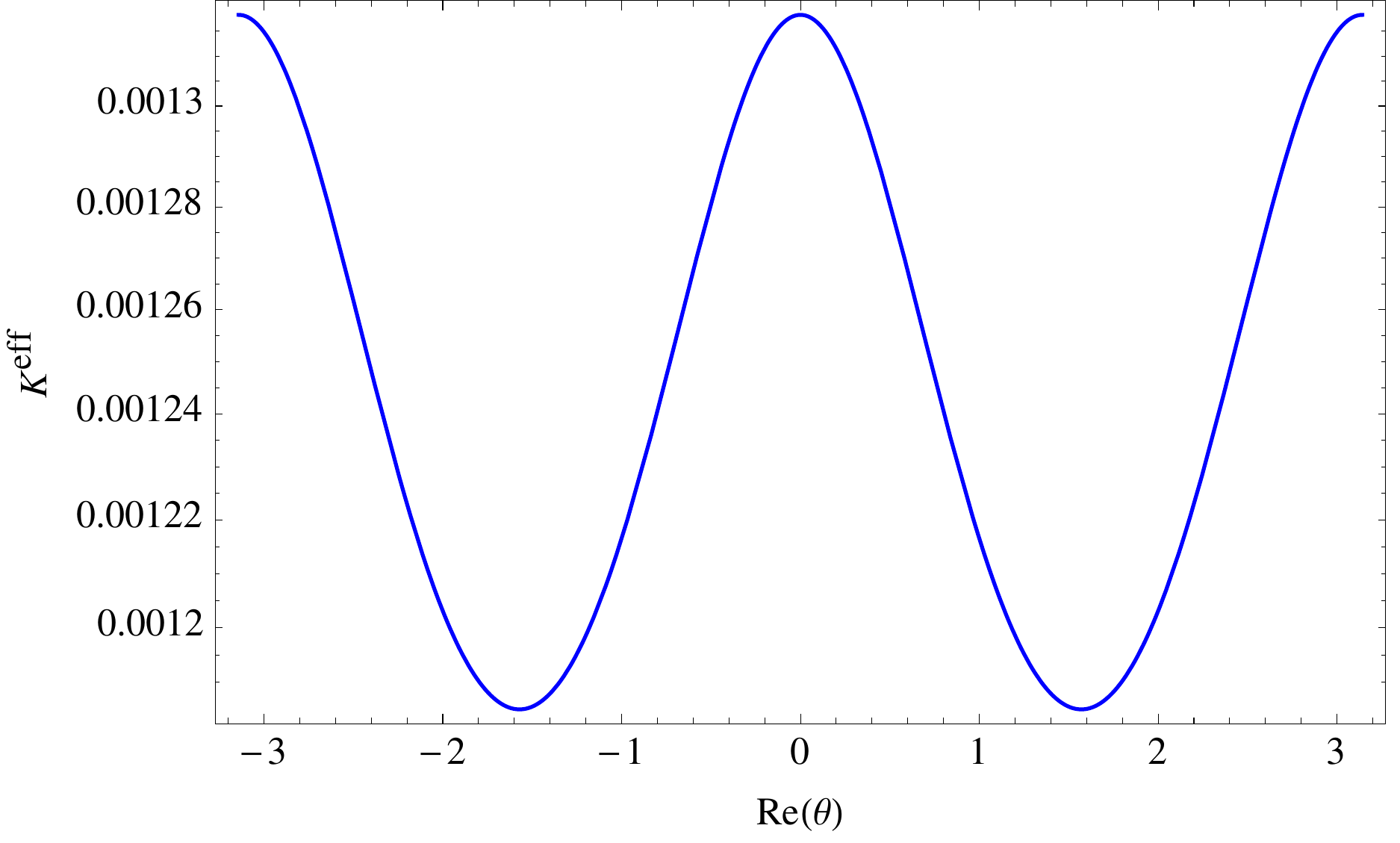}
	\end{minipage}
	\caption{The washout parameter as a function of the $\Re \lbrack \protect%
		\theta ]$. Left panel: $\Im \lbrack \protect\theta ]=0.005$, right panel: $%
		\Im \lbrack \protect\theta ]=1.5$. }
	\label{washout1}
\end{figure}

Before counting flavored $\text{CP}-$asymmetry
parameters from the decay of each heavy pseudo-Dirac neutrinos, we must
rotate the sterile neutrinos $\nu _{aR},N_{aR}$, into their mass basis. As
previously 
mentioned, the sterile neutrinos form three pairs of quasi-degenerate
pseudo-Dirac fermions due to the small $\mu -$parameter, which is generated
at two loop level. The eigenstates $\left( N_{aR}^{+},N_{aR}^{-}\right) $,
corresponding to the eigenvalues $\left( M_{\nu }^{+},-M_{\nu }^{-}\right) $%
, are related to $\left( \nu _{Ra},N_{Ra}\right) $ through: 
\begin{equation}
N_{aR}^{+}=\frac{1}{\sqrt{2}}\left( \nu _{aR}+N_{aR}\right) ,\hspace*{0.3cm}%
\hspace*{0.3cm}N_{aR}^{-}=\frac{i}{\sqrt{2}}\left( \nu _{aR}-N_{aR}\right) .
\end{equation}%
Henceforth, the Yukawa interactions of $\nu _{aR},N_{aR}$ can be modified
and written on the new basis as follows 
\begin{equation}
-\mathcal{L}_{Y}^{l}\ni \sum_{i=1}^{3}\sum_{a=1}^{3}y_{ia}^{\nu }\overline{l}%
_{iL}\widetilde{\phi }_{2}\frac{\left( N_{aR}^{+}-iN_{aR}^{-}\right) }{\sqrt{%
2}}+\sum_{a=1}^{3}\sum_{n=1}^{2}\left( x_{N}\right) _{aj}\frac{\left( 
\overline{N^{+}}_{aR}+i\overline{N^{-}}_{aR}\right) }{\sqrt{2}}\Psi
_{nR}^{C}\varphi _{1}  \label{yu1}
\end{equation}%
The lepton asymmetry is generated from the decay of the lightest pair of
pseudo-Dirac neutrinos, called $(N_{\pm })$, and it can be enhanced due to a
resonance effect \cite{Lepto1, Lepto2, Dib:2019jod}. If $(N_{\pm })$ couples
only with a SM lepton, the washout factor is determined from the inverse
decay of the SM lepton and Higgs into the pair of pseudo-Dirac neutrinos.
Since the washout factor has a quadratic suppression with the $\mu -$%
parameter \cite{Lepto3, Lepto4}, the smallness of $\mu $ can naturally
suppress the washout factor. However, in our model, the $\mu -$parameter can
be small in a technically natural way since it is generated at the two-loop
level. 
Therefore, in addition to the inverse decay $lH^{\pm }\rightarrow N_{\pm
}\rightarrow lH^{^{\pm }}$, the model creates new washout processes: $%
lH^{^{\pm }}\rightarrow N_{\pm }\rightarrow \Psi _{R}\varphi _{1}$ (see Eq.(%
\ref{yu1})). In the high-temperature region (temperature larger than the
inverse see-saw scale), new washout processes can be avoided if the Yukawa
couplings $(x_{aj}^{N})$ are very suppressed. This is unreasonable, because
the $\mu -$parameter is generated at two loop level, as shown 
in Eq.(\ref{mu1a}). If the temperature of the Universe drops below the
see-saw scale, the inverse decays producing $N_{aR}$ fall out of thermal
equilibrium, and thermal leptogenesis can happen. Assuming that the fermions 
$\Psi _{nR}$ are heavier than the lightest pseudo-Dirac pair, $(N^{\pm })$,
the lepton asymmetry generates via the decay of the lightest pair of
pseudo-Dirac $(N_{\pm }\equiv N_{1}^{\pm })$ to the SM Higgs and lepton
doublets and has the following form 
\begin{equation}
\epsilon _{\pm }=\sum_{\alpha =1}^{3}\frac{\left\{ \Gamma \left( N_{\pm
}\rightarrow l_{\alpha }H^{+}\right) -\Gamma \left( N_{\pm }\rightarrow \bar{%
l}_{\alpha }H^{-}\right) \right\} }{\left\{ \Gamma \left( N_{\pm
}\rightarrow l_{\alpha }H^{+}\right) +\Gamma \left( N_{\pm }\rightarrow \bar{%
l}_{\alpha }H^{-}\right) \right\} }\simeq \frac{\func{Im}\left\{ \left( %
\left[ \left( y_{N_{+}}\right) ^{\dagger }\left( y_{N_{-}}\right) \right]
^{2}\right) _{11}\right\} }{8\pi A_{\pm }}\frac{r}{r^{2}+\frac{\Gamma _{\mp
}^{2}}{m_{N_{\mp }}^{2}}},
\end{equation}%
where $y_{N_{\pm }}=\frac{y^{\left( \nu \right) }}{\sqrt{2}}\left( 1\pm 
\frac{1}{4}M^{-1}\mu \right) ,$ $r\equiv \frac{m_{N_{+}}^{2}-m_{N_{-}}^{2}}{%
m_{N_{+}}m_{N_{-}}},\hspace*{0.3cm}A_{\pm }=\left( (y^{\nu })^{\dag }y^{\nu
}\right) _{11},\hspace*{0.3cm}\Gamma _{\pm }\equiv \frac{A_{\pm }m_{N_{\pm }}%
}{8\pi }$. In the weak and strong washout region, the approximate baryon
asymmetry is estimated as 
\begin{eqnarray}
\eta _{B} &=&\frac{\epsilon _{N_{\pm }}}{g_{\ast }}\hspace*{0.3cm}\text{for}%
\hspace*{0.3cm}K_{N_{\pm }}^{eff}\ll 1,  \notag \\
\eta _{B} &=&\frac{0.3\epsilon _{N_{\pm }}}{g_{\ast }K_{N_{\pm }}^{eff}(\ln
K_{N_{\pm }}^{eff})^{0.6}}\hspace*{0.3cm}\text{for}\hspace*{0.3cm}K_{N_{\pm
}}^{eff}\gg 1.  \label{app1}
\end{eqnarray}%
Here, $g_{\ast }\simeq 118$ is the number of relativistic degrees of
freedom. In the leptogenesis epoch, the effective washout parameter is
defined as 
\begin{equation}
K_{N_{\pm }}^{eff}\simeq \left( \frac{\Gamma _{+}+\Gamma _{-}}{H}\right)
\left( \frac{m_{N^{+}}-m_{N^{-}}}{\Gamma _{\pm }}\right) ^{2}
\end{equation}%
where $\text{H}=\sqrt{\frac{4\pi ^{3}g_{\ast }}{45}}\frac{T^{2}}{M_{Pl}}$ is
the Hubble constant. For a successful leptogenesis and inverse see-saw
mechanism, we have to {choose} the parameter space including four Yukawa
couplings $y^{\nu },y^{N},x_{N},x_{\Psi }$, two VEVs $v_{2},v_{\sigma }$,
and as well as the masses of $\Omega _{n},\Psi _{n},\Im \lbrack \varphi
_{i}],\Re \lbrack \varphi _{i}]$. To study this result more quantitatively
it is convenient to use the Casas-Ibarra parametrization \cite%
{Casas,Das:2012ze} of the Yukawa coupling $y^{\nu }$, as follows 
\begin{equation}
y^{\nu }=\frac{v_{\sigma }}{v_{2}}\left( \text{U}_{\text{PMNS}}\text{M}_{\nu
}^{\frac{1}{2}}\text{R}\mu ^{-\frac{1}{2}}y^{N}\right) .
\end{equation}%
In this parametrization, R is a complex orthogonal matrix that has the
general form 
\begin{equation}
R=\left( 
\begin{array}{ccc}
c_{y}c_{z} & -s_{x}c_{z}s_{y}-c_{x}s_{z} & s_{x}s_{z}-c_{x}s_{y}c_{z} \\ 
c_{y}s_{z} & c_{x}c_{z}-s_{x}s_{y}s_{z} & -c_{z}s_{x}-c_{x}s_{y}s_{z} \\ 
s_{y} & s_{x}c_{y} & c_{x}c_{y}%
\end{array}%
\right) ,
\end{equation}%
where $c_{x}=\cos x,s_{x}=\sin x$ and so on, with $x,y,z\in C$. Here $\text{U%
}_{\text{PMNS}}$ is the Pontecorvo-Maki-Nakagawa-Sakata mixing matrix for
the lepton sector, while $\text{M}_{\nu }=\text{Diag}(m_{\nu _{1}},m_{\nu
_{2}},m_{\nu _{3}})$ is a diagonal light active neutrino mass matrix. The
current best fit values for the light neutrino masses, mixing angles and the
CP violation phase of $\text{U}_{\text{PMNS}}$ are given in \cite%
{deSalas:2020pgw}. In order to reduce the number of free parameters, we need
to take some assumptions. Therefore, we fix the three complex mixing angles $%
x=0,y=z=\Re \lbrack \theta ]+i\Im \lbrack \theta ]$. Notice that the values
of the Yukawas $y_{ij}^{\nu }$ are sensitive to the $\Im \lbrack \theta ]$
but not sensitive to $\Re \lbrack \theta ]$. Additionally, we assume that $%
Y^{(N)}$ is a diagonal matrix and is fixed as $Y^{(N)}=\text{Diag}(\text{0.5}%
,\text{0.9i},\text{1.8})$. Two VEVs are chosen as $v_{2}=24.6\text{GeV}$, $%
v_{\sigma }\simeq 5\times 10^{3}\text{GeV }$. These choices guarantee that
the gauge interactions of $N_{1}^{\pm }$ decouple at the leptogenesis epoch 
\cite{Mohapatra, SFKing}. In the perturbative region, the Yukawa coupling
given in $(\ref{yu1})$ has to satisfy the condition: $(y^{\nu
})_{ij}^{2}<4\pi $. Thus, the large-value domain of $\Im \lbrack \theta ]$
is inhibited, namely $-2<\Im \lbrack \theta ]<2$. With the above choices,
the mass of light neutrinos can approach the experimental limit \cite%
{deSalas:2020pgw} when $\mu $ satisfies: $\mu \simeq 1\text{keV}.$ \newline
\hspace*{0.3cm} Figs.(\ref{washout}) show the strong sensitivity of the
washout parameter with respect to $\Im \lbrack \theta ]$, especially in the
region of small values of $\Im \lbrack \theta ]$. The strong washout regime, 
$K_{eff}\gg 1$, corresponds to small values of $\Im \lbrack \theta ]$ (left
panel), while the weak washout regime, $K_{eff}\ll 1$ corresponds to large
values of $\Im \lbrack \theta ]$ (right panel). For fixed values of $\Im
\lbrack \theta ]$, the washout parameter oscillates over $\Re \lbrack \theta
]$ (see Figs. (\ref{washout1})), and the amplitude of oscillation increases
as the value of $\Im \lbrack \theta ]$ decreases. The washout effect is not
affected much by $\Re \lbrack \theta ]$. 
\begin{figure}[tbp]
	\centering
	\begin{minipage}{0.5\textwidth}
		{\label{a}\includegraphics[width=\textwidth]{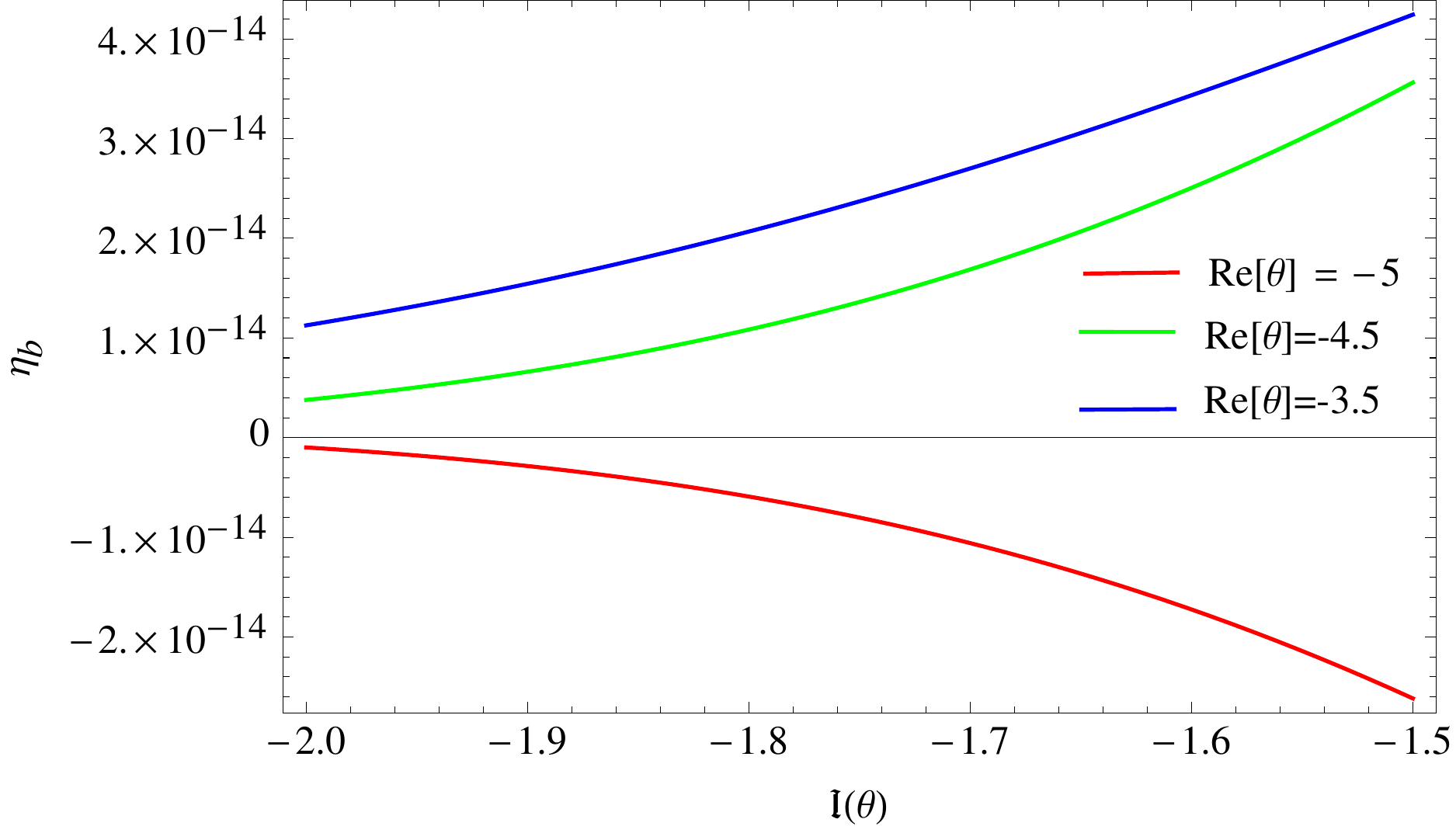}}
	\end{minipage}\hfill 
	\begin{minipage}{0.5\textwidth}
		\includegraphics[width=\textwidth]{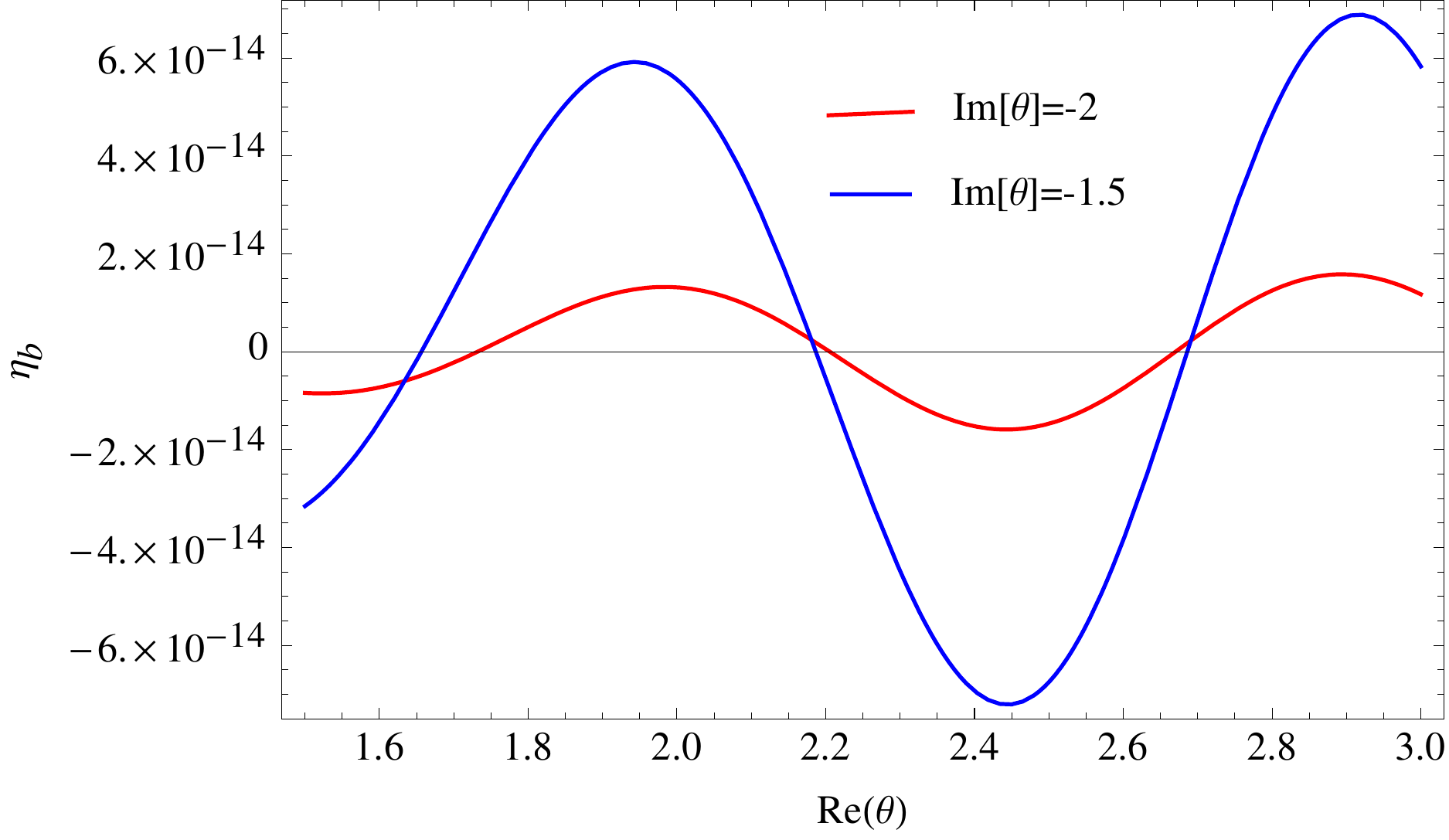}
	\end{minipage}
	\caption{Estimating the baryon asymmetry in the weak-washout regime. Left
		panel:Log-Plot of the baryon asymmetry as a function of the $\Im \lbrack 
		\protect\theta ]$. Right panel: Log-Plot the baryon asymmetry as a function
		of the $\Re \lbrack \protect\theta ]$. }
	\label{Asymmetry1}
\end{figure}
\begin{figure}[]
	\centering
	\begin{minipage}{0.5\textwidth}
		{\label{a}\includegraphics[width=\textwidth]{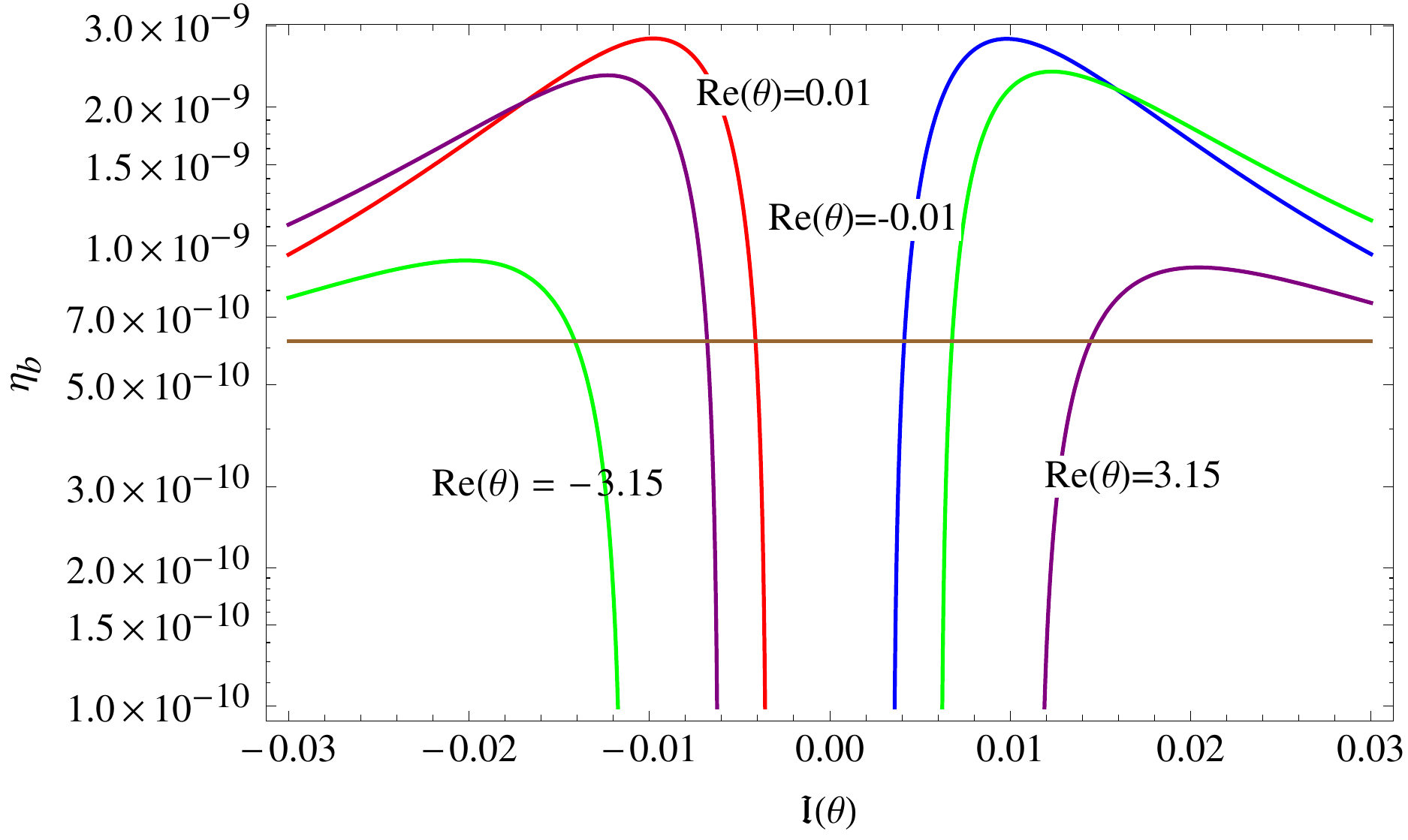}}
	\end{minipage}\hfill 
	\begin{minipage}{0.5\textwidth}
		\includegraphics[width=\textwidth]{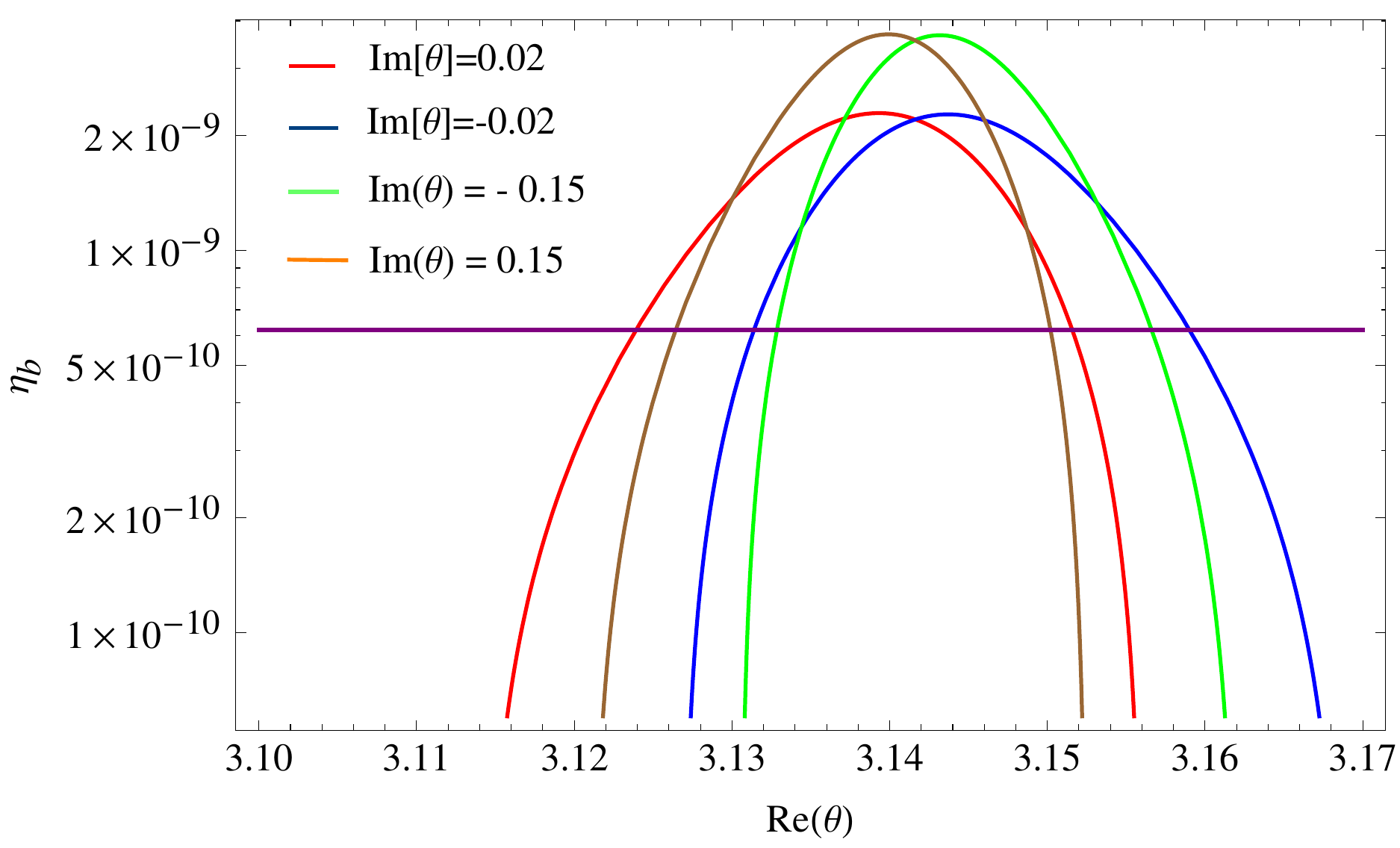}
	\end{minipage}
	\caption{Estimating the baryon asymmetry in the strong-washout regime. Left
		panel: Plot of the baryon asymmetry as a function of the $\Im[ \protect\theta%
		]$. Right panel: Plot of the baryon asymmetry as a function of the $\Re[ 
		\protect\theta]$. }
	\label{Asymmetry2}
\end{figure}
Figs. (\ref{Asymmetry1}) show the generated baryon asymmetry from a
numerical study of the approximate results of the Boltzmann equation, given
in Eq.(\ref{app1}). In the weak washout regime, the predicted baryon
asymmetry barely reaches the observed values $\eta _{B}\simeq 6.08\times
10^{-10}.$

In the strong-washout regime, the baryon asymmetry generation can reach the
observed value, for the small phase entries of the $\text{R}$ matrix. For
small values of $\Im \lbrack \theta ]$, we obtain the strong washout effect
for every choice of $\Re \lbrack \theta ]$ (see the left panel of Fig.(\ref%
{washout})). However, not every value of $\Re \lbrack \theta ]$ predicts a
sufficient amount of baryon asymmetries for the Universe. The right-panel of
Figs. (\ref{Asymmetry2}) shows the amount of baryon asymmetry as a function
of $\Re \lbrack \theta ]$, which seems to drastically change under
variations of $\Re \lbrack \theta ]$. Fixing $\Re \lbrack \theta ]$, the
behavior of the baryon asymmetry as a function of the imaginary part of the
complex angle $\theta $ appears in the left panel of Figs. (\ref{Asymmetry2}%
). It drastically decreases as $\Im \lbrack \theta ]$ increases. We conclude
that in the present model leptogenesis is viable with a strong-washout
regime if we include the small phase of the $\text{R}$ matrix. The amount of
baryon asymmetry oscillates according to $\theta $, and the oscillation
amplitude can reach the observed value.

\newpage

\section{Conclusions}

\label{conclusions} We have built a 2HDM theory in which both particle
content and symmetry are enlarged. We added several gauge singlet scalars
and electrically charged vector-like fermions, as well as right-handed
Majorana neutrinos, with the SM gauge symmetry being supplemented by a $%
U(1)_{X}\times Z_{2}\times Z_{4}$ family symmetry. We have built a
renormalizable theory based on the given particle content, where to the best
of our knowledge, for the first time an inverse seesaw mechanism produces
the SM fermion mass hierarchy. The nonuniversal $U(1)_{X}$ gauge symmetry
and the discrete $Z_{4}$ symmetry are spontaneously broken, whereas the $%
Z_{2}$ symmetry is preserved, thus allowing to have stable scalar and
fermionic dark matter candidates. Our proposed theory is consistent with the
observed SM fermion mass hierarchy, the tiny values for the light active
neutrino masses, the lepton and baryon asymmetries of the Universe, the dark
matter relic density, the meson oscillation experimental data as well as the
muon and electron anomalous magnetic moments. \newline

\section*{Acknowledgments}

A.E.C.H and I.S. are supported by ANID-Chile FONDECYT 1210378, ANID-Chile
FONDECYT 1180232, ANID-Chile FONDECYT 3150472, ANID PIA/APOYO AFB180002 and
Milenio-ANID-ICN2019\_044. D.T.Huong acknowledges the financial support of
the Vietnam Academy of Science and Technology under Grant No.
NVCC05.13/21-21, and the International Centre of Physics at the Institute of
Physics, Vietnam Academy of Science and Technology with Grant number
CIP.2021.02. \newline

\centerline{\bf{REFERENCES}}\vspace{-0.4cm} 
\bibliographystyle{utphys}
\bibliography{Refs2HDMU1X}

\providecommand{\href}[2]{#2}\begingroup\raggedright\begin{thebibliography}{100}

\bibitem{Davidson:1987mh}
A.~Davidson and K.~C. Wali, ``{Universal Seesaw Mechanism?},''
\href{http://dx.doi.org/10.1103/PhysRevLett.59.393}{{\em Phys. Rev. Lett.}
  {\bfseries 59} (1987) 393}.

\bibitem{Davidson:1989bx}
A.~Davidson, S.~Ranfone, and K.~C. Wali, ``{Quark Masses and Mixing Angles From
  Universal Seesaw Mechanism},''
  \href{http://dx.doi.org/10.1103/PhysRevD.41.208}{{\em Phys. Rev. D}
  {\bfseries 41} (1990) 208}.

\bibitem{Davidson:1987tr}
A.~Davidson and K.~C. Wali, ``{Family Mass Hierarchy From Universal Seesaw
  Mechanism},'' \href{http://dx.doi.org/10.1103/PhysRevLett.60.1813}{{\em Phys.
  Rev. Lett.} {\bfseries 60} (1988) 1813}.

\bibitem{Balakrishna:1988ks}
B.~S. Balakrishna, A.~L. Kagan, and R.~N. Mohapatra, ``{Quark Mixings and Mass
  Hierarchy From Radiative Corrections},''
\href{http://dx.doi.org/10.1016/0370-2693(88)91676-0}{{\em Phys. Lett.}
  {\bfseries B205} (1988) 345--352}.

\bibitem{Ma:1988fp}
E.~Ma, ``{Radiative Quark and Lepton Masses Through Soft Supersymmetry
  Breaking},''
\href{http://dx.doi.org/10.1103/PhysRevD.39.1922}{{\em Phys. Rev.} {\bfseries
  D39} (1989) 1922}.

\bibitem{Ma:1989ys}
E.~Ma, D.~Ng, J.~T. Pantaleone, and G.-G. Wong, ``{One Loop Induced Fermion
  Masses and Exotic Interactions in a Standard Model Context},''
\href{http://dx.doi.org/10.1103/PhysRevD.40.1586}{{\em Phys. Rev.} {\bfseries
  D40} (1989) 1586}.

\bibitem{Ma:1990ce}
E.~Ma, ``{Hierarchical Radiative Quark and Lepton Mass Matrices},''
\href{http://dx.doi.org/10.1103/PhysRevLett.64.2866}{{\em Phys. Rev. Lett.}
  {\bfseries 64} (1990) 2866--2869}.

\bibitem{Ma:1998dn}
E.~Ma, ``{Pathways to naturally small neutrino masses},''
  \href{http://dx.doi.org/10.1103/PhysRevLett.81.1171}{{\em Phys. Rev. Lett.}
  {\bfseries 81} (1998) 1171--1174},
\href{http://arxiv.org/abs/hep-ph/9805219}{{\ttfamily arXiv:hep-ph/9805219
  [hep-ph]}}.

\bibitem{Kitabayashi:2000nq}
T.~Kitabayashi and M.~Yasue, ``{Radiatively induced neutrino masses and
  oscillations in an SU(3)(L) x U(1)(N) gauge model},''
  \href{http://dx.doi.org/10.1103/PhysRevD.63.095002}{{\em Phys. Rev.}
  {\bfseries D63} (2001) 095002},
\href{http://arxiv.org/abs/hep-ph/0010087}{{\ttfamily arXiv:hep-ph/0010087
  [hep-ph]}}.

\bibitem{Ma:2006km}
E.~Ma, ``{Verifiable radiative seesaw mechanism of neutrino mass and dark
  matter},'' \href{http://dx.doi.org/10.1103/PhysRevD.73.077301}{{\em Phys.
  Rev.} {\bfseries D73} (2006) 077301},
\href{http://arxiv.org/abs/hep-ph/0601225}{{\ttfamily arXiv:hep-ph/0601225
  [hep-ph]}}.

\bibitem{Dong:2006gx}
P.~V. Dong, D.~T. Huong, T.~T. Huong, and H.~N. Long, ``{Fermion masses in the
  economical 3-3-1 model},''
  \href{http://dx.doi.org/10.1103/PhysRevD.74.053003}{{\em Phys. Rev.}
  {\bfseries D74} (2006) 053003},
\href{http://arxiv.org/abs/hep-ph/0607291}{{\ttfamily arXiv:hep-ph/0607291
  [hep-ph]}}.

\bibitem{Chang:2006aa}
D.~Chang and H.~N. Long, ``{Interesting radiative patterns of neutrino mass in
  an SU(3)(C) x SU(3)(L) x U(1)(X) model with right-handed neutrinos},''
  \href{http://dx.doi.org/10.1103/PhysRevD.73.053006}{{\em Phys. Rev.}
  {\bfseries D73} (2006) 053006},
\href{http://arxiv.org/abs/hep-ph/0603098}{{\ttfamily arXiv:hep-ph/0603098
  [hep-ph]}}.

\bibitem{Gu:2007ug}
P.-H. Gu and U.~Sarkar, ``{Radiative Neutrino Mass, Dark Matter and
  Leptogenesis},'' \href{http://dx.doi.org/10.1103/PhysRevD.77.105031}{{\em
  Phys. Rev.} {\bfseries D77} (2008) 105031},
\href{http://arxiv.org/abs/0712.2933}{{\ttfamily arXiv:0712.2933 [hep-ph]}}.

\bibitem{Ma:2008cu}
E.~Ma and D.~Suematsu, ``{Fermion Triplet Dark Matter and Radiative Neutrino
  Mass},'' \href{http://dx.doi.org/10.1142/S021773230903059X}{{\em Mod. Phys.
  Lett.} {\bfseries A24} (2009) 583--589},
\href{http://arxiv.org/abs/0809.0942}{{\ttfamily arXiv:0809.0942 [hep-ph]}}.

\bibitem{Sierra:2008wj}
D.~Aristizabal~Sierra, J.~Kubo, D.~Restrepo, D.~Suematsu, and O.~Zapata,
  ``{Radiative seesaw: Warm dark matter, collider and lepton flavour violating
  signals},'' \href{http://dx.doi.org/10.1103/PhysRevD.79.013011}{{\em Phys.
  Rev.} {\bfseries D79} (2009) 013011},
\href{http://arxiv.org/abs/0808.3340}{{\ttfamily arXiv:0808.3340 [hep-ph]}}.

\bibitem{Nardi:2011jp}
E.~Nardi, D.~Restrepo, and M.~Velasquez, ``{Neutrino Masses in $SU(5)\times
  U(1)_F$ with Adjoint Flavons},''
  \href{http://dx.doi.org/10.1140/epjc/s10052-012-1941-1}{{\em Eur. Phys. J.}
  {\bfseries C72} (2012) 1941},
\href{http://arxiv.org/abs/1108.0722}{{\ttfamily arXiv:1108.0722 [hep-ph]}}.

\bibitem{Huong:2012pg}
D.~T. Huong, L.~T. Hue, M.~C. Rodriguez, and H.~N. Long, ``{Supersymmetric
  reduced minimal 3-3-1 model},''
  \href{http://dx.doi.org/10.1016/j.nuclphysb.2013.01.016}{{\em Nucl. Phys.}
  {\bfseries B870} (2013) 293--322},
\href{http://arxiv.org/abs/1210.6776}{{\ttfamily arXiv:1210.6776 [hep-ph]}}.

\bibitem{Restrepo:2013aga}
D.~Restrepo, O.~Zapata, and C.~E. Yaguna, ``{Models with radiative neutrino
  masses and viable dark matter candidates},''
  \href{http://dx.doi.org/10.1007/JHEP11(2013)011}{{\em JHEP} {\bfseries 11}
  (2013) 011},
\href{http://arxiv.org/abs/1308.3655}{{\ttfamily arXiv:1308.3655 [hep-ph]}}.

\bibitem{Ma:2013yga}
E.~Ma, I.~Picek, and B.~Radovčić, ``{New Scotogenic Model of Neutrino Mass
  with $U(1)_D$ Gauge Interaction},''
  \href{http://dx.doi.org/10.1016/j.physletb.2013.09.049}{{\em Phys. Lett.}
  {\bfseries B726} (2013) 744--746},
\href{http://arxiv.org/abs/1308.5313}{{\ttfamily arXiv:1308.5313 [hep-ph]}}.

\bibitem{Ma:2013mga}
E.~Ma, ``{Radiative Origin of All Quark and Lepton Masses through Dark Matter
  with Flavor Symmetry},''
  \href{http://dx.doi.org/10.1103/PhysRevLett.112.091801}{{\em Phys. Rev.
  Lett.} {\bfseries 112} (2014) 091801},
\href{http://arxiv.org/abs/1311.3213}{{\ttfamily arXiv:1311.3213 [hep-ph]}}.

\bibitem{Hernandez:2013mcf}
A.~E. Carcamo~Hernandez, R.~Martinez, and F.~Ochoa, ``{Radiative seesaw-type
  mechanism of quark masses in $SU(3)_C \otimes SU(3)_L \otimes U(1)_X$},''
  \href{http://dx.doi.org/10.1103/PhysRevD.87.075009}{{\em Phys. Rev.}
  {\bfseries D87} no.~7, (2013) 075009},
\href{http://arxiv.org/abs/1302.1757}{{\ttfamily arXiv:1302.1757 [hep-ph]}}.

\bibitem{Hernandez:2013dta}
A.~E. Carcamo~Hernandez, I.~de~Medeiros~Varzielas, S.~G. Kovalenko, H.~Päs,
  and I.~Schmidt, ``{Lepton masses and mixings in an $A_4$ multi-Higgs model
  with a radiative seesaw mechanism},''
  \href{http://dx.doi.org/10.1103/PhysRevD.88.076014}{{\em Phys. Rev.}
  {\bfseries D88} no.~7, (2013) 076014},
\href{http://arxiv.org/abs/1307.6499}{{\ttfamily arXiv:1307.6499 [hep-ph]}}.

\bibitem{Okada:2013iba}
H.~Okada and K.~Yagyu, ``{Radiative generation of lepton masses},''
  \href{http://dx.doi.org/10.1103/PhysRevD.89.053008}{{\em Phys. Rev.}
  {\bfseries D89} no.~5, (2014) 053008},
\href{http://arxiv.org/abs/1311.4360}{{\ttfamily arXiv:1311.4360 [hep-ph]}}.

\bibitem{Sierra:2014rxa}
D.~Aristizabal~Sierra, A.~Degee, L.~Dorame, and M.~Hirsch, ``{Systematic
  classification of two-loop realizations of the Weinberg operator},''
  \href{http://dx.doi.org/10.1007/JHEP03(2015)040}{{\em JHEP} {\bfseries 03}
  (2015) 040},
\href{http://arxiv.org/abs/1411.7038}{{\ttfamily arXiv:1411.7038 [hep-ph]}}.

\bibitem{Campos:2014lla}
M.~D. Campos, A.~E. Cárcamo~Hernández, S.~Kovalenko, I.~Schmidt, and
  E.~Schumacher, ``{Fermion masses and mixings in an $SU(5)$ grand unified
  model with an extra flavor symmetry},''
  \href{http://dx.doi.org/10.1103/PhysRevD.90.016006}{{\em Phys. Rev.}
  {\bfseries D90} no.~1, (2014) 016006},
\href{http://arxiv.org/abs/1403.2525}{{\ttfamily arXiv:1403.2525 [hep-ph]}}.

\bibitem{Boucenna:2014ela}
S.~M. Boucenna, S.~Morisi, and J.~W.~F. Valle, ``{Radiative neutrino mass in
  3-3-1 scheme},'' \href{http://dx.doi.org/10.1103/PhysRevD.90.013005}{{\em
  Phys. Rev.} {\bfseries D90} no.~1, (2014) 013005},
\href{http://arxiv.org/abs/1405.2332}{{\ttfamily arXiv:1405.2332 [hep-ph]}}.

\bibitem{Hernandez:2015hrt}
A.~E. Cárcamo~Hernández, ``{A novel and economical explanation for SM fermion
  masses and mixings},''
  \href{http://dx.doi.org/10.1140/epjc/s10052-016-4351-y}{{\em Eur. Phys. J.}
  {\bfseries C76} no.~9, (2016) 503},
\href{http://arxiv.org/abs/1512.09092}{{\ttfamily arXiv:1512.09092 [hep-ph]}}.

\bibitem{Aranda:2015xoa}
A.~Aranda and E.~Peinado, ``{A new radiative neutrino mass generation mechanism
  with higher dimensional scalar representations and custodial symmetry},''
  \href{http://dx.doi.org/10.1016/j.physletb.2016.01.007}{{\em Phys. Lett.}
  {\bfseries B754} (2016) 11--13},
\href{http://arxiv.org/abs/1508.01200}{{\ttfamily arXiv:1508.01200 [hep-ph]}}.

\bibitem{Restrepo:2015ura}
D.~Restrepo, A.~Rivera, M.~Sánchez-Peláez, O.~Zapata, and W.~Tangarife,
  ``{Radiative Neutrino Masses in the Singlet-Doublet Fermion Dark Matter Model
  with Scalar Singlets},''
  \href{http://dx.doi.org/10.1103/PhysRevD.92.013005}{{\em Phys. Rev.}
  {\bfseries D92} no.~1, (2015) 013005},
\href{http://arxiv.org/abs/1504.07892}{{\ttfamily arXiv:1504.07892 [hep-ph]}}.

\bibitem{Longas:2015sxk}
R.~Longas, D.~Portillo, D.~Restrepo, and O.~Zapata, ``{The Inert Zee Model},''
  \href{http://dx.doi.org/10.1007/JHEP03(2016)162}{{\em JHEP} {\bfseries 03}
  (2016) 162},
\href{http://arxiv.org/abs/1511.01873}{{\ttfamily arXiv:1511.01873 [hep-ph]}}.

\bibitem{Fraser:2015zed}
S.~Fraser, E.~Ma, and M.~Zakeri, ``{Verifiable Associated Processes from
  Radiative Lepton Masses with Dark Matter},''
  \href{http://dx.doi.org/10.1103/PhysRevD.93.115019}{{\em Phys. Rev.}
  {\bfseries D93} no.~11, (2016) 115019},
\href{http://arxiv.org/abs/1511.07458}{{\ttfamily arXiv:1511.07458 [hep-ph]}}.

\bibitem{Fraser:2015mhb}
S.~Fraser, C.~Kownacki, E.~Ma, and O.~Popov, ``{Type II Radiative Seesaw Model
  of Neutrino Mass with Dark Matter},''
  \href{http://dx.doi.org/10.1103/PhysRevD.93.013021}{{\em Phys. Rev.}
  {\bfseries D93} no.~1, (2016) 013021},
\href{http://arxiv.org/abs/1511.06375}{{\ttfamily arXiv:1511.06375 [hep-ph]}}.

\bibitem{Okada:2015bxa}
H.~Okada, N.~Okada, and Y.~Orikasa, ``{Radiative seesaw mechanism in a minimal
  3-3-1 model},'' \href{http://dx.doi.org/10.1103/PhysRevD.93.073006}{{\em
  Phys. Rev.} {\bfseries D93} no.~7, (2016) 073006},
\href{http://arxiv.org/abs/1504.01204}{{\ttfamily arXiv:1504.01204 [hep-ph]}}.

\bibitem{Wang:2015saa}
W.~Wang and Z.-L. Han, ``{Radiative linear seesaw model, dark matter, and
  $U(1)_{B-L}$},'' \href{http://dx.doi.org/10.1103/PhysRevD.92.095001}{{\em
  Phys. Rev.} {\bfseries D92} (2015) 095001},
\href{http://arxiv.org/abs/1508.00706}{{\ttfamily arXiv:1508.00706 [hep-ph]}}.

\bibitem{Sierra:2016qfa}
D.~Aristizabal~Sierra, C.~Simoes, and D.~Wegman, ``{Closing in on minimal dark
  matter and radiative neutrino masses},''
  \href{http://dx.doi.org/10.1007/JHEP06(2016)108}{{\em JHEP} {\bfseries 06}
  (2016) 108},
\href{http://arxiv.org/abs/1603.04723}{{\ttfamily arXiv:1603.04723 [hep-ph]}}.

\bibitem{Arbelaez:2016mhg}
C.~Arbeláez, A.~E. Cárcamo~Hernández, S.~Kovalenko, and I.~Schmidt,
  ``{Radiative Seesaw-type Mechanism of Fermion Masses and Non-trivial Quark
  Mixing},'' \href{http://dx.doi.org/10.1140/epjc/s10052-017-4948-9}{{\em Eur.
  Phys. J.} {\bfseries C77} no.~6, (2017) 422},
\href{http://arxiv.org/abs/1602.03607}{{\ttfamily arXiv:1602.03607 [hep-ph]}}.

\bibitem{Nomura:2016emz}
T.~Nomura and H.~Okada, ``{Radiatively induced Quark and Lepton Mass Model},''
  \href{http://dx.doi.org/10.1016/j.physletb.2016.08.023}{{\em Phys. Lett.}
  {\bfseries B761} (2016) 190--196},
\href{http://arxiv.org/abs/1606.09055}{{\ttfamily arXiv:1606.09055 [hep-ph]}}.

\bibitem{Nomura:2016fzs}
T.~Nomura and H.~Okada, ``{Four-loop Neutrino Model Inspired by Diphoton Excess
  at 750 GeV},'' \href{http://dx.doi.org/10.1016/j.physletb.2016.02.022}{{\em
  Phys. Lett.} {\bfseries B755} (2016) 306--311},
\href{http://arxiv.org/abs/1601.00386}{{\ttfamily arXiv:1601.00386 [hep-ph]}}.

\bibitem{Kownacki:2016hpm}
C.~Kownacki and E.~Ma, ``{Gauge $U(1)$ dark symmetry and radiative light
  fermion masses},''
  \href{http://dx.doi.org/10.1016/j.physletb.2016.06.024}{{\em Phys. Lett.}
  {\bfseries B760} (2016) 59--62},
\href{http://arxiv.org/abs/1604.01148}{{\ttfamily arXiv:1604.01148 [hep-ph]}}.

\bibitem{Kownacki:2016pmx}
C.~Kownacki, E.~Ma, N.~Pollard, and M.~Zakeri, ``{Generalized Gauge U(1) Family
  Symmetry for Quarks and Leptons},''
  \href{http://dx.doi.org/10.1016/j.physletb.2017.01.013}{{\em Phys. Lett.}
  {\bfseries B766} (2017) 149--152},
\href{http://arxiv.org/abs/1611.05017}{{\ttfamily arXiv:1611.05017 [hep-ph]}}.

\bibitem{Nomura:2016ezz}
T.~Nomura, H.~Okada, and N.~Okada, ``{A Colored KNT Neutrino Model},''
  \href{http://dx.doi.org/10.1016/j.physletb.2016.09.038}{{\em Phys. Lett.}
  {\bfseries B762} (2016) 409--414},
\href{http://arxiv.org/abs/1608.02694}{{\ttfamily arXiv:1608.02694 [hep-ph]}}.

\bibitem{Camargo-Molina:2016yqm}
J.~E. Camargo-Molina, A.~P. Morais, A.~Ordell, R.~Pasechnik, M.~O.~P. Sampaio,
  and J.~Wessén, ``{Reviving trinification models through an E6 -extended
  supersymmetric GUT},''
  \href{http://dx.doi.org/10.1103/PhysRevD.95.075031}{{\em Phys. Rev.}
  {\bfseries D95} no.~7, (2017) 075031},
\href{http://arxiv.org/abs/1610.03642}{{\ttfamily arXiv:1610.03642 [hep-ph]}}.

\bibitem{Camargo-Molina:2016bwm}
J.~E. Camargo-Molina, A.~P. Morais, R.~Pasechnik, and J.~Wessén, ``{On a
  radiative origin of the Standard Model from Trinification},''
  \href{http://dx.doi.org/10.1007/JHEP09(2016)129}{{\em JHEP} {\bfseries 09}
  (2016) 129},
\href{http://arxiv.org/abs/1606.03492}{{\ttfamily arXiv:1606.03492 [hep-ph]}}.

\bibitem{vonderPahlen:2016cbw}
F.~von~der Pahlen, G.~Palacio, D.~Restrepo, and O.~Zapata, ``{Radiative Type
  III Seesaw Model and its collider phenomenology},''
  \href{http://dx.doi.org/10.1103/PhysRevD.94.033005}{{\em Phys. Rev.}
  {\bfseries D94} no.~3, (2016) 033005},
\href{http://arxiv.org/abs/1605.01129}{{\ttfamily arXiv:1605.01129 [hep-ph]}}.

\bibitem{Bonilla:2016diq}
C.~Bonilla, E.~Ma, E.~Peinado, and J.~W.~F. Valle, ``{Two-loop Dirac neutrino
  mass and WIMP dark matter},''
  \href{http://dx.doi.org/10.1016/j.physletb.2016.09.027}{{\em Phys. Lett.}
  {\bfseries B762} (2016) 214--218},
\href{http://arxiv.org/abs/1607.03931}{{\ttfamily arXiv:1607.03931 [hep-ph]}}.

\bibitem{Gu:2016xno}
P.-H. Gu, ``{High-scale leptogenesis with three-loop neutrino mass generation
  and dark matter},'' \href{http://dx.doi.org/10.1007/JHEP04(2017)159}{{\em
  JHEP} {\bfseries 04} (2017) 159},
\href{http://arxiv.org/abs/1611.03256}{{\ttfamily arXiv:1611.03256 [hep-ph]}}.

\bibitem{Das:2017ski}
A.~Das, T.~Nomura, H.~Okada, and S.~Roy, ``{Generation of a radiative neutrino
  mass in the linear seesaw framework, charged lepton flavor violation, and
  dark matter},'' \href{http://dx.doi.org/10.1103/PhysRevD.96.075001}{{\em
  Phys. Rev. D} {\bfseries 96} no.~7, (2017) 075001},
  \href{http://arxiv.org/abs/1704.02078}{{\ttfamily arXiv:1704.02078
  [hep-ph]}}.

\bibitem{Nomura:2017emk}
T.~Nomura and H.~Okada, ``{Loop induced type-II seesaw model and GeV dark
  matter with $U(1)_{B-L}$ gauge symmetry},''
  \href{http://dx.doi.org/10.1016/j.physletb.2017.10.033}{{\em Phys. Lett.}
  {\bfseries B774} (2017) 575--581},
\href{http://arxiv.org/abs/1704.08581}{{\ttfamily arXiv:1704.08581 [hep-ph]}}.

\bibitem{Nomura:2017vzp}
T.~Nomura and H.~Okada, ``{Radiative neutrino mass in an alternative
  $U(1)_{B-L}$ gauge symmetry},''
  \href{http://dx.doi.org/10.1016/j.nuclphysb.2019.02.025}{{\em Nucl. Phys.}
  {\bfseries B941} (2019) 586--599},
\href{http://arxiv.org/abs/1705.08309}{{\ttfamily arXiv:1705.08309 [hep-ph]}}.

\bibitem{Nomura:2017tzj}
T.~Nomura and H.~Okada, ``{A model with isospin doublet $U(1)_D$ gauge
  symmetry},'' \href{http://dx.doi.org/10.1142/S0217751X18500896}{{\em Int. J.
  Mod. Phys.} {\bfseries A33} no.~14n15, (2018) 1850089},
\href{http://arxiv.org/abs/1706.05268}{{\ttfamily arXiv:1706.05268 [hep-ph]}}.

\bibitem{Wang:2017mcy}
W.~Wang, R.~Wang, Z.-L. Han, and J.-Z. Han, ``{The $B-L$ Scotogenic Models for
  Dirac Neutrino Masses},''
  \href{http://dx.doi.org/10.1140/epjc/s10052-017-5446-9}{{\em Eur. Phys. J.}
  {\bfseries C77} no.~12, (2017) 889},
\href{http://arxiv.org/abs/1705.00414}{{\ttfamily arXiv:1705.00414 [hep-ph]}}.

\bibitem{Bernal:2017xat}
N.~Bernal, A.~E. Cárcamo~Hernández, I.~de~Medeiros~Varzielas, and
  S.~Kovalenko, ``{Fermion masses and mixings and dark matter constraints in a
  model with radiative seesaw mechanism},''
  \href{http://dx.doi.org/10.1007/JHEP05(2018)053}{{\em JHEP} {\bfseries 05}
  (2018) 053},
\href{http://arxiv.org/abs/1712.02792}{{\ttfamily arXiv:1712.02792 [hep-ph]}}.

\bibitem{CarcamoHernandez:2017kra}
A.~E. Cárcamo~Hernández and H.~N. Long, ``{A highly predictive $A_{4}$
  flavour 3-3-1 model with radiative inverse seesaw mechanism},''
  \href{http://dx.doi.org/10.1088/1361-6471/aaace7}{{\em J. Phys.} {\bfseries
  G45} no.~4, (2018) 045001},
\href{http://arxiv.org/abs/1705.05246}{{\ttfamily arXiv:1705.05246 [hep-ph]}}.

\bibitem{CarcamoHernandez:2017cwi}
A.~E. Cárcamo~Hernández, S.~Kovalenko, H.~N. Long, and I.~Schmidt, ``{A
  variant of 3-3-1 model for the generation of the SM fermion mass and mixing
  pattern},'' \href{http://dx.doi.org/10.1007/JHEP07(2018)144}{{\em JHEP}
  {\bfseries 07} (2018) 144},
\href{http://arxiv.org/abs/1705.09169}{{\ttfamily arXiv:1705.09169 [hep-ph]}}.

\bibitem{Ma:2017kgb}
E.~Ma and U.~Sarkar, ``{Radiative Left-Right Dirac Neutrino Mass},''
  \href{http://dx.doi.org/10.1016/j.physletb.2017.08.071}{{\em Phys. Lett.}
  {\bfseries B776} (2018) 54--57},
\href{http://arxiv.org/abs/1707.07698}{{\ttfamily arXiv:1707.07698 [hep-ph]}}.

\bibitem{Cepedello:2017eqf}
R.~Cepedello, M.~Hirsch, and J.~C. Helo, ``{Loop neutrino masses from $d = 7$
  operator},'' \href{http://dx.doi.org/10.1007/JHEP07(2017)079}{{\em JHEP}
  {\bfseries 07} (2017) 079},
\href{http://arxiv.org/abs/1705.01489}{{\ttfamily arXiv:1705.01489 [hep-ph]}}.

\bibitem{Dev:2018pjn}
A.~Dev and R.~N. Mohapatra, ``{Natural Alignment of Quark Flavors and
  Radiatively Induced Quark Mixings},''
  \href{http://dx.doi.org/10.1103/PhysRevD.98.073002}{{\em Phys. Rev.}
  {\bfseries D98} no.~7, (2018) 073002},
\href{http://arxiv.org/abs/1804.01598}{{\ttfamily arXiv:1804.01598 [hep-ph]}}.

\bibitem{CarcamoHernandez:2018hst}
A.~E. Cárcamo~Hernández, S.~Kovalenko, J.~W.~F. Valle, and C.~A.
  Vaquera-Araujo, ``{Neutrino predictions from a left-right symmetric flavored
  extension of the standard model},''
  \href{http://dx.doi.org/10.1007/JHEP02(2019)065}{{\em JHEP} {\bfseries 02}
  (2019) 065},
\href{http://arxiv.org/abs/1811.03018}{{\ttfamily arXiv:1811.03018 [hep-ph]}}.

\bibitem{Rojas:2018wym}
N.~Rojas, R.~Srivastava, and J.~W.~F. Valle, ``{Simplest Scoto-Seesaw
  Mechanism},'' \href{http://dx.doi.org/10.1016/j.physletb.2018.12.014}{{\em
  Phys. Lett.} {\bfseries B789} (2019) 132--136},
\href{http://arxiv.org/abs/1807.11447}{{\ttfamily arXiv:1807.11447 [hep-ph]}}.

\bibitem{Nomura:2018vfz}
T.~Nomura and H.~Okada, ``{Zee-Babu type model with $U(1)_{L_\mu - L_\tau}$
  gauge symmetry},'' \href{http://dx.doi.org/10.1103/PhysRevD.97.095023}{{\em
  Phys. Rev.} {\bfseries D97} no.~9, (2018) 095023},
\href{http://arxiv.org/abs/1803.04795}{{\ttfamily arXiv:1803.04795 [hep-ph]}}.

\bibitem{Reig:2018mdk}
M.~Reig, D.~Restrepo, J.~W.~F. Valle, and O.~Zapata, ``{Bound-state dark matter
  and Dirac neutrino masses},''
  \href{http://dx.doi.org/10.1103/PhysRevD.97.115032}{{\em Phys. Rev.}
  {\bfseries D97} no.~11, (2018) 115032},
\href{http://arxiv.org/abs/1803.08528}{{\ttfamily arXiv:1803.08528 [hep-ph]}}.

\bibitem{Bernal:2018aon}
N.~Bernal, D.~Restrepo, C.~Yaguna, and O.~Zapata, ``{Two-component dark matter
  and a massless neutrino in a new $B-L$ model},''
  \href{http://dx.doi.org/10.1103/PhysRevD.99.015038}{{\em Phys. Rev.}
  {\bfseries D99} no.~1, (2019) 015038},
\href{http://arxiv.org/abs/1808.03352}{{\ttfamily arXiv:1808.03352 [hep-ph]}}.

\bibitem{Calle:2018ovc}
J.~Calle, D.~Restrepo, C.~E. Yaguna, and O.~Zapata, ``{Minimal radiative Dirac
  neutrino mass models},''
  \href{http://dx.doi.org/10.1103/PhysRevD.99.075008}{{\em Phys. Rev.}
  {\bfseries D99} no.~7, (2019) 075008},
\href{http://arxiv.org/abs/1812.05523}{{\ttfamily arXiv:1812.05523 [hep-ph]}}.

\bibitem{Aranda:2018lif}
A.~Aranda, C.~Bonilla, and E.~Peinado, ``{Dynamical generation of neutrino mass
  scales},'' \href{http://dx.doi.org/10.1016/j.physletb.2019.01.068}{{\em Phys.
  Lett.} {\bfseries B792} (2019) 40--42},
\href{http://arxiv.org/abs/1808.07727}{{\ttfamily arXiv:1808.07727 [hep-ph]}}.

\bibitem{Cepedello:2018rfh}
R.~Cepedello, R.~M. Fonseca, and M.~Hirsch, ``{Systematic classification of
  three-loop realizations of the Weinberg operator},''
  \href{http://dx.doi.org/10.1007/JHEP10(2018)197,
  10.1007/JHEP06(2019)034}{{\em JHEP} {\bfseries 10} (2018) 197},
  \href{http://arxiv.org/abs/1807.00629}{{\ttfamily arXiv:1807.00629
  [hep-ph]}}.
[erratum: JHEP06,034(2019)].

\bibitem{CarcamoHernandez:2018vdj}
A.~E. Cárcamo~Hernández, J.~Vignatti, and A.~Zerwekh, ``{Generating lepton
  masses and mixings with a heavy vector doublet},''
  \href{http://dx.doi.org/10.1088/1361-6471/ab4499}{{\em J. Phys.} {\bfseries
  G46} no.~11, (2019) 115007},
\href{http://arxiv.org/abs/1807.05321}{{\ttfamily arXiv:1807.05321 [hep-ph]}}.

\bibitem{Ma:2018zuj}
E.~Ma, ``{$U(1)_\chi$, Seesaw Dark Matter, and Higgs Decay},''
\href{http://arxiv.org/abs/1810.06506}{{\ttfamily arXiv:1810.06506 [hep-ph]}}.

\bibitem{Ma:2018bow}
E.~Ma, ``{$U(1)_\chi$ and Seesaw Dirac Neutrinos},''
\href{http://arxiv.org/abs/1811.09645}{{\ttfamily arXiv:1811.09645 [hep-ph]}}.

\bibitem{Li:2018aov}
S.-P. Li, X.-Q. Li, and Y.-D. Yang, ``{Muon $g-2$ in a $U(1)$-symmetric
  Two-Higgs-Doublet Model},''
  \href{http://dx.doi.org/10.1103/PhysRevD.99.035010}{{\em Phys. Rev. D}
  {\bfseries 99} no.~3, (2019) 035010},
  \href{http://arxiv.org/abs/1808.02424}{{\ttfamily arXiv:1808.02424
  [hep-ph]}}.

\bibitem{Arnan:2019uhr}
P.~Arnan, A.~Crivellin, M.~Fedele, and F.~Mescia, ``{Generic Loop Effects of
  New Scalars and Fermions in $b\to s\ell^+\ell^-$, $(g-2)_\mu$ and a
  Vector-like $4^{\rm th}$ Generation},''
  \href{http://dx.doi.org/10.1007/JHEP06(2019)118}{{\em JHEP} {\bfseries 06}
  (2019) 118}, \href{http://arxiv.org/abs/1904.05890}{{\ttfamily
  arXiv:1904.05890 [hep-ph]}}.

\bibitem{Ma:2019byo}
E.~Ma, ``{Two-loop $Z_4$ Dirac neutrino masses and mixing, with
  self-interacting dark matter},''
  \href{http://dx.doi.org/10.1016/j.nuclphysb.2019.114725}{{\em Nucl. Phys.}
  {\bfseries B946} (2019) 114725},
\href{http://arxiv.org/abs/1907.04665}{{\ttfamily arXiv:1907.04665 [hep-ph]}}.

\bibitem{Ma:2019iwj}
E.~Ma, ``{Scotogenic cobimaximal Dirac neutrino mixing from $\Delta (27)$ and
  $U(1)_\chi $},'' \href{http://dx.doi.org/10.1140/epjc/s10052-019-7440-x}{{\em
  Eur. Phys. J.} {\bfseries C79} no.~11, (2019) 903},
\href{http://arxiv.org/abs/1905.01535}{{\ttfamily arXiv:1905.01535 [hep-ph]}}.

\bibitem{CarcamoHernandez:2019xkb}
A.~E. Cárcamo~Hernández, S.~Kovalenko, R.~Pasechnik, and I.~Schmidt,
  ``{Phenomenology of an extended IDM with loop-generated fermion mass
  hierarchies},'' \href{http://dx.doi.org/10.1140/epjc/s10052-019-7101-0}{{\em
  Eur. Phys. J.} {\bfseries C79} no.~7, (2019) 610},
\href{http://arxiv.org/abs/1901.09552}{{\ttfamily arXiv:1901.09552 [hep-ph]}}.

\bibitem{Ma:2019yfo}
E.~Ma, ``{Scotogenic $U(1)_\chi$ Dirac neutrinos},''
  \href{http://dx.doi.org/10.1016/j.physletb.2019.05.006}{{\em Phys. Lett.}
  {\bfseries B793} (2019) 411--414},
\href{http://arxiv.org/abs/1901.09091}{{\ttfamily arXiv:1901.09091 [hep-ph]}}.

\bibitem{Nomura:2019yft}
T.~Nomura and H.~Okada, ``{A two loop induced neutrino mass model with modular
  $A_4$ symmetry},''
  \href{http://dx.doi.org/10.1016/j.nuclphysb.2021.115372}{{\em Nucl. Phys.}
  {\bfseries B966} (2021) 115372},
\href{http://arxiv.org/abs/1906.03927}{{\ttfamily arXiv:1906.03927 [hep-ph]}}.

\bibitem{Nomura:2019vqc}
T.~Nomura and H.~Okada, ``{A radiative neutrino mass model with hidden gauge
  symmetry inducing semi-annihilating dark matter},''
\href{http://arxiv.org/abs/1904.13066}{{\ttfamily arXiv:1904.13066 [hep-ph]}}.

\bibitem{Nomura:2019jxj}
T.~Nomura and H.~Okada, ``{A modular $A_4$ symmetric model of dark matter and
  neutrino},'' \href{http://dx.doi.org/10.1016/j.physletb.2019.134799}{{\em
  Phys. Lett.} {\bfseries B797} (2019) 134799},
\href{http://arxiv.org/abs/1904.03937}{{\ttfamily arXiv:1904.03937 [hep-ph]}}.

\bibitem{CentellesChulia:2019gic}
S.~Centelles~Chuliá, R.~Cepedello, E.~Peinado, and R.~Srivastava,
  ``{Scotogenic dark symmetry as a residual subgroup of Standard Model
  symmetries},'' \href{http://dx.doi.org/10.1088/1674-1137/44/8/083110}{{\em
  Chin. Phys.} {\bfseries C44} no.~8, (2020) 083110},
\href{http://arxiv.org/abs/1901.06402}{{\ttfamily arXiv:1901.06402 [hep-ph]}}.

\bibitem{Bonilla:2018ynb}
C.~Bonilla, S.~Centelles-Chuliá, R.~Cepedello, E.~Peinado, and R.~Srivastava,
  ``{Dark matter stability and Dirac neutrinos using only Standard Model
  symmetries},'' \href{http://dx.doi.org/10.1103/PhysRevD.101.033011}{{\em
  Phys. Rev.} {\bfseries D101} no.~3, (2020) 033011},
\href{http://arxiv.org/abs/1812.01599}{{\ttfamily arXiv:1812.01599 [hep-ph]}}.

\bibitem{Pramanick:2019oxb}
S.~Pramanick, ``{Scotogenic S3 symmetric generation of realistic neutrino
  mixing},'' \href{http://dx.doi.org/10.1103/PhysRevD.100.035009}{{\em Phys.
  Rev.} {\bfseries D100} no.~3, (2019) 035009},
\href{http://arxiv.org/abs/1904.07558}{{\ttfamily arXiv:1904.07558 [hep-ph]}}.

\bibitem{Arbelaez:2019wyz}
C.~Arbeláez, A.~E. Cárcamo~Hernández, R.~Cepedello, M.~Hirsch, and
  S.~Kovalenko, ``{Radiative type-I seesaw neutrino masses},''
  \href{http://dx.doi.org/10.1103/PhysRevD.100.115021}{{\em Phys. Rev.}
  {\bfseries D100} no.~11, (2019) 115021},
\href{http://arxiv.org/abs/1910.04178}{{\ttfamily arXiv:1910.04178 [hep-ph]}}.

\bibitem{Avila:2019hhv}
I.~M. Ávila, V.~De~Romeri, L.~Duarte, and J.~W.~F. Valle, ``{Phenomenology of
  scotogenic scalar dark matter},''
  \href{http://dx.doi.org/10.1140/epjc/s10052-020-08480-z}{{\em Eur. Phys. J.}
  {\bfseries C80} no.~10, (2020) 908},
\href{http://arxiv.org/abs/1910.08422}{{\ttfamily arXiv:1910.08422 [hep-ph]}}.

\bibitem{CarcamoHernandez:2019cbd}
A.~E. Cárcamo~Hernández, S.~Kovalenko, R.~Pasechnik, and I.~Schmidt,
  ``{Sequentially loop-generated quark and lepton mass hierarchies in an
  extended Inert Higgs Doublet model},''
  \href{http://dx.doi.org/10.1007/JHEP06(2019)056}{{\em JHEP} {\bfseries 06}
  (2019) 056},
\href{http://arxiv.org/abs/1901.02764}{{\ttfamily arXiv:1901.02764 [hep-ph]}}.

\bibitem{CarcamoHernandez:2019lhv}
A.~E. Cárcamo~Hernández, D.~T. Huong, and H.~N. Long, ``{Minimal model for
  the fermion flavor structure, mass hierarchy, dark matter, leptogenesis, and
  the electron and muon anomalous magnetic moments},''
  \href{http://dx.doi.org/10.1103/PhysRevD.102.055002}{{\em Phys. Rev.}
  {\bfseries D102} no.~5, (2020) 055002},
\href{http://arxiv.org/abs/1910.12877}{{\ttfamily arXiv:1910.12877 [hep-ph]}}.

\bibitem{Arbelaez:2019ofg}
C.~Arbeláez, A.~E. Cárcamo~Hernández, R.~Cepedello, S.~Kovalenko, and
  I.~Schmidt, ``{Sequentially loop suppressed fermion masses from a single
  discrete symmetry},'' \href{http://dx.doi.org/10.1007/JHEP06(2020)043}{{\em
  JHEP} {\bfseries 06} (2020) 043},
\href{http://arxiv.org/abs/1911.02033}{{\ttfamily arXiv:1911.02033 [hep-ph]}}.

\bibitem{CarcamoHernandez:2020pnh}
A.~E. Cárcamo~Hernández, L.~T. Hue, S.~Kovalenko, and H.~N. Long, ``{An
  extended 3-3-1 model with two scalar triplets and linear seesaw mechanism},''
\href{http://arxiv.org/abs/2001.01748}{{\ttfamily arXiv:2001.01748 [hep-ph]}}.

\bibitem{CarcamoHernandez:2020ney}
A.~E. Cárcamo~Hernández, C.~O. Dib, and U.~J. Saldaña-Salazar, ``{When $\tan
  \beta$ meets all the mixing angles},''
  \href{http://dx.doi.org/10.1016/j.physletb.2020.135750}{{\em Phys. Lett.}
  {\bfseries B809} (2020) 135750},
\href{http://arxiv.org/abs/2001.07140}{{\ttfamily arXiv:2001.07140 [hep-ph]}}.

\bibitem{CarcamoHernandez:2020pxw}
A.~E. Cárcamo~Hernández, Y.~Hidalgo~Velásquez, S.~Kovalenko, H.~N. Long,
  N.~A. Pérez-Julve, and V.~V. Vien, ``{Fermion spectrum and $g-2$ anomalies
  in a low scale 3-3-1 model},''
  \href{http://dx.doi.org/10.1140/epjc/s10052-021-08974-4.,
  10.1140/epjc/s10052-021-08974-4}{{\em Eur. Phys. J.} {\bfseries C81} no.~2,
  (2021) 191},
\href{http://arxiv.org/abs/2002.07347}{{\ttfamily arXiv:2002.07347 [hep-ph]}}.

\bibitem{CarcamoHernandez:2020owa}
A.~E. Cárcamo~Hernández, D.~T. Huong, S.~Kovalenko, A.~P. Morais,
  R.~Pasechnik, and I.~Schmidt, ``{How low-scale trinification sheds light in
  the flavor hierarchies, neutrino puzzle, dark matter, and leptogenesis},''
  \href{http://dx.doi.org/10.1103/PhysRevD.102.095003}{{\em Phys. Rev.}
  {\bfseries D102} no.~9, (2020) 095003},
\href{http://arxiv.org/abs/2004.11450}{{\ttfamily arXiv:2004.11450 [hep-ph]}}.

\bibitem{CarcamoHernandez:2020ehn}
A.~E. Cárcamo~Hernández, J.~W.~F. Valle, and C.~A. Vaquera-Araujo, ``{Simple
  theory for scotogenic dark matter with residual matter-parity},''
  \href{http://dx.doi.org/10.1016/j.physletb.2020.135757}{{\em Phys. Lett.}
  {\bfseries B809} (2020) 135757},
\href{http://arxiv.org/abs/2006.06009}{{\ttfamily arXiv:2006.06009 [hep-ph]}}.

\bibitem{Hernandez:2021uxx}
A.~E.~C. Hernández and I.~Schmidt, ``{A renormalizable left-right symmetric
  model with low scale seesaw mechanisms},''
\href{http://arxiv.org/abs/2101.02718}{{\ttfamily arXiv:2101.02718 [hep-ph]}}.

\bibitem{CarcamoHernandez:2021iat}
A.~E. C\'arcamo~Hern\'andez, C.~Espinoza, J.~Carlos G\'omez-Izquierdo, and
  M.~Mondrag\'on, ``{Fermion masses and mixings, dark matter, leptogenesis and
  $g-2$ muon anomaly in an extended 2HDM with inverse seesaw},''
  \href{http://arxiv.org/abs/2104.02730}{{\ttfamily arXiv:2104.02730
  [hep-ph]}}.

\bibitem{CarcamoHernandez:2021qhf}
A.~E. C\'arcamo~Hern\'andez, S.~Kovalenko, M.~Maniatis, and I.~Schmidt,
  ``{Fermion mass hierarchy and g-2 anomalies in an extended 3HDM Model},''
  \href{http://arxiv.org/abs/2104.07047}{{\ttfamily arXiv:2104.07047
  [hep-ph]}}.

\bibitem{CarcamoHernandez:2021tlv}
A.~E. C\'arcamo~Hern\'andez, S.~Kovalenko, F.~S. Queiroz, and Y.~S. Villamizar,
  ``{An extended 3-3-1 model with radiative linear seesaw mechanism},''
  \href{http://arxiv.org/abs/2105.01731}{{\ttfamily arXiv:2105.01731
  [hep-ph]}}.

\bibitem{Abada:2021yot}
A.~Abada, N.~Bernal, A.~E.~C. Hern\'andez, X.~Marcano, and G.~Piazza, ``{Gauged
  inverse seesaw from dark matter},''
  \href{http://dx.doi.org/10.1140/epjc/s10052-021-09535-5}{{\em Eur. Phys. J.
  C} {\bfseries 81} no.~8, (2021) 758},
  \href{http://arxiv.org/abs/2107.02803}{{\ttfamily arXiv:2107.02803
  [hep-ph]}}.

\bibitem{Bonilla:2021ize}
C.~Bonilla, A.~E.~C. Hern\'andez, J.~a. Gon\c{c}alves, F.~F. Freitas, A.~P.
  Morais, and R.~Pasechnik, ``{Collider signatures of vector-like fermions from
  a flavor symmetric 2HDM},'' \href{http://arxiv.org/abs/2107.14165}{{\ttfamily
  arXiv:2107.14165 [hep-ph]}}.

\bibitem{Hernandez:2021zje}
A.~E.~C. Hern\'andez, C.~Hati, S.~Kovalenko, J.~W.~F. Valle, and C.~A.
  Vaquera-Araujo, ``{Scotogenic neutrino masses with gauged matter parity and
  gauge coupling unification},''
  \href{http://arxiv.org/abs/2109.05029}{{\ttfamily arXiv:2109.05029
  [hep-ph]}}.

\bibitem{Kiritsis:2002aj}
E.~Kiritsis and P.~Anastasopoulos, ``{The Anomalous magnetic moment of the muon
  in the D-brane realization of the standard model},''
  \href{http://dx.doi.org/10.1088/1126-6708/2002/05/054}{{\em JHEP} {\bfseries
  05} (2002) 054}, \href{http://arxiv.org/abs/hep-ph/0201295}{{\ttfamily
  arXiv:hep-ph/0201295}}.

\bibitem{Appelquist:2004mn}
T.~Appelquist, M.~Piai, and R.~Shrock, ``{Lepton dipole moments in extended
  technicolor models},''
  \href{http://dx.doi.org/10.1016/j.physletb.2004.04.062}{{\em Phys. Lett.}
  {\bfseries B593} (2004) 175--180},
\href{http://arxiv.org/abs/hep-ph/0401114}{{\ttfamily arXiv:hep-ph/0401114
  [hep-ph]}}.

\bibitem{Giudice:2012ms}
G.~F. Giudice, P.~Paradisi, and M.~Passera, ``{Testing new physics with the
  electron g-2},'' \href{http://dx.doi.org/10.1007/JHEP11(2012)113}{{\em JHEP}
  {\bfseries 11} (2012) 113},
\href{http://arxiv.org/abs/1208.6583}{{\ttfamily arXiv:1208.6583 [hep-ph]}}.

\bibitem{Omura:2015nja}
Y.~Omura, E.~Senaha, and K.~Tobe, ``{Lepton-flavor-violating Higgs decay $h \to
  \mu\tau$ and muon anomalous magnetic moment in a general two Higgs doublet
  model},'' \href{http://dx.doi.org/10.1007/JHEP05(2015)028}{{\em JHEP}
  {\bfseries 05} (2015) 028}, \href{http://arxiv.org/abs/1502.07824}{{\ttfamily
  arXiv:1502.07824 [hep-ph]}}.

\bibitem{Falkowski:2018dsl}
A.~Falkowski, S.~F. King, E.~Perdomo, and M.~Pierre, ``{Flavourful $Z'$ portal
  for vector-like neutrino Dark Matter and $R_{K^{(*)}}$},''
  \href{http://dx.doi.org/10.1007/JHEP08(2018)061}{{\em JHEP} {\bfseries 08}
  (2018) 061},
\href{http://arxiv.org/abs/1803.04430}{{\ttfamily arXiv:1803.04430 [hep-ph]}}.

\bibitem{Crivellin:2018qmi}
A.~Crivellin, M.~Hoferichter, and P.~Schmidt-Wellenburg, ``{Combined
  explanations of $(g-2)_{\mu,e}$ and implications for a large muon EDM},''
  \href{http://dx.doi.org/10.1103/PhysRevD.98.113002}{{\em Phys. Rev.}
  {\bfseries D98} no.~11, (2018) 113002},
\href{http://arxiv.org/abs/1807.11484}{{\ttfamily arXiv:1807.11484 [hep-ph]}}.

\bibitem{Allanach:2015gkd}
B.~Allanach, F.~S. Queiroz, A.~Strumia, and S.~Sun, ``{$Z^{\prime}$ models for
  the LHCb and $g-2$ muon anomalies},''
  \href{http://dx.doi.org/10.1103/PhysRevD.93.055045,
  10.1103/PhysRevD.95.119902}{{\em Phys. Rev.} {\bfseries D93} no.~5, (2016)
  055045}, \href{http://arxiv.org/abs/1511.07447}{{\ttfamily arXiv:1511.07447
  [hep-ph]}}.
[Erratum: Phys. Rev.D95,no.11,119902(2017)].

\bibitem{Padley:2015uma}
B.~P. Padley, K.~Sinha, and K.~Wang, ``{Natural Supersymmetry, Muon $g-2$, and
  the Last Crevices for the Top Squark},''
  \href{http://dx.doi.org/10.1103/PhysRevD.92.055025}{{\em Phys. Rev.}
  {\bfseries D92} no.~5, (2015) 055025},
\href{http://arxiv.org/abs/1505.05877}{{\ttfamily arXiv:1505.05877 [hep-ph]}}.

\bibitem{Chen:2016dip}
C.-H. Chen, T.~Nomura, and H.~Okada, ``{Explanation of $B \to K^{(*)} \ell^+
  \ell^-$ and muon $g-2$, and implications at the LHC},''
  \href{http://dx.doi.org/10.1103/PhysRevD.94.115005}{{\em Phys. Rev.}
  {\bfseries D94} no.~11, (2016) 115005},
\href{http://arxiv.org/abs/1607.04857}{{\ttfamily arXiv:1607.04857 [hep-ph]}}.

\bibitem{Raby:2017igl}
S.~Raby and A.~Trautner, ``{Vectorlike chiral fourth family to explain muon
  anomalies},'' \href{http://dx.doi.org/10.1103/PhysRevD.97.095006}{{\em Phys.
  Rev.} {\bfseries D97} no.~9, (2018) 095006},
\href{http://arxiv.org/abs/1712.09360}{{\ttfamily arXiv:1712.09360 [hep-ph]}}.

\bibitem{Chiang:2017tai}
C.-W. Chiang, H.~Okada, and E.~Senaha, ``{Dark matter, muon $g-2$, electric
  dipole moments, and $Z\to \ell_i^+ \ell_j^-$ in a one-loop induced neutrino
  model},'' \href{http://dx.doi.org/10.1103/PhysRevD.96.015002}{{\em Phys.
  Rev.} {\bfseries D96} no.~1, (2017) 015002},
\href{http://arxiv.org/abs/1703.09153}{{\ttfamily arXiv:1703.09153 [hep-ph]}}.

\bibitem{Chen:2017hir}
C.-H. Chen, T.~Nomura, and H.~Okada, ``{Excesses of muon $g-2$,
  $R_{D^{(\ast)}}$, and $R_K$ in a leptoquark model},''
  \href{http://dx.doi.org/10.1016/j.physletb.2017.10.005}{{\em Phys. Lett.}
  {\bfseries B774} (2017) 456--464},
\href{http://arxiv.org/abs/1703.03251}{{\ttfamily arXiv:1703.03251 [hep-ph]}}.

\bibitem{Megias:2017dzd}
E.~Megias, M.~Quiros, and L.~Salas, ``{$g_\mu-2$ from Vector-Like Leptons in
  Warped Space},'' \href{http://dx.doi.org/10.1007/JHEP05(2017)016}{{\em JHEP}
  {\bfseries 05} (2017) 016}, \href{http://arxiv.org/abs/1701.05072}{{\ttfamily
  arXiv:1701.05072 [hep-ph]}}.

\bibitem{Davoudiasl:2018fbb}
H.~Davoudiasl and W.~J. Marciano, ``{Tale of two anomalies},''
  \href{http://dx.doi.org/10.1103/PhysRevD.98.075011}{{\em Phys. Rev.}
  {\bfseries D98} no.~7, (2018) 075011},
\href{http://arxiv.org/abs/1806.10252}{{\ttfamily arXiv:1806.10252 [hep-ph]}}.

\bibitem{Liu:2018xkx}
J.~Liu, C.~E.~M. Wagner, and X.-P. Wang, ``{A light complex scalar for the
  electron and muon anomalous magnetic moments},''
  \href{http://dx.doi.org/10.1007/JHEP03(2019)008}{{\em JHEP} {\bfseries 03}
  (2019) 008},
\href{http://arxiv.org/abs/1810.11028}{{\ttfamily arXiv:1810.11028 [hep-ph]}}.

\bibitem{Nomura:2019btk}
T.~Nomura and H.~Okada, ``{Muon anomalous magnetic moment, $Z$ boson decays,
  and collider physics in multicharged particles},''
  \href{http://dx.doi.org/10.1103/PhysRevD.101.015021}{{\em Phys. Rev.}
  {\bfseries D101} no.~1, (2020) 015021},
\href{http://arxiv.org/abs/1903.05958}{{\ttfamily arXiv:1903.05958 [hep-ph]}}.

\bibitem{Kawamura:2019rth}
J.~Kawamura, S.~Raby, and A.~Trautner, ``{Complete vectorlike fourth family and
  new $U(1)^{\prime}$ for muon anomalies},''
  \href{http://dx.doi.org/10.1103/PhysRevD.100.055030}{{\em Phys. Rev.}
  {\bfseries D100} no.~5, (2019) 055030},
\href{http://arxiv.org/abs/1906.11297}{{\ttfamily arXiv:1906.11297 [hep-ph]}}.

\bibitem{Bauer:2019gfk}
M.~Bauer, M.~Neubert, S.~Renner, M.~Schnubel, and A.~Thamm, ``{Axionlike
  Particles, Lepton-Flavor Violation, and a New Explanation of $a_\mu$ and
  $a_e$},'' \href{http://dx.doi.org/10.1103/PhysRevLett.124.211803}{{\em Phys.
  Rev. Lett.} {\bfseries 124} no.~21, (2020) 211803},
\href{http://arxiv.org/abs/1908.00008}{{\ttfamily arXiv:1908.00008 [hep-ph]}}.

\bibitem{Botella:2018gzy}
F.~J. Botella, F.~Cornet-Gomez, and M.~Nebot, ``{Flavor conservation in
  two-Higgs-doublet models},''
  \href{http://dx.doi.org/10.1103/PhysRevD.98.035046}{{\em Phys. Rev.}
  {\bfseries D98} no.~3, (2018) 035046},
\href{http://arxiv.org/abs/1803.08521}{{\ttfamily arXiv:1803.08521 [hep-ph]}}.

\bibitem{Han:2018znu}
X.-F. Han, T.~Li, L.~Wang, and Y.~Zhang, ``{Simple interpretations of lepton
  anomalies in the lepton-specific inert two-Higgs-doublet model},''
  \href{http://dx.doi.org/10.1103/PhysRevD.99.095034}{{\em Phys. Rev.}
  {\bfseries D99} no.~9, (2019) 095034},
\href{http://arxiv.org/abs/1812.02449}{{\ttfamily arXiv:1812.02449 [hep-ph]}}.

\bibitem{Wang:2018hnw}
L.~Wang, J.~M. Yang, M.~Zhang, and Y.~Zhang, ``{Revisiting lepton-specific 2HDM
  in light of muon $g−2$ anomaly},''
  \href{http://dx.doi.org/10.1016/j.physletb.2018.11.045}{{\em Phys. Lett.}
  {\bfseries B788} (2019) 519--529},
\href{http://arxiv.org/abs/1809.05857}{{\ttfamily arXiv:1809.05857 [hep-ph]}}.

\bibitem{Dutta:2018fge}
B.~Dutta and Y.~Mimura, ``{Electron $g-2$ with flavor violation in MSSM},''
  \href{http://dx.doi.org/10.1016/j.physletb.2018.12.070}{{\em Phys. Lett.}
  {\bfseries B790} (2019) 563--567},
\href{http://arxiv.org/abs/1811.10209}{{\ttfamily arXiv:1811.10209 [hep-ph]}}.

\bibitem{Badziak:2019gaf}
M.~Badziak and K.~Sakurai, ``{Explanation of electron and muon g − 2
  anomalies in the MSSM},''
  \href{http://dx.doi.org/10.1007/JHEP10(2019)024}{{\em JHEP} {\bfseries 10}
  (2019) 024},
\href{http://arxiv.org/abs/1908.03607}{{\ttfamily arXiv:1908.03607 [hep-ph]}}.

\bibitem{Endo:2019bcj}
M.~Endo and W.~Yin, ``{Explaining electron and muon $g-2$ anomaly in SUSY
  without lepton-flavor mixings},''
  \href{http://dx.doi.org/10.1007/JHEP08(2019)122}{{\em JHEP} {\bfseries 08}
  (2019) 122},
\href{http://arxiv.org/abs/1906.08768}{{\ttfamily arXiv:1906.08768 [hep-ph]}}.

\bibitem{Hiller:2019mou}
G.~Hiller, C.~Hormigos-Feliu, D.~F. Litim, and T.~Steudtner, ``{Anomalous
  magnetic moments from asymptotic safety},''
  \href{http://dx.doi.org/10.1103/PhysRevD.102.071901}{{\em Phys. Rev.}
  {\bfseries D102} no.~7, (2020) 071901},
\href{http://arxiv.org/abs/1910.14062}{{\ttfamily arXiv:1910.14062 [hep-ph]}}.

\bibitem{CarcamoHernandez:2019ydc}
A.~E. Cárcamo~Hernández, S.~F. King, H.~Lee, and S.~J. Rowley, ``{Is it
  possible to explain the muon and electron $g-2$ in a $Z^{\prime}$ model?},''
  \href{http://dx.doi.org/10.1103/PhysRevD.101.115016}{{\em Phys. Rev.}
  {\bfseries D101} no.~11, (2020) 115016},
\href{http://arxiv.org/abs/1910.10734}{{\ttfamily arXiv:1910.10734 [hep-ph]}}.

\bibitem{Kawamura:2019hxp}
J.~Kawamura, S.~Raby, and A.~Trautner, ``{Complete vectorlike fourth family
  with $U(1)^{\prime}$: A global analysis},''
  \href{http://dx.doi.org/10.1103/PhysRevD.101.035026}{{\em Phys. Rev.}
  {\bfseries D101} no.~3, (2020) 035026},
\href{http://arxiv.org/abs/1911.11075}{{\ttfamily arXiv:1911.11075 [hep-ph]}}.

\bibitem{Sabatta:2019nfg}
D.~Sabatta, A.~S. Cornell, A.~Goyal, M.~Kumar, B.~Mellado, and X.~Ruan,
  ``{Connecting muon anomalous magnetic moment and multi-lepton anomalies at
  LHC},'' \href{http://dx.doi.org/10.1088/1674-1137/44/6/063103}{{\em Chin.
  Phys.} {\bfseries C44} no.~6, (2020) 063103},
\href{http://arxiv.org/abs/1909.03969}{{\ttfamily arXiv:1909.03969 [hep-ph]}}.

\bibitem{Chen:2020tfr}
K.-F. Chen, C.-W. Chiang, and K.~Yagyu, ``{An explanation for the muon and
  electron $g − 2$ anomalies and dark matter},''
  \href{http://dx.doi.org/10.1007/JHEP09(2020)119}{{\em JHEP} {\bfseries 09}
  (2020) 119}, \href{http://arxiv.org/abs/2006.07929}{{\ttfamily
  arXiv:2006.07929 [hep-ph]}}.

\bibitem{Iguro:2019sly}
S.~Iguro, Y.~Omura, and M.~Takeuchi, ``{Testing the 2HDM explanation of the
  muon g \textendash{} 2 anomaly at the LHC},''
  \href{http://dx.doi.org/10.1007/JHEP11(2019)130}{{\em JHEP} {\bfseries 11}
  (2019) 130}, \href{http://arxiv.org/abs/1907.09845}{{\ttfamily
  arXiv:1907.09845 [hep-ph]}}.

\bibitem{Li:2020dbg}
S.-P. Li, X.-Q. Li, Y.-Y. Li, Y.-D. Yang, and X.~Zhang, ``{Power-aligned 2HDM:
  a correlative perspective on $(g-2)_{e,\mu}$},''
  \href{http://dx.doi.org/10.1007/JHEP01(2021)034}{{\em JHEP} {\bfseries 01}
  (2021) 034}, \href{http://arxiv.org/abs/2010.02799}{{\ttfamily
  arXiv:2010.02799 [hep-ph]}}.

\bibitem{Arbelaez:2020rbq}
C.~Arbeláez, R.~Cepedello, R.~M. Fonseca, and M.~Hirsch, ``{$(g-2)$ anomalies
  and neutrino mass},''
  \href{http://dx.doi.org/10.1103/PhysRevD.102.075005}{{\em Phys. Rev.}
  {\bfseries D102} no.~7, (2020) 075005},
\href{http://arxiv.org/abs/2007.11007}{{\ttfamily arXiv:2007.11007 [hep-ph]}}.

\bibitem{Hiller:2020fbu}
G.~Hiller, C.~Hormigos-Feliu, D.~F. Litim, and T.~Steudtner, ``{Model Building
  from Asymptotic Safety with Higgs and Flavor Portals},''
  \href{http://dx.doi.org/10.1103/PhysRevD.102.095023}{{\em Phys. Rev.}
  {\bfseries D102} no.~9, (2020) 095023},
\href{http://arxiv.org/abs/2008.08606}{{\ttfamily arXiv:2008.08606 [hep-ph]}}.

\bibitem{Jana:2020pxx}
S.~Jana, V.~P. K., and S.~Saad, ``{Resolving electron and muon $g-2$ within the
  2HDM},'' \href{http://dx.doi.org/10.1103/PhysRevD.101.115037}{{\em Phys.
  Rev.} {\bfseries D101} no.~11, (2020) 115037},
\href{http://arxiv.org/abs/2003.03386}{{\ttfamily arXiv:2003.03386 [hep-ph]}}.

\bibitem{deJesus:2020ngn}
A.~S. de~Jesus, S.~Kovalenko, C.~A. de~S.~Pires, F.~S. Queiroz, and Y.~S.
  Villamizar, ``{Dead or alive? Implications of the muon anomalous magnetic
  moment for 3-3-1 models},''
  \href{http://dx.doi.org/10.1016/j.physletb.2020.135689}{{\em Phys. Lett.}
  {\bfseries B809} (2020) 135689},
\href{http://arxiv.org/abs/2003.06440}{{\ttfamily arXiv:2003.06440 [hep-ph]}}.

\bibitem{deJesus:2020upp}
A.~S. De~Jesus, S.~Kovalenko, F.~S. Queiroz, C.~Siqueira, and K.~Sinha,
  ``{Vectorlike leptons and inert scalar triplet: Lepton flavor violation,
  $g-2$, and collider searches},''
  \href{http://dx.doi.org/10.1103/PhysRevD.102.035004}{{\em Phys. Rev.}
  {\bfseries D102} no.~3, (2020) 035004},
\href{http://arxiv.org/abs/2004.01200}{{\ttfamily arXiv:2004.01200 [hep-ph]}}.

\bibitem{Hati:2020fzp}
C.~Hati, J.~Kriewald, J.~Orloff, and A.~M. Teixeira, ``{Anomalies in $^8$Be
  nuclear transitions and $(g-2)_{e,\mu}$: towards a minimal combined
  explanation},'' \href{http://dx.doi.org/10.1007/JHEP07(2020)235}{{\em JHEP}
  {\bfseries 07} (2020) 235},
\href{http://arxiv.org/abs/2005.00028}{{\ttfamily arXiv:2005.00028 [hep-ph]}}.

\bibitem{Botella:2020xzf}
F.~J. Botella, F.~Cornet-Gomez, and M.~Nebot, ``{Electron and muon $g-2$
  anomalies in general flavour conserving two Higgs doublets models},''
  \href{http://dx.doi.org/10.1103/PhysRevD.102.035023}{{\em Phys. Rev.}
  {\bfseries D102} no.~3, (2020) 035023},
\href{http://arxiv.org/abs/2006.01934}{{\ttfamily arXiv:2006.01934 [hep-ph]}}.

\bibitem{Dorsner:2020aaz}
I.~Doršner, S.~Fajfer, and S.~Saad, ``{$\mu \to e \gamma$ selecting scalar
  leptoquark solutions for the $(g-2)_{e,\mu}$ puzzles},''
  \href{http://dx.doi.org/10.1103/PhysRevD.102.075007}{{\em Phys. Rev.}
  {\bfseries D102} no.~7, (2020) 075007},
\href{http://arxiv.org/abs/2006.11624}{{\ttfamily arXiv:2006.11624 [hep-ph]}}.

\bibitem{Calibbi:2020emz}
L.~Calibbi, M.~L. López-Ibáñez, A.~Melis, and O.~Vives, ``{Muon and electron
  $g−2$ and lepton masses in flavor models},''
  \href{http://dx.doi.org/10.1007/JHEP06(2020)087}{{\em JHEP} {\bfseries 06}
  (2020) 087},
\href{http://arxiv.org/abs/2003.06633}{{\ttfamily arXiv:2003.06633 [hep-ph]}}.

\bibitem{Dinh:2020pqn}
L.~T. Hue, P.~N. Thanh, and T.~D. Tham, ``{Anomalous Magnetic Dipole Moment
  \((g-2)\mu\) in 3-3-1 Model with Inverse Seesaw Neutrinos},''
\href{http://dx.doi.org/10.15625/0868-3166/30/3/14963}{{\em Commun.in Phys.}
  {\bfseries 30} no.~3, (2020) 221--230}.

\bibitem{Jana:2020joi}
S.~Jana, P.~K. Vishnu, W.~Rodejohann, and S.~Saad, ``{Dark matter assisted
  lepton anomalous magnetic moments and neutrino masses},''
  \href{http://dx.doi.org/10.1103/PhysRevD.102.075003}{{\em Phys. Rev.}
  {\bfseries D102} no.~7, (2020) 075003},
\href{http://arxiv.org/abs/2008.02377}{{\ttfamily arXiv:2008.02377 [hep-ph]}}.

\bibitem{Chun:2020uzw}
E.~J. Chun and T.~Mondal, ``{Explaining $g-2$ anomalies in two Higgs doublet
  model with vector-like leptons},''
  \href{http://dx.doi.org/10.1007/JHEP11(2020)077}{{\em JHEP} {\bfseries 11}
  (2020) 077},
\href{http://arxiv.org/abs/2009.08314}{{\ttfamily arXiv:2009.08314 [hep-ph]}}.

\bibitem{Chua:2020dya}
C.-K. Chua, ``{Data-driven study of the implications of anomalous magnetic
  moments and lepton flavor violating processes of $e$, $\mu$ and $\tau$},''
  \href{http://dx.doi.org/10.1103/PhysRevD.102.055022}{{\em Phys. Rev.}
  {\bfseries D102} no.~5, (2020) 055022},
\href{http://arxiv.org/abs/2004.11031}{{\ttfamily arXiv:2004.11031 [hep-ph]}}.

\bibitem{Daikoku:2020nhr}
Y.~Daikoku and H.~Okada, ``{Lepton Anomalous Magnetic Moments in an $S_4$
  Flavor-Symmetric Extra U(1) Model},''
\href{http://arxiv.org/abs/2011.10374}{{\ttfamily arXiv:2011.10374 [hep-ph]}}.

\bibitem{Banerjee:2020zvi}
H.~Banerjee, B.~Dutta, and S.~Roy, ``{Supersymmetric gauged $
  \mathrm{U}{(1)}_{L_{\mu }-{L}_{\tau }} $ model for electron and muon $(g −
  2)$ anomaly},'' \href{http://dx.doi.org/10.1007/JHEP03(2021)211}{{\em JHEP}
  {\bfseries 03} (2021) 211},
\href{http://arxiv.org/abs/2011.05083}{{\ttfamily arXiv:2011.05083 [hep-ph]}}.

\bibitem{Chen:2020jvl}
C.-H. Chen and T.~Nomura, ``{Electron and muon $g-2$, radiative neutrino mass,
  and $\ell' \to \ell \gamma$ in a $U(1)_{e-\mu}$ model},''
  \href{http://dx.doi.org/10.1016/j.nuclphysb.2021.115314}{{\em Nucl. Phys.}
  {\bfseries B964} (2021) 115314},
\href{http://arxiv.org/abs/2003.07638}{{\ttfamily arXiv:2003.07638 [hep-ph]}}.

\bibitem{Bigaran:2020jil}
I.~Bigaran and R.~R. Volkas, ``{Getting chirality right: Single scalar
  leptoquark solutions to the $(g-2)_{e,\mu}$ puzzle},''
  \href{http://dx.doi.org/10.1103/PhysRevD.102.075037}{{\em Phys. Rev.}
  {\bfseries D102} no.~7, (2020) 075037},
\href{http://arxiv.org/abs/2002.12544}{{\ttfamily arXiv:2002.12544 [hep-ph]}}.

\bibitem{Kawamura:2020qxo}
J.~Kawamura, S.~Okawa, and Y.~Omura, ``{Current status and muon $g − 2$
  explanation of lepton portal dark matter},''
  \href{http://dx.doi.org/10.1007/JHEP08(2020)042}{{\em JHEP} {\bfseries 08}
  (2020) 042},
\href{http://arxiv.org/abs/2002.12534}{{\ttfamily arXiv:2002.12534 [hep-ph]}}.

\bibitem{Endo:2020mev}
M.~Endo, S.~Iguro, and T.~Kitahara, ``{Probing $e\mu$ flavor-violating ALP at
  Belle II},'' \href{http://dx.doi.org/10.1007/JHEP06(2020)040}{{\em JHEP}
  {\bfseries 06} (2020) 040},
\href{http://arxiv.org/abs/2002.05948}{{\ttfamily arXiv:2002.05948 [hep-ph]}}.

\bibitem{Iguro:2020rby}
S.~Iguro, Y.~Omura, and M.~Takeuchi, ``{Probing $\mu\tau$ flavor-violating
  solutions for the muon $g-2$ anomaly at Belle II},''
  \href{http://dx.doi.org/10.1007/JHEP09(2020)144}{{\em JHEP} {\bfseries 09}
  (2020) 144}, \href{http://arxiv.org/abs/2002.12728}{{\ttfamily
  arXiv:2002.12728 [hep-ph]}}.

\bibitem{Yin:2020afe}
W.~Yin and M.~Yamaguchi, ``{Muon $g-2$ at multi-TeV muon collider},''
  \href{http://arxiv.org/abs/2012.03928}{{\ttfamily arXiv:2012.03928
  [hep-ph]}}.

\bibitem{Chen:2021rnl}
N.~Chen, B.~Wang, and C.-Y. Yao, ``{The collider tests of a leptophilic scalar
  for the anomalous magnetic moments},''
\href{http://arxiv.org/abs/2102.05619}{{\ttfamily arXiv:2102.05619 [hep-ph]}}.

\bibitem{Athron:2021iuf}
P.~Athron, C.~Balázs, D.~H. Jacob, W.~Kotlarski, D.~Stöckinger, and
  H.~Stöckinger-Kim, ``{New physics explanations of $a_\mu$ in light of the
  FNAL muon $g-2$ measurement},''
\href{http://arxiv.org/abs/2104.03691}{{\ttfamily arXiv:2104.03691 [hep-ph]}}.

\bibitem{Arcadi:2021cwg}
G.~Arcadi, L.~Calibbi, M.~Fedele, and F.~Mescia, ``{Muon $g-2$ and
  $B$-anomalies from Dark Matter},''
  \href{http://arxiv.org/abs/2104.03228}{{\ttfamily arXiv:2104.03228
  [hep-ph]}}.

\bibitem{Das:2021zea}
P.~Das, M.~Kumar~Das, and N.~Khan, ``{The FIMP-WIMP dark matter and Muon g-2 in
  the extended singlet scalar model},''
  \href{http://arxiv.org/abs/2104.03271}{{\ttfamily arXiv:2104.03271
  [hep-ph]}}.

\bibitem{Yin:2021yqy}
W.~Yin and W.~Yin, ``{Radiative lepton mass and muon $g-2$ with suppressed
  lepton flavor and CP violations},''
  \href{http://arxiv.org/abs/2103.14234}{{\ttfamily arXiv:2103.14234
  [hep-ph]}}.

\bibitem{Yin:2021mls}
W.~Yin, ``{Muon $g-2$ Anomaly in Anomaly Mediation},''
  \href{http://arxiv.org/abs/2104.03259}{{\ttfamily arXiv:2104.03259
  [hep-ph]}}.

\bibitem{Chiang:2021pma}
C.-W. Chiang and K.~Yagyu, ``{Radiative Seesaw Mechanism for Charged
  Leptons},'' \href{http://arxiv.org/abs/2104.00890}{{\ttfamily
  arXiv:2104.00890 [hep-ph]}}.

\bibitem{Escribano:2021css}
P.~Escribano, J.~Terol-Calvo, and A.~Vicente, ``{$\boldsymbol{(g-2)_{e,\mu}}$
  in an extended inverse type-III seesaw},''
  \href{http://arxiv.org/abs/2104.03705}{{\ttfamily arXiv:2104.03705
  [hep-ph]}}.

\bibitem{Zhang:2021gun}
H.-B. Zhang, C.-X. Liu, J.-L. Yang, and T.-F. Feng, ``{Muon anomalous magnetic
  dipole moment in the $\mu\nu$SSM},''
  \href{http://arxiv.org/abs/2104.03489}{{\ttfamily arXiv:2104.03489
  [hep-ph]}}.

\bibitem{Yang:2021duj}
J.-L. Yang, H.-B. Zhang, C.-X. Liu, X.-X. Dong, and T.-F. Feng, ``{Muon $(g-2)$
  in the B-LSSM},'' \href{http://arxiv.org/abs/2104.03542}{{\ttfamily
  arXiv:2104.03542 [hep-ph]}}.

\bibitem{Li:2021lnz}
T.~Li, M.~A. Schmidt, C.-Y. Yao, and M.~Yuan, ``{Charged lepton flavor
  violation in light of the muon magnetic moment anomaly and colliders},''
  \href{http://arxiv.org/abs/2104.04494}{{\ttfamily arXiv:2104.04494
  [hep-ph]}}.

\bibitem{Hernandez:2021tii}
A.~E.~C. Hern\'andez, S.~F. King, and H.~Lee, ``{Fermion mass hierarchies from
  vector-like families with an extended 2HDM and a possible explanation for the
  electron and muon anomalous magnetic moments},''
  \href{http://arxiv.org/abs/2101.05819}{{\ttfamily arXiv:2101.05819
  [hep-ph]}}.

\bibitem{Hernandez:2021mxo}
A.~E.~C. Hern\'andez, H.~N. Long, M.~L. Mora-Urrutia, N.~H. Thao, and V.~V.
  Vien, ``{Fermion masses and mixings and $g-2$ muon anomaly in a 3-3-1 model
  with $D_4$ family symmetry},''
  \href{http://arxiv.org/abs/2104.04559}{{\ttfamily arXiv:2104.04559
  [hep-ph]}}.

\bibitem{Saez:2021qta}
B.~D. S\'aez and K.~Ghorbani, ``{Singlet Scalars as Dark Matter and the Muon
  g-2 Anomaly},'' \href{http://arxiv.org/abs/2107.08945}{{\ttfamily
  arXiv:2107.08945 [hep-ph]}}.

\bibitem{Yu:2021suw}
B.~Yu and S.~Zhou, ``{General Remarks on the One-loop Contributions to the Muon
  Anomalous Magnetic Moment},''
  \href{http://arxiv.org/abs/2106.11291}{{\ttfamily arXiv:2106.11291
  [hep-ph]}}.

\bibitem{Chowdhury:2021tnm}
T.~A. Chowdhury and S.~Saad, ``{Non-Abelian Vector Dark Matter and Lepton
  $g-2$},'' \href{http://arxiv.org/abs/2107.11863}{{\ttfamily arXiv:2107.11863
  [hep-ph]}}.

\bibitem{Zhang:2021dgl}
D.~Zhang, ``{Radiative neutrino masses, lepton flavor mixing and muon g
  \ensuremath{-} 2 in a leptoquark model},''
  \href{http://dx.doi.org/10.1007/JHEP07(2021)069}{{\em JHEP} {\bfseries 07}
  (2021) 069}, \href{http://arxiv.org/abs/2105.08670}{{\ttfamily
  arXiv:2105.08670 [hep-ph]}}.

\bibitem{Perez-Martinez:2021zjj}
R.~Perez-Martinez, S.~Ramos-Sanchez, and P.~K.~S. Vaudrevange, ``{Landscape of
  promising nonsupersymmetric string models},''
  \href{http://dx.doi.org/10.1103/PhysRevD.104.046026}{{\em Phys. Rev. D}
  {\bfseries 104} no.~4, (2021) 046026},
  \href{http://arxiv.org/abs/2105.03460}{{\ttfamily arXiv:2105.03460
  [hep-th]}}.

\bibitem{Jueid:2021avn}
A.~Jueid, J.~Kim, S.~Lee, and J.~Song, ``{Type-X two Higgs doublet model in
  light of the muon $\mathbf{g-2}$: confronting Higgs and collider data},''
  \href{http://arxiv.org/abs/2104.10175}{{\ttfamily arXiv:2104.10175
  [hep-ph]}}.

\bibitem{Abi:2021gix}
{\bfseries Muon g-2} Collaboration, B.~Abi {\em et~al.}, ``{Measurement of the
  Positive Muon Anomalous Magnetic Moment to 0.46 ppm},''
  \href{http://dx.doi.org/10.1103/PhysRevLett.126.141801}{{\em Phys. Rev.
  Lett.} {\bfseries 126} no.~14, (2021) 141801},
\href{http://arxiv.org/abs/2104.03281}{{\ttfamily arXiv:2104.03281 [hep-ex]}}.

\bibitem{Xing:2020ijf}
Z.-z. Xing, ``{Flavor structures of charged fermions and massive neutrinos},''
  \href{http://dx.doi.org/10.1016/j.physrep.2020.02.001}{{\em Phys. Rept.}
  {\bfseries 854} (2020) 1--147},
  \href{http://arxiv.org/abs/1909.09610}{{\ttfamily arXiv:1909.09610
  [hep-ph]}}.

\bibitem{Zyla:2020zbs}
{\bfseries Particle Data Group} Collaboration, P.~Zyla {\em et~al.}, ``{Review
  of Particle Physics},'' \href{http://dx.doi.org/10.1093/ptep/ptaa104}{{\em
  PTEP} {\bfseries 2020} no.~8, (2020) 083C01}.

\bibitem{Sinervo:2779463}
{\bfseries ATLAS Collaboration} Collaboration, P.~Sinervo, ``{Vector-like Quark
  Searched at ATLAS},''. \url{https://cds.cern.ch/record/2779463}.

\bibitem{Sanyal:2019xcp}
P.~Sanyal, ``{Limits on the Charged Higgs Parameters in the Two Higgs Doublet
  Model using CMS $\sqrt{s}=13$ TeV Results},''
  \href{http://dx.doi.org/10.1140/epjc/s10052-019-7431-y}{{\em Eur. Phys. J. C}
  {\bfseries 79} no.~11, (2019) 913},
  \href{http://arxiv.org/abs/1906.02520}{{\ttfamily arXiv:1906.02520
  [hep-ph]}}.

\bibitem{CMS:2020osd}
{\bfseries CMS} Collaboration, A.~M. Sirunyan {\em et~al.}, ``{Search for a
  light charged Higgs boson in the H$^\pm$ $\to $ cs channel in proton-proton
  collisions at $\sqrt{s} =$ 13 TeV},''
  \href{http://dx.doi.org/10.1103/PhysRevD.102.072001}{{\em Phys. Rev. D}
  {\bfseries 102} no.~7, (2020) 072001},
  \href{http://arxiv.org/abs/2005.08900}{{\ttfamily arXiv:2005.08900
  [hep-ex]}}.

\bibitem{Kajiyama:2013rla}
Y.~Kajiyama, H.~Okada, and T.~Toma, ``{Multicomponent dark matter particles in
  a two-loop neutrino model},''
  \href{http://dx.doi.org/10.1103/PhysRevD.88.015029}{{\em Phys. Rev. D}
  {\bfseries 88} no.~1, (2013) 015029},
  \href{http://arxiv.org/abs/1303.7356}{{\ttfamily arXiv:1303.7356 [hep-ph]}}.

\bibitem{Hagiwara:2011af}
K.~Hagiwara, R.~Liao, A.~D. Martin, D.~Nomura, and T.~Teubner, ``{$(g-2)_\mu$
  and $\alpha(M^2_Z)$ re-evaluated using new precise data},''
  \href{http://dx.doi.org/10.1088/0954-3899/38/8/085003}{{\em J. Phys.}
  {\bfseries G38} (2011) 085003},
\href{http://arxiv.org/abs/1105.3149}{{\ttfamily arXiv:1105.3149 [hep-ph]}}.

\bibitem{Davier:2017zfy}
M.~Davier, A.~Hoecker, B.~Malaescu, and Z.~Zhang, ``{Reevaluation of the
  hadronic vacuum polarisation contributions to the Standard Model predictions
  of the muon $g-2$ and ${\alpha (m_Z^2)}$ using newest hadronic cross-section
  data},'' \href{http://dx.doi.org/10.1140/epjc/s10052-017-5161-6}{{\em Eur.
  Phys. J.} {\bfseries C77} no.~12, (2017) 827},
\href{http://arxiv.org/abs/1706.09436}{{\ttfamily arXiv:1706.09436 [hep-ph]}}.

\bibitem{Nomura:2018lsx}
T.~Nomura and H.~Okada, ``{One-loop neutrino mass model without any additional
  symmetries},'' \href{http://dx.doi.org/10.1016/j.dark.2019.100359}{{\em Phys.
  Dark Univ.} {\bfseries 26} (2019) 100359},
\href{http://arxiv.org/abs/1808.05476}{{\ttfamily arXiv:1808.05476 [hep-ph]}}.

\bibitem{Blum:2018mom}
{\bfseries RBC, UKQCD} Collaboration, T.~Blum, P.~A. Boyle, V.~Gülpers,
  T.~Izubuchi, L.~Jin, C.~Jung, A.~Jüttner, C.~Lehner, A.~Portelli, and J.~T.
  Tsang, ``{Calculation of the hadronic vacuum polarization contribution to the
  muon anomalous magnetic moment},''
  \href{http://dx.doi.org/10.1103/PhysRevLett.121.022003}{{\em Phys. Rev.
  Lett.} {\bfseries 121} no.~2, (2018) 022003},
\href{http://arxiv.org/abs/1801.07224}{{\ttfamily arXiv:1801.07224 [hep-lat]}}.

\bibitem{Keshavarzi:2018mgv}
A.~Keshavarzi, D.~Nomura, and T.~Teubner, ``{Muon $g-2$ and $\alpha(M_Z^2)$: a
  new data-based analysis},''
  \href{http://dx.doi.org/10.1103/PhysRevD.97.114025}{{\em Phys. Rev.}
  {\bfseries D97} no.~11, (2018) 114025},
\href{http://arxiv.org/abs/1802.02995}{{\ttfamily arXiv:1802.02995 [hep-ph]}}.

\bibitem{Aoyama:2020ynm}
T.~Aoyama {\em et~al.}, ``{The anomalous magnetic moment of the muon in the
  Standard Model},''
  \href{http://dx.doi.org/10.1016/j.physrep.2020.07.006}{{\em Phys. Rept.}
  {\bfseries 887} (2020) 1--166},
\href{http://arxiv.org/abs/2006.04822}{{\ttfamily arXiv:2006.04822 [hep-ph]}}.

\bibitem{Parker:2018vye}
R.~H. Parker, C.~Yu, W.~Zhong, B.~Estey, and H.~Müller, ``{Measurement of the
  fine-structure constant as a test of the Standard Model},''
  \href{http://dx.doi.org/10.1126/science.aap7706}{{\em Science} {\bfseries
  360} (2018) 191},
\href{http://arxiv.org/abs/1812.04130}{{\ttfamily arXiv:1812.04130
  [physics.atom-ph]}}.

\bibitem{Morel:2020dww}
L.~Morel, Z.~Yao, P.~Cladé, and S.~Guellati-Khélifa, ``{Determination of the
  fine-structure constant with an accuracy of 81 parts per trillion},''
\href{http://dx.doi.org/10.1038/s41586-020-2964-7}{{\em Nature} {\bfseries 588}
  no.~7836, (2020) 61--65}.

\bibitem{Bennett:2006fi}
{\bfseries Muon g-2} Collaboration, G.~W. Bennett {\em et~al.}, ``{Final Report
  of the Muon E821 Anomalous Magnetic Moment Measurement at BNL},''
  \href{http://dx.doi.org/10.1103/PhysRevD.73.072003}{{\em Phys. Rev. D}
  {\bfseries 73} (2006) 072003},
  \href{http://arxiv.org/abs/hep-ex/0602035}{{\ttfamily arXiv:hep-ex/0602035}}.

\bibitem{Diaz:2002uk}
R.~A. Diaz, R.~Martinez, and J.~A. Rodriguez, ``{Phenomenology of lepton flavor
  violation in 2HDM(3) from (g-2)(mu) and leptonic decays},''
  \href{http://dx.doi.org/10.1103/PhysRevD.67.075011}{{\em Phys. Rev.}
  {\bfseries D67} (2003) 075011},
\href{http://arxiv.org/abs/hep-ph/0208117}{{\ttfamily arXiv:hep-ph/0208117
  [hep-ph]}}.

\bibitem{Jegerlehner:2009ry}
F.~Jegerlehner and A.~Nyffeler, ``{The Muon g-2},''
  \href{http://dx.doi.org/10.1016/j.physrep.2009.04.003}{{\em Phys. Rept.}
  {\bfseries 477} (2009) 1--110},
\href{http://arxiv.org/abs/0902.3360}{{\ttfamily arXiv:0902.3360 [hep-ph]}}.

\bibitem{Kelso:2014qka}
C.~Kelso, H.~N. Long, R.~Martinez, and F.~S. Queiroz, ``{Connection of
  $g-2_{\mu}$, electroweak, dark matter, and collider constraints on 331
  models},'' \href{http://dx.doi.org/10.1103/PhysRevD.90.113011}{{\em Phys.
  Rev.} {\bfseries D90} no.~11, (2014) 113011},
\href{http://arxiv.org/abs/1408.6203}{{\ttfamily arXiv:1408.6203 [hep-ph]}}.

\bibitem{Lindner:2016bgg}
M.~Lindner, M.~Platscher, and F.~S. Queiroz, ``{A Call for New Physics : The
  Muon Anomalous Magnetic Moment and Lepton Flavor Violation},''
  \href{http://dx.doi.org/10.1016/j.physrep.2017.12.001}{{\em Phys. Rept.}
  {\bfseries 731} (2018) 1--82},
\href{http://arxiv.org/abs/1610.06587}{{\ttfamily arXiv:1610.06587 [hep-ph]}}.

\bibitem{Kowalska:2017iqv}
K.~Kowalska and E.~M. Sessolo, ``{Expectations for the muon g-2 in simplified
  models with dark matter},''
  \href{http://dx.doi.org/10.1007/JHEP09(2017)112}{{\em JHEP} {\bfseries 09}
  (2017) 112},
\href{http://arxiv.org/abs/1707.00753}{{\ttfamily arXiv:1707.00753 [hep-ph]}}.

\bibitem{Hernandez:2015rfa}
A.~E. Carcamo~Hernandez, S.~Kovalenko, and I.~Schmidt, ``{Precision
  measurements constraints on the number of Higgs doublets},''
  \href{http://dx.doi.org/10.1103/PhysRevD.91.095014}{{\em Phys. Rev.}
  {\bfseries D91} (2015) 095014},
\href{http://arxiv.org/abs/1503.03026}{{\ttfamily arXiv:1503.03026 [hep-ph]}}.

\bibitem{Dedes:2002er}
A.~Dedes and A.~Pilaftsis, ``{Resummed Effective Lagrangian for Higgs Mediated
  FCNC Interactions in the CP Violating MSSM},''
  \href{http://dx.doi.org/10.1103/PhysRevD.67.015012}{{\em Phys. Rev. D}
  {\bfseries 67} (2003) 015012},
  \href{http://arxiv.org/abs/hep-ph/0209306}{{\ttfamily arXiv:hep-ph/0209306}}.

\bibitem{Aranda:2012bv}
A.~Aranda, C.~Bonilla, and J.~L. Diaz-Cruz, ``{Three generations of Higgses and
  the cyclic groups},''
  \href{http://dx.doi.org/10.1016/j.physletb.2012.09.011}{{\em Phys. Lett. B}
  {\bfseries 717} (2012) 248--251},
  \href{http://arxiv.org/abs/1204.5558}{{\ttfamily arXiv:1204.5558 [hep-ph]}}.

\bibitem{Khalil:2013ixa}
S.~Khalil and S.~Salem, ``{Enhancement of $H \to \gamma\gamma$ in $SU(5)$ model
  with 45$_{H^1}$ plet},''
  \href{http://dx.doi.org/10.1016/j.nuclphysb.2013.08.016}{{\em Nucl. Phys. B}
  {\bfseries 876} (2013) 473--492},
  \href{http://arxiv.org/abs/1304.3689}{{\ttfamily arXiv:1304.3689 [hep-ph]}}.

\bibitem{Queiroz:2016gif}
F.~S. Queiroz, C.~Siqueira, and J.~W.~F. Valle, ``{Constraining Flavor Changing
  Interactions from LHC Run-2 Dilepton Bounds with Vector Mediators},''
  \href{http://dx.doi.org/10.1016/j.physletb.2016.10.057}{{\em Phys. Lett. B}
  {\bfseries 763} (2016) 269--274},
  \href{http://arxiv.org/abs/1608.07295}{{\ttfamily arXiv:1608.07295
  [hep-ph]}}.

\bibitem{Buras:2016dxz}
A.~J. Buras and F.~De~Fazio, ``{331 Models Facing the Tensions in $\Delta F=2$
  Processes with the Impact on $\varepsilon^\prime/\varepsilon$,
  $B_s\to\mu^+\mu^-$ and $B\to K^*\mu^+\mu^-$},''
  \href{http://dx.doi.org/10.1007/JHEP08(2016)115}{{\em JHEP} {\bfseries 08}
  (2016) 115}, \href{http://arxiv.org/abs/1604.02344}{{\ttfamily
  arXiv:1604.02344 [hep-ph]}}.

\bibitem{Ferreira:2017tvy}
P.~M. Ferreira, I.~P. Ivanov, E.~Jim\'enez, R.~Pasechnik, and H.~Ser\^odio,
  ``{CP4 miracle: shaping Yukawa sector with CP symmetry of order four},''
  \href{http://dx.doi.org/10.1007/JHEP01(2018)065}{{\em JHEP} {\bfseries 01}
  (2018) 065}, \href{http://arxiv.org/abs/1711.02042}{{\ttfamily
  arXiv:1711.02042 [hep-ph]}}.

\bibitem{Duy:2020hhk}
N.~T. Duy, T.~Inami, and D.~T. Huong, ``{Physical constraints derived from FCNC
  in the 3-3-1-1 model},'' \href{http://arxiv.org/abs/2009.09698}{{\ttfamily
  arXiv:2009.09698 [hep-ph]}}.

\bibitem{HuongRelic1}
N.~T. D. N. T. N. L. D.~T. D.~T.~Huong, P. V.~Dong, ``{Investigation of Dark
  Matter in the $3-2-3-1$ Model},''
  \href{http://dx.doi.org/10.1103/PhysRevD.98.055033}{{\em Phys. Rev. D}
  {\bfseries 98} (2018) 055033},
\href{http://arxiv.org/abs/1802.10402}{{\ttfamily arXiv:1802.10402 [hep-ph]}}.

\bibitem{HuongRelic2}
{\bfseries Planck} Collaboration, N.~Aghanim {\em et~al.}, ``{Planck 2018
  results. VI. Cosmological parameters},''
  \href{http://dx.doi.org/10.1051/0004-6361/201833910}{{\em Astron. Astrophys.}
  {\bfseries 641} (2020) A6}, \href{http://arxiv.org/abs/1807.06209}{{\ttfamily
  arXiv:1807.06209 [astro-ph.CO]}}.

\bibitem{Directsearch}
P.~V. Dong, C.~S. Kim, D.~V. Soa, and N.~T. Thuy, ``{Investigation of Dark
  Matter in Minimal 3-3-1 Models},''
  \href{http://dx.doi.org/10.1103/PhysRevD.91.115019}{{\em Phys. Rev. D}
  {\bfseries 91} no.~11, (2015) 115019},
  \href{http://arxiv.org/abs/1501.04385}{{\ttfamily arXiv:1501.04385
  [hep-ph]}}.

\bibitem{XENON1T}
{\bfseries XENON} Collaboration, E.~Aprile {\em et~al.}, ``{Dark Matter Search
  Results from a One Ton-Year Exposure of XENON1T},''
  \href{http://dx.doi.org/10.1103/PhysRevLett.121.111302}{{\em Phys. Rev.
  Lett.} {\bfseries 121} no.~11, (2018) 111302},
  \href{http://arxiv.org/abs/1805.12562}{{\ttfamily arXiv:1805.12562
  [astro-ph.CO]}}.

\bibitem{Queiroz}
S.~Profumo and F.~S. Queiroz, ``{Constraining the $Z'$ mass in 331 models using
  direct dark matter detection},''
  \href{http://dx.doi.org/10.1140/epjc/s10052-014-2960-x}{{\em Eur. Phys. J. C}
  {\bfseries 74} no.~7, (2014) 2960},
  \href{http://arxiv.org/abs/1307.7802}{{\ttfamily arXiv:1307.7802 [hep-ph]}}.

\bibitem{Huong2019vej}
D.~T. Huong, D.~N. Dinh, L.~D. Thien, and P.~Van~Dong, ``{Dark matter and
  flavor changing in the flipped 3-3-1 model},''
  \href{http://dx.doi.org/10.1007/JHEP08(2019)051}{{\em JHEP} {\bfseries 08}
  (2019) 051}, \href{http://arxiv.org/abs/1906.05240}{{\ttfamily
  arXiv:1906.05240 [hep-ph]}}.

\bibitem{Lepto1}
A.~Pilaftsis, ``{CP violation and baryogenesis due to heavy Majorana
  neutrinos},'' \href{http://dx.doi.org/10.1103/PhysRevD.56.5431}{{\em Phys.
  ReV. D} {\bfseries 56} (1997) 5431},
\href{http://arxiv.org/abs/9707235}{{\ttfamily arXiv:9707235 [hep-ph]}}.

\bibitem{Lepto2}
U.~S. Pei-Hong~Gu, ``{Leptogenesis with Linear, Inverse or Double Seesaw},''
  \href{http://dx.doi.org/10.1016/j.physletb.2010.09.062}{{\em Phys.Lett.B}
  {\bfseries 09} (2010) 5062},
\href{http://arxiv.org/abs/1007.2323}{{\ttfamily arXiv:1007.2323 [hep-ph]}}.

\bibitem{Dib:2019jod}
C.~Dib, S.~Kovalenko, I.~Schmidt, and A.~Smetana, ``{Low-scale seesaw from
  neutrino condensation},''
  \href{http://dx.doi.org/10.1016/j.nuclphysb.2019.114910}{{\em Nucl. Phys. B}
  {\bfseries 952} (2020) 114910},
  \href{http://arxiv.org/abs/1904.06280}{{\ttfamily arXiv:1904.06280
  [hep-ph]}}.

\bibitem{Lepto3}
T.~H. S.~Blanchet and F.~X. Josse-Michaux, ``{Reconciling leptogenesis with
  observable $\mu \to e \gamma$ rates},''
  \href{http://dx.doi.org/10.1007/JHEP04(2010)023}{{\em JHEP} {\bfseries 04}
  (2010) 023},
\href{http://arxiv.org/abs/0912.3153}{{\ttfamily arXiv:0912.3153 [hep-ph]}}.

\bibitem{Lepto4}
R.~N.~M. Steve~Blanchet, P. S. Bhupal~Dev, ``{Leptogenesis with TeV Scale
  Inverse Seesaw in SO(10)},''
  \href{http://dx.doi.org/10.1103/PhysRevD.82.115025}{{\em Phys. Rev. D}
  {\bfseries 82} (2010) 115025},
\href{http://arxiv.org/abs/1010.1471}{{\ttfamily arXiv:1010.1471 [hep-ph]}}.

\bibitem{Casas}
R.~R.~V. Matthew J.~Dolan, Tomasz P.~Dutka, ``{Dirac-Phase Thermal Leptogenesis
  in the extended Type-I Seesaw Model},''
  \href{http://dx.doi.org/10.1088/1475-7516/2018/06/012}{{\em JCAP} {\bfseries
  06} (2018) 012},
\href{http://arxiv.org/abs/1802.08373}{{\ttfamily arXiv:1802.08373 [hep-ph]}}.

\bibitem{Das:2012ze}
A.~Das and N.~Okada, ``{Inverse seesaw neutrino signatures at the LHC and
  ILC},'' \href{http://dx.doi.org/10.1103/PhysRevD.88.113001}{{\em Phys. Rev.
  D} {\bfseries 88} (2013) 113001},
  \href{http://arxiv.org/abs/1207.3734}{{\ttfamily arXiv:1207.3734 [hep-ph]}}.

\bibitem{deSalas:2020pgw}
P.~F. de~Salas, D.~V. Forero, S.~Gariazzo, P.~Mart\'\i{}nez-Mirav\'e, O.~Mena,
  C.~A. Ternes, M.~T\'ortola, and J.~W.~F. Valle, ``{2020 global reassessment
  of the neutrino oscillation picture},''
  \href{http://dx.doi.org/10.1007/JHEP02(2021)071}{{\em JHEP} {\bfseries 02}
  (2021) 071}, \href{http://arxiv.org/abs/2006.11237}{{\ttfamily
  arXiv:2006.11237 [hep-ph]}}.

\bibitem{Mohapatra}
R.~N. Mohapatra and X.~Zhang, ``{Electroweak baryogenesis in
  left-right-symmetric models},''
  \href{http://dx.doi.org/10.1103/PhysRevD.46.5331}{{\em Phys. Rev. D}
  {\bfseries 46} (1992) 5331}.

\bibitem{SFKing}
S.~F. King and T.~Yanagida, ``{Testing the see-saw mechanism at collider
  energies},'' \href{http://dx.doi.org/10.1143/PTP.114.1035}{{\em Prog. Theor.
  Phys} {\bfseries 114} (2006) 1035},
\href{http://arxiv.org/abs/0411030}{{\ttfamily arXiv:0411030 [hep-ph]}}.

\end{thebibliography}\endgroup

\end{document}